\documentclass[12pt]{article}

\usepackage{tikz}
\usepackage{empheq}

\usetikzlibrary{3d,calc}

\AtBeginDocument{
	\addtocontents{toc}{\normalsize}
}
\pdfoutput=1

\usepackage{amsmath,amssymb,amsfonts,amsthm,amscd,mathrsfs}
\usepackage{MnSymbol}
\usepackage{esvect}
\usepackage{xcolor}
\definecolor{darkblue}{rgb}{0.1,0.1,.7}
\usepackage[]{version}
\usepackage[]{graphicx}
\usepackage[]{latexsym}
\usepackage{geometry}
\usepackage[all,cmtip]{xy}

\usepackage[margin=10pt,font=small,labelfont=bf]{caption}
\usepackage{ifthen}
\usepackage{soul}
\usepackage{tikz}
\usepackage{array,mathrsfs,amsfonts,yfonts,dsfont,bbm,colonequals}
\usepackage[nodisplayskipstretch]{setspace}
\usepackage{dsfont}
\usepackage{cite}
\usepackage{xspace}
\usepackage{empheq}
\usepackage{extarrows}
\usepackage{braket}
\usepackage[normalem]{ulem}

\usepackage{xcolor}
\usepackage[colorlinks, linkcolor=darkblue, citecolor=darkblue, urlcolor=darkblue, linktocpage]{hyperref} 

\usepackage{subcaption}
\usepackage[utf8]{inputenc}
\usepackage{setspace}
\usepackage{float, graphicx}
\usetikzlibrary{shapes.geometric}
\tikzset{
  vtx/.style={
    circle,
    draw=blue,
    fill=blue,
    inner sep=1pt
  },
  wcirc/.style={
    circle,
    draw=white,
    fill=white,
    inner sep=2pt
  },
  bcirc/.style={
    circle,
    draw=black,
    fill=black,
    inner sep=1pt
  },
  dcirc/.style={
    circle,
    draw=blue,
    fill=blue,
    inner sep=1pt
  },
  rcirc/.style={
    circle,
    draw=red,
    fill=red,
    inner sep=1pt
  },
  phi/.style={
    thick
  },
  sigma/.style={
    thick,
    dashed
  },
  vl1/.style={
    thick,
    blue
  },
  vl2/.style={
    thick,
    dashed,
    blue
  },
  valign/.style={
    baseline={([yshift=-.55ex]current bounding box.center)}
  }
}

\numberwithin{equation}{section}

\newcommand{\normd}{C_{\psi}}
\newcommand{\dino}{\psi}

\newcommand{\hdO}{\Delta_{\hat{\mathcal O}}}
\newcommand{\hcO}{{\hat{\mathcal O}}}
\newcommand{\cO}{\mathcal O}

\newcommand{\reef}[1]{(\ref{#1})}
\newcommand{\be}{\begin{equation}}
\newcommand{\ee}{\end{equation}}
\newcommand{\bea}{\begin{eqnarray}}
\newcommand{\eea}{\end{eqnarray}}
\newcommand{\ba}{\begin{equation}\begin{aligned}}
\newcommand{\ea}{\end{aligned}\end{equation}}

\newcommand{\ud}{\mathrm d}

\newcommand{\Df}{{\Delta_\phi}}

\def\mk{{\mathcal{K}}}

\def\md{{\mathcal{D}}}

\newcommand{\veps}{\varepsilon}
\newcommand{\blambda}{\bar{\lambda}}
\newcommand{\op}{\mathcal{O}}
\newcommand{\disp}{D}
\newcommand{\tilt}{\mathcal{T}}
\newcommand{\bulk}{\mathcal{X}}

\usepackage{cancel}

\newcommand{\normt}{C_t}
\newcommand{\normD}{C_D}
\newcommand{\normO}{C_\op}

\newcommand*\pFqskip{8mu}
\catcode`,\active

\newcommand*\pFqt{\begingroup
        \catcode`\,\active
        \def ,{\mskip\pFqskip\relax}%
        \dopFqt
}

\catcode`\,12

\catcode`\,12
\def\dopFqt#1#2#3#4#5{%
        {}_{#1}\tilde F_{#2}\biggl[\genfrac..{0pt}{}{#3}{#4};#5\biggr]%
        \endgroup
}

\begin{document}
	
	\vspace*{-.6in} \thispagestyle{empty}
	\begin{flushright}
		\end{flushright}
	%%
	%% Title
	%%
	\vspace{1cm} {\Large
		\begin{center}
			{\bf Consequences of symmetry-breaking on conformal defect data}\\
	\end{center}}
	\vspace{1cm}
	%%%
	%% Authors
	%%%
	\begin{center}
		{\bf Bastien Girault$^t$, Miguel F.~Paulos$^\psi$, Philine van Vliet$^D$ }\\[1cm] 
		{
			\small
			{\em Laboratoire de Physique, \'Ecole Normale Sup\'erieure, \\
   Universit{\'e} PSL, CNRS, Sorbonne Universit{\'e}, Universit{\'e} Paris Cit{\'e}, \\
24 rue Lhomond, F-75005 Paris, France}

			\normalsize
		}
		
	\end{center}
	
	\begin{center}
		{ 
        ${}^t$\texttt{\  bastien.girault@ens.fr}\\ 
        ${}^\psi$ \texttt{\  miguel.paulos@ens.fr}\\ 
        ${}^D$\texttt{\  philine.vanvliet@ens.fr}
		}
		\\
	\end{center}
	
	\vspace{8mm}
	
	\begin{abstract}
 \vspace{2mm}
   \noindent Conformal defects spontaneously break part of the symmetry algebra of a bulk CFT. We show that the broken Ward identities imply very general sum rules on the defect CFT data as well as on the DOE data of bulk operators, which we call defect soft theorems. Our derivation is elementary, allowing us to easily reproduce and generalize constraints on displacement and tilt operators previously obtained in the literature as well as a plethora of new ones, including constraints on bulk-defect correlation functions. For line defects we rewrite constraints in dispersive sum rule form, showing they lead to exact, optimal bounds on the OPE data of the defect. We test these sum rules in concrete perturbative examples, finding perfect agreement with existing calculations and making new predictions for various dCFT data.   
	\end{abstract}
	\vspace{2in}

	\newpage
	
	{
		\setlength{\parskip}{0.05in}
		\tableofcontents
		\renewcommand{\baselinestretch}{1.0}\normalsize
	}
		
	\setlength{\parskip}{0.1in}
 \setlength{\abovedisplayskip}{15pt}
 \setlength{\belowdisplayskip}{15pt}
 \setlength{\belowdisplayshortskip}{15pt}
 \setlength{\abovedisplayshortskip}{15pt}

\bigskip \bigskip
\section{Introduction}

    Symmetries play a central role in modern physics, allowing us to construct and classify theories, constrain observables, and to explain a wide range of phenomena. But the way in which such symmetries are broken is equally important, and comes with its own set of constraints. 
    A famous example arises in QCD (with zero mass quarks), where the chiral symmetry of the theory  is spontaneously broken. This leads to exactly massless particles -- pions -- associated to the lack of conservation of the the axial current,
\begin{equation}
    \partial^{\mu} J_{\mu}^{\tilde{a}}(x) \propto \pi^{\tilde{a}} (x)\:.
\end{equation}
Furthermore it is well known that this leads to relation on scattering amplitudes of pions at zero momentum, such as the existence of the Adler zero~\cite{Adler65} and its generalization for multiple pions in the form of soft pion constraints~\cite{Weinberg66}. 

In this work we point out that these results have a direct analog for conformal field theories in the presence of conformal defects \cite{McAvity:1993ue,Billo:2016cpy}. 
 The presence of such defects partially breaks the bulk (super)conformal group and any additional bulk flavor symmetry. 
 This breaking manifests as non-conservation of the stress tensor or symmetry currents in the bulk and leads to ``Nambu-Goldstone" fields on the defect with protected scaling dimension. Schematically,
\ba
\partial_\mu T^{\mu a} &=\delta_{\mathcal D} D^a\,,\\
\partial_\mu J^{\mu,i} &=\delta_{\mathcal D} t^{i}\,,\\
\partial_\mu \mathcal J^{\mu}_\alpha &=\delta_{\mathcal D} \psi_\alpha \,,
\ea
with the breaking terms localized on the defect by the delta function $\delta_\mathcal D$. Here  $D^a,t^i$ are the so-called displacement and tilt operators respectively, associated to breaking of translation and flavor symmetries, and $\psi_{\alpha}$ what we name the \textit{displacino} which corresponds to the breaking of supersymmetry.

In this work we show it is possible to derive ``defect soft theorems'' associated to these defect operators. These follow very directly from the broken Ward identities above and take the form of integrated correlator constraints where the correlator needs to at least contain one of these protected defect operators, but can contain an arbitrary number of bulk- and defect operators in addition. This framework leads to two general identities, equations \reef{eq:id1} and \reef{eq:id2}, which essentially arise by considering insertions of one or two broken charges into a correlation function of bulk operators. These identities can be applied to very general patterns of symmetry breaking by conformal defects, and for any correlation function of bulk and defect operators. We show that they can be used in an elementary fashion to derive integral constraints both on defect four point functions involving protected operators as well as on `bulk form factors', namely bulk-defect-defect three-point functions. While most of these results are new, special cases of our defect soft theorems were obtained in the literature using different methods, allowing us to cross-check our derivations. 
\footnote{
For a four-point function of defect operators with a tilt, understood as a marginal operator on the defect, the ``double soft'' integrated correlator constraints were first derived in 
\cite{Kutasov:1988xb,Friedan:2012hi} for 1d and 2d CFTs, and the single soft in \cite{Behan:2017mwi,Bashmakov:2017rko} for any $d$.
For the displacement, defect four-point function constraints were derived for line defects Chern-Simons theories in \cite{Gabai:2022vri} and recently for line defects in general CFTs in \cite{Gabai:2025zcs} in the case of line defects. Relations for bulk-defect two-point functions have been found for the tilt \cite{Padayasi:2021sik} and the displacement \cite{Billo:2016cpy}. See also \cite{Belton:2025hbu,Drukker:2022pxk,Cavaglia:2022yvv,Cavaglia:2022qpg,Ferrando:2025ufj} for related works.} 

Having derived these constraints, we move on to another of our key results. We show that for form factors in general dimension, as well as for line defect correlators involving identical operators, we can derive from them certain ``dispersive'' sum rules on the CFT data by using dispersive representations of the correlators. These improved sum rules immediately lead to rigorous and tight bounds on defect data:
\vspace{0.3cm}

\setlength\fboxrule{1.2pt}
\setlength{\fboxsep}{8pt} 
{\hspace{-18pt}\noindent\fbox{\parbox{\textwidth-18pt}{
{\bf Spectral bounds on line defects}
\begin{itemize}
\item Conformal line defects must have an operator of dimension $\Delta$ either a singlet or traceless symmetric tensor under transverse rotations satisfying
\ba
\Delta\in [1,2] \cup [3,4]\:.
\ea
\item If the defect breaks a global symmetry then there must exist an even parity operator with dimension $\Delta$ satisfying
\ba
\Delta\in [1,2]\:.
\ea
\end{itemize}
\vspace{-0.3cm}
}}}
   \vspace{0.3cm} 
\noindent

We verify our sum rules in the specific example of the pinning line defect of the bulk $O(N)$ CFT. This line defect, also known as a localized magnetic field line defect \cite{Allais:2014fqa,Cuomo:2021kfm}, flows to a defect CFT in the IR which breaks the $O(N)$ global symmetry to an $O(N-1)$ symmetry on the defect, resulting in the appearance of tilt operators as well as displacement operators. It~is one of the simplest non-trivial defects breaking global and conformal symmetry one can write down, and has been widely studied in the literature in the $d = 4-\varepsilon$ expansion \cite{Allais:2014fqa,Cuomo:2021kfm,Gimenez-Grau:2022czc,Nishioka:2022qmj,Gimenez-Grau:2022ebb,Bianchi:2022sbz,Belton:2025hbu}, using the numerical conformal bootstrap \cite{Gimenez-Grau:2022czc}, and through Monte Carlo simulations \cite{allais2014,Parisen_Toldin_2017}. Here we will use this example to check our sum rules and conversely derive new, higher-order predictions in the $d=4-\varepsilon$ expansion for certain defect CFT data. In order to do this, we will supplement the results in the literature with newly derived results for bulk-defect correlators to first order in $\varepsilon$. 

Although we have specialized to constraints on four-point functions involving these special defect operators, as well as on correlation functions containing one or two defect operators and a bulk operator, there are several generalisations that would be important to work out in the near future, most notably on higher point defect correlators as well as form factors involving the bulk stress-tensor. More generally we expect that the results given in this paper will be useful to bootstrap numerically a plethora of interesting defect CFTs, such as the pinning defect considered here, the $\mathcal N=4$ 1/2 BPS line defect \cite{Cooke:2017qgm,Giombi:2017cqn,Liendo:2018ukf,Cavaglia:2021bnz,Artico_2025}, the 3d Ising monodromy defect \cite{Gaiotto:2013nva}, and line defects corresponding to flux tubes in 3d Yang-Mills theory in AdS \cite{Gabai:2025hwf}.

The plan of the paper is as follows. In section 2 we discuss  our conventions for defect kinematics and derive the basic soft and double soft theorems in their general form. Sections 3, 4 and 5 then apply these general identities to the cases of tilts, displacements and the displacino respectively, deriving a multitude integral constraints on defect correlators and bulk-defect-defect form factors. In section 6 we specialize to line defects and show how integrated constraints can be turned into dispersive sum rules on defect data by working with `master functionals', or equivalently using the Polyakov/local block expansions. These improved sum rules immediately imply the simple but tight bounds evoked above.  Section 7 is concerned with perturbative computations in the context of the pinning line defect of the $O(N)$ CFT, while in section 8 we check both that this data satisfies our sum rules, which are then used to make make new perturbative predictions. Several technical appendices complete this work.

\vspace{0.3cm}\noindent
{\bf Note:} While this paper was being completed we became aware of work in progress by Drukker, Kravchuk and Kong \cite{Drukker:2025dfm} and by Belton and Kong \cite{Belton:2025ief} using different methods whose results partially overlap with ours. The paper \cite{Gabai:2025hwf} also appeared where the displacement sum rules of \cite{Gabai:2025zcs} were rederived using Ward identity techniques after private communication.

\section{Defects and broken symmetries}
\subsection{Kinematics and correlators}
In this section we introduce our setup and notation. We consider conformal defects extended along $p$ dimensions in a $d$-dimensional CFT. Spacetime is described by the Euclidean metric
\ba
\ud s^2=\sum_{m=1}^p \ud x^m \ud x^m+\sum_{a=1}^{d-p} \ud y^a \ud y^a \:.
\ea
We take the defect to be located at $y=0$. The presence of the defect breaks the bulk $d$-dimensional conformal symmetry onto a subgroup. Furthermore, if the bulk CFT has an additional global symmetry $G$ then the defect can either preserve or break this symmetry. We have
\begin{equation}
    SO(d+1,1)\times G \to SO(p+1,1) \times SO(d-p) \times \hat{G}\:.
\end{equation}
Here $SO(p+1,1)$ is the $p$ dimensional group of conformal transformations preserved by the defect, while the $SO(d-p)$ factor corresponds to rotations transverse to the defect, which we call the transverse spin. Note that it is possible for defects to break this $SO(d-p)$ symmetry, or only preserve it in a combination with another global symmetry. Here for simplicity we do not consider such cases. We also assume defects preserve parity and inversion symmetry, although generalizations should be straightforward.

\subsubsection*{Defect correlators}
The defect supports local operators $\hat{\cO}(x)$ (denoted with a hat). They carry quantum numbers corresponding to conformal transformations on the defect, transverse spin, and may also be in non-trivial representations of the preserved symmetry group $\hat{G}$. Correlators of defect operators behave as those in an ordinary CFT and in particular they satisfy an OPE,\footnote{Our definitions of $\lambda$ and $\mu$ are such that they do not depend on the choice of normalisation of defect/bulk operators, but here we give them with unit two point function.}
\ba
    \hat{\cO}_a\times \hat{\cO}_b=\sum_{c} {\lambda}_{abc} \hat{\cO}_c \:.
\ea
In our conventions, four-point defect correlators are generically written as 
\ba
    &\langle \hat\cO_1(x_1) \hat\cO_2 (x_2) \hat\cO_3 (x_3) \hat\cO_4(x_4)\rangle=\mk_{1234} (x) {\mathcal{G}}_{1234}(z,\bar z)\:, \\
    &\mk_{1234} (x) = \frac{1}{{x}_{13}^{\Delta_1 + \Delta_3} {x}_{24}^{\Delta_2 + \Delta_4}} \left(\frac{x_{34}}{x_{14}}\right)^{\Delta_{13}} \left(\frac{{x}_{14}}{{x}_{12}} \right)^{\Delta_{24}}\:, \quad x_{ij}=x_i-x_j,\quad \Delta_{ij} = \Delta_i - \Delta_j\:, \label{eq:corr}
\ea
where as usual
\ba
    z\bar z=\frac{x_{12}^2 x_{34}^2}{x_{13}^2 x_{24}^2}\,, \qquad (1-z)(1-\bar z)=\frac{x_{14}^2 x_{23}^2}{x_{13}^2 x_{24}^2}\:. \label{eq:zzbardef}
\ea
We will also be interested in the case where defect operators are charged under an $O(n)$ type symmetry. This could either correspond to transverse rotations or an internal symmetry. In particular, when all operators live in the fundamental representation $\hat{\mathcal  O}\to \hat{\mathcal O}_a$ we write
\ba
\label{eq:4pt-channel-split}
{\mathcal{G}}\to \delta_{ab}\delta_{cd}\, {\mathcal{G}}_S+\frac 12\left(\delta_{ac}\delta_{bd}+\delta_{ad}\delta_{bc}-\frac{2}{n}\delta_{ab}\delta_{cd}\right) {\mathcal{G}}_T+\frac 12\left(\delta_{ad}\delta_{bc}-\delta_{ac}\delta_{bd}\right){\mathcal{G}}_A
\ea
with $S,T,A$ corresponding to exchanges of operators in the OPE in the singlet, traceless symmetric and antisymmetric irreps of $O(n)$.
In our conventions the crossing equations take the form:
\ba
    {\mathcal{G}}_{1234}(z,\bar z)&=\left|\frac{z}{1-z}\right|^{\Delta_{24}}{\mathcal G}_{4123}(1-z,1-\bar z) \:,\\
    \Leftrightarrow 
    \sum_{\hcO} {\lambda}_{12\hcO} \lambda_{43\hcO} G^{12,34}_{\Delta_\hcO,{j}_\hcO}(z,\bar z)&=\left|\frac{z}{1-z}\right|^{\Delta_{24}}\,\sum_{\hcO} \lambda_{41\hcO} \lambda_{32\hcO} G^{41,23}_{\Delta_\hcO,{j}_\hcO}(1-z,1-\bar z)\:\,,
\ea
where in the second line we used the OPE to express the correlator in terms of conformal blocks, which depend on the conformal dimension ${\Delta}$ of the operator exchanged in the OPE, as well as its spin ${j}$. For $O(n)$ invariant correlators as above it is useful to introduce the combinations:
\ba
\mathcal G^{(1)}_{+}:=\mathcal G_{aaaa}&=\mathcal G_S+\left(\frac{n-1}n\right)\, \mathcal G_T\,, \\
\mathcal G^{(2)}_{+}:=\,\mathcal G_{abab}&=\frac 12 \mathcal G_T-\frac 12\mathcal G_A\,, \\ \mathcal G_{-}:=\mathcal G_{aabb}-\mathcal G_{abba}&=\mathcal G_S-\left(\frac{n+2}{2n}\right)\mathcal G_T-\frac 12 \mathcal G_A
\ea
which are (anti)crossing symmetric depending on the $(-)+$ label.

In applications we will often focus on the special case of a conformal line defect. In this case the defect conformal group amounts to $SL(2,\mathds R)$ and defect four-point functions now depend only on a single cross-ratio,
\ba
{\mathcal{G}}(z,z)\equiv {\mathcal{G}}(z)\,, \qquad z=\bar z=\frac{x_{12} x_{34}}{x_{13} x_{24}}
\ea
with 1d conformal blocks:
\begin{equation}
    G_{\Delta}^{1d; 12,34} = z^{\Delta-\Delta_1-\Delta_4} \,   _2 F_1 \Big(\Delta - \Delta_{12}, \Delta + \Delta_{34}, 2 \Delta; z \Big)\:, \quad \Delta_{ij} = \Delta_i - \Delta_j\:.
\end{equation}
Note that there is no spin in $1d$.

\subsubsection*{Bulk insertions}   
In this work we will consider, besides correlators of only defect operators, also correlators involving one bulk insertion. It is worth pointing out that we can always choose to work with bulk insertions that have definite quantum numbers under transverse rotations by Kaluza-Klein reduction. The reason this is a useful thing to do is that in any correlator involving only one bulk together with a finite number of defect operators with definite transverse spin charges, only a finite number of such KK modes actually contribute. For instance, if the two defect operators have transverse spin zero, then only the s-wave mode of the bulk operator is relevant, i.e. $\int \ud n\, \Psi(\hat x,r,n)$.

Bulk CFT operators~$\Psi$ can be represented in terms of defect operators when brought sufficiently close to the defect using the defect operator expansion (DOE):
\ba
    |y|^{\Delta_{\Psi}} \Psi(x,y)\underset{|y|\to 0}= \sum_{\hcO} |y|^{\Delta_{\hcO}} \mu^{\Psi}_{\hcO}\, \hcO(x)+\ldots
\ea
where the dots include sums over operators with higher charges under transverse rotations. The DOE fixes correlators with one bulk and one defect insertion. For instance, if the latter has no transverse spin we have
\ba
\label{eq:bulk-defect}
\langle \hcO(x_1) \Psi(x_2,y)\rangle=\mu^{\Psi}_{\hcO}\,K^{\Psi}_{\hcO}(y,x_1,x_2)\,,\qquad K^{\Psi}_{\hcO}(y,x_1,x_2):=\frac{|y|^{\Delta_\hcO-\Delta_{\Psi}}}{(y^2+x_{12}^2)^{\Delta_{\hcO}}}\, .
\ea
Another case which will be useful later is when the bulk field is a current. Note that in this case the bulk current does not have canonical normalization, and we write
\ba
\label{eq:JOm}
\langle J_\mu(y,x_1) \hcO_m(x_2) \rangle=\mathbb T_{\mu m}^{J\op_m}\,K^{J}_{\hcO}(y,x_1,x_2)\sqrt{C_J} \mu^{J}_{\cO}\:,
\ea
with $C_J$ the normalization of the current bulk two-point function\footnote{Concretely, such that at short distances we have 
\ba
\langle J_{\mu}(x) J_\nu(0)\rangle =\frac{C_J}{x^{2(d-1)}}\, \left(\eta_{\mu \nu}-2\frac{x_\mu x_\nu}{x^2}\right)
\ea
} and the structure\footnote{This is $Q_\text{BD}^0$ in the convention of~\cite{Billo:2016cpy}.}
\begin{equation}
\label{eq:structure-JOm}
    \mathbb T^{J\op_m}_{mn}= \delta_{mn}-2\frac{x_{12m}\,x_{12n}}{y^2+x_{12}^2}\:, \qquad \mathbb T^{J \op_m}_{an} = -2\frac{y_{a}\,x_{12n}}{y^2+x_{12}^2}\:.
\end{equation}

In the case of a one-dimensional defect we should think of $\hcO_m$ as a pseudoscalar, i.e. odd under defect parity. Exact conservation precludes a non-zero two point function between the current and a defect scalar, but there is an exception if the defect breaks the symmetry associated to the current as we will see in section \ref{sec:tiltSRApp}.

We will also consider a class of  correlators with one bulk and two defect insertions, which we will call {\em form factors}. A generic form factor is written
\ba
&\langle \hat{\cO}_1({x}_1) \hat{\cO}_2({x}_2) \Psi({x}_3,y)\rangle = \mathcal K^{\Psi}_{\hat\cO_1 \hat\cO_2}(x_i,y)
    \mathcal H^{\Psi}_{\hat\cO_1\hat\cO_2}(w)\:, \\
&\mathcal K^{\Psi}_{\hat\cO_1 \hat\cO_2}(x_i,y):=
\frac{|y|^{-\Delta_\Psi}}{{x}_{12}^{\Delta_1+\Delta_2}}\,
    \left(
    \frac {y^2+{x}_{23}^2}{y^2+{x}_{13}^2}
    \right)^{\frac{\Delta_{12}}2} \,,\label{eq:formf}
\ea
with the cross-ratio
\ba
\label{eq:cross-ratio}
 w=\frac{y^2 {x}_{12}^2}{(y^2+{x}_{13}^2)(y^2+{x}_{23}^2)}\:.
\ea
The DOE now implies
\ba
\label{eq:formf-blockexp}
    \mathcal H^{\Psi}_{\hcO_1\hcO_2}(w)=\sum_{c} \mu^{\Psi}_{c} \lambda_{12c}\, H_{\Delta_c}(w)\:,
\ea
where $H_{\Delta} (w)$ are conformal blocks, which for a defect of dimension $p$, are given by
\ba
    H_{\Delta}(w)= w^{\frac{\Delta}{2}} \,_2 F_1 \left(\frac{\Delta + \Delta_{12}}{2} , \frac{\Delta - \Delta_{12}}{2}; \Delta + 1 - \frac{p}{2}; w \right)\:.\label{eq:BDDblocks}
\ea
Note that the blocks are independent of the bulk dimension and codimension. 

When the bulk operator is a conserved current (of conformal dimension $\Delta_J=d-1$) we now have 
\ba
\langle \hat\cO_1( x_1) \hat\cO_2( x_2) J_\mu(x_3,y)\rangle= \mathcal K^{J}_{\hat\cO_1 \hat\cO_2}(x_i,y)
    \left[ \mathbb{T}_\mu^{(+)}
    \mathcal H^{J,(+)}_{\hat\cO_1\hat\cO_2}(w)+\mathbb{T}_\mu^{(-)}
    \mathcal H^{J,(-)}_{\hat\cO_1\hat\cO_2}(w)\right]\:.\label{eq:formfv}
\ea
The two possible structures are
\begin{equation}
\label{eq:structure-JOO}
    \mathbb{T}_m^{(\pm)} = -|y|\left(\frac{ (x_{13})_{m}}{y^2+ x_{13}^2}\pm\frac{ (x_{23})_{m}}{y^2+ x_{23}^2}\right)\:, \quad \mathbb{T}_{a}^{(\pm)} = y_a|y|\left(\frac{1}{y^2+ x_{13}^2}\pm\frac{1}{y^2+ x_{23}^2}-\frac{\delta_{\pm,+}}{y^2}\right)\:,
\end{equation}
The DOE implies
\ba
    \mathcal H^{J,(\pm)}_{\hat\cO_1\hat\cO_2}(w)=\sum_{c} \mu^{J,(\pm)}_{c} \lambda_{12c}\, H^{(\pm)}_{\Delta_c}(w)\:,
\ea
and the conformal blocks are modified to
\begin{align}
    H^{(+)}_\Delta(w) &= w^\frac{\Delta+1}2 {}_2F_1\Big(\tfrac{\Delta+\Delta_{12}+1}2,\tfrac{\Delta-\Delta_{12}+1}2;\Delta+2-\tfrac p2;w\Big) \:, \\
    H^{(-)}_\Delta(w) &= \frac12 w^\frac{\Delta-1}2 \left[{}_2F_1\Big(\tfrac{\Delta+\Delta_{12}+1}2,\tfrac{\Delta-\Delta_{12}-1}2;\Delta+1-\tfrac p2;w\Big) + (1\leftrightarrow2) \right]\:.
\end{align}
For a conserved current, these two parts are not independent, but satisfy the relation
\begin{equation}
    \left(2(w-1)w\, \frac{\partial}{\partial w}+p\right)\mathcal H^{J,(+)}_{\hat\cO_1\hat\cO_2}(w) = \Delta_{12}\, w\,\mathcal H^{J,(-)}_{\hat\cO_1\hat\cO_2}(w)\:.
\end{equation}
The homogeneous solution of this equation (so with no RHS) is
\begin{equation}
    \mathcal H^{J,(+),0}_{\hat\cO_1\hat\cO_2}(w)= \mathcal N\left(\frac{w}{1-w}\right)^\frac p2 \:.
\end{equation}
This solution actually produces a delta function source localized on the~defect:
\ba
\langle \hat\op_1( x_1)  \hat\op_2( x_2) \partial_\mu J^\mu( x_3,y)\rangle\supset \delta^{(d-p)}(y)\, \frac{\mathcal N\, V_{d-p-1}}{ x_{12}^{ \Delta_1+ \Delta_2-p} x_{13}^{ \Delta_1+p- \Delta_2} x_{23}^{ \Delta_2+p- \Delta_1}}\:.
\ea
Thus for a conserved current we should set $\mathcal N=0$. As we will see in the section \ref{sec:tiltSRApp} as well as in the next subsection, if the global symmetry associated to the current is broken by the defect then conservation is modified as $\partial J=\delta^{(d-p)}\, t$ for an operator $t$ called the tilt of protected dimension $p$. The above is perfectly consistent with this, setting:
\ba
\mathcal N=\lambda_{\hcO_1 \hcO_2 t} / V_{d-p-1}\,,
\ea
with $V_k$ the volume of a $k$-dimensional sphere.

\subsection{Broken Ward identities}

Denote the continuous symmetry group of the bulk CFT as $G$ and the corresponding symmetry generators by $Q^{S}$, with algebra
\begin{equation}
  [Q^S,Q^T]=f^{ST}_{\ \ \ U}\, Q^U\:.  
\end{equation}
The algebra includes the conformal generators as well as generators for any additional continuous internal symmetry that the CFT may have.
The charges evaluated on an oriented hypersurface $\Sigma$ are obtained by integrating some current $\mathcal J^A$,
\ba
Q^S_{\Sigma}=\int_\Sigma *\mathcal J^S \:. \label{eq:qj}
\ea
 We denote the variation of a generic field $\Phi$ under the action of a generator as 
\begin{equation}
    [Q^S,\Phi]=\delta_{S} \Phi\: \qquad \Rightarrow \qquad \partial_\mu \mathcal J^{S,\mu}(x) \Phi(0)=\delta^{(d)}(x)\, \delta_S \Phi(0)\:.
\end{equation}
In the presence of the defect the symmetry is necessarily broken. Let us denote the broken generators by $Q^{\tilde s}$ and the unbroken ones as $Q^{s}$. By definition the latter form a closed subalgebra. 
We choose a basis of broken generators such that
\begin{equation}
\begin{aligned}
    [Q^{\tilde s},Q^{t}]&=f^{\tilde s t}_{\ \  \tilde u} Q^{\tilde u}\\
    [Q^{\tilde s},Q^{\tilde t}]&=f^{\tilde s \tilde t}_{\ \  u} Q^u+f^{\tilde s \tilde t}_{\ \  \tilde u} Q^{\tilde u}\:.
\end{aligned}
\end{equation}
We then have
\ba
\label{eq:WIcurrent1}
\partial_\mu \mathcal J^{\tilde s,\mu}( x,y)=\delta^{(d-p)}(y) \tilt^{\tilde s}(x)\:,
\ea
as well as
\ba
    \lim_{|y|\to 0} |y|^{d-p-2} y^a \mathcal J^{\tilde s}_{a}( x,y)=\, \tilt^{\tilde s}({x})/V_{d-p-1} \label{eq:push}
\ea
for some operator $\tilt^{\tilde s}$ of protected (defect) scaling dimension.

We denote the vacuum in the presence of the defect as $|0\rangle_\md$. We then have generically:
\ba
    Q^s\ket{0}_{\md}=0\,, \qquad Q^{\tilde s}\ket{0}_{\md} \neq 0\:.
\ea
We will be interested in computing charges on hypersurfaces which cross the defect, see figure \ref{contour}. To make the charge operator well defined a priori requires modifying its expression \reef{eq:qj} by possible relevant defect operators with appropriate quantum numbers. This can always be done so below for simplicity we will simply assume no such operators are present. Using the integrated form of the broken Ward identity \eqref{eq:WIcurrent1}, we can write 
\ba
    Q^{\tilde s}\ket{0}_{\md} = \int_{
     x^0<T} \ud^p x\, \ud^{d-p}y\, \partial_{\mu} \mathcal J^{\tilde s,\mu}\ket{0}_{\md} = \int_{ x^0<T}\ud^p  x\, {\tilt}^{\tilde s}( x)\ket{0}_{\md}\:, \label{eq:qond}
\ea
where we computed the charge on a hypersurface of fixed time $ x^0=T$. We can deduce the action of the unbroken generators on the tilt:
\begin{equation}
    [Q^t,{\tilt}^{\tilde s}]=f^{t\tilde s}_{\ \  \tilde u} {\tilt}^{\tilde u}\:,
\end{equation}
In general we cannot say how the broken generators act on defect operators, but we do know how they act on bulk insertions as this is unmodified by the presence of the defect. Consider now the following correlator:
\ba
\langle  \bulk_L Q^{\tilde s} \bulk_R \rangle, 
\ea
Here $\bulk_{L,R}$ represent collections of bulk operator insertions to the past and future of time~$T$ where we insert $Q^{\tilde a}$, e.g. $\bulk_{L}= \op_1 (x_1) \ldots \op_i (x_n)$ with $x_i^0>T$.
Expressing this charge in terms of the integrated current on a fixed time contour we can either close this contour to the future or the past. Equating these two and using the broken Ward identity give us our: 
\begin{figure}[t]
\centering
\begin{tikzpicture}[x={(0.95cm,0)},y={(0cm,1cm)},z={(0.5cm,0.5cm)}]

% Time axis (shifted left)
\draw[thick] (0,2,0) -- (0,5,0) node[above] {$x^0$};
\draw[thick] (0,-0.5,0) -- (0,1,0);
\draw[->] (0,4.5,0) -- (0,5,0);

\fill[green!15,opacity=0.5] (2,1,0) ellipse (1.8cm and 0.6cm);
\node at (4.2,1,0) {$\bulk_R$};

% Crosses in Chi_R
\node at (2.7,0.9,-0.4) {$\times$};
\node at (1.0,1.1,-0.2) {$\times$};

% Slanted transverse plane at x^0=T
\filldraw[fill=blue!10,opacity=0.5]
   (-3.3,2,-1.5) -- (3.3,2,-1.5) -- (3.3,2,1.5) -- (-3.3,2,1.5) -- cycle;
\node at (-2.3,2,0) {$x^0=T$};

% Shaded ovals for Chi_L and Chi_R
\fill[red!15,opacity=0.5] (2,3,0) ellipse (1.8cm and 0.6cm);
\node at (4.2,3,0) {$\bulk_L$};

% Crosses in Chi_L
\node at (2.3,3.2,0.2) {$\times$};
\node at (1.3,2.9,-0.3) {$\times$};
\node at (2.9,3.1,-0.5) {$\times$};
\end{tikzpicture}
\caption{Insertion of charge operator along the codimension-1 hypersurface $x^0=T$. }
\label{contour}
\end{figure}
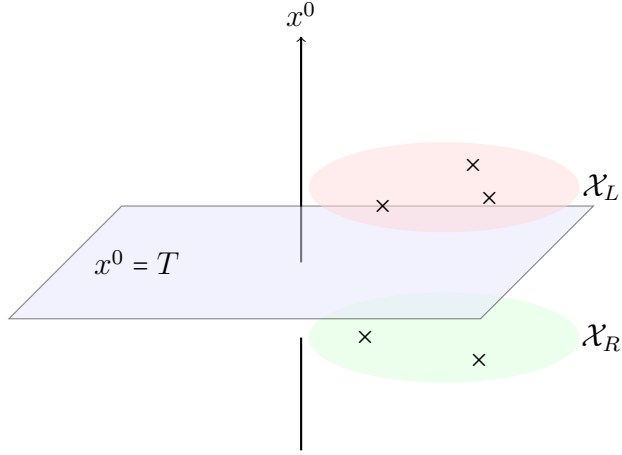

\setlength\fboxrule{1.2pt}
\setlength{\fboxsep}{8pt} 
{\hspace{-18pt}\noindent\fbox{\parbox{\textwidth-18pt}{
{\bf First identity (defect soft theorem):}
     \begin{equation}\label{eq:id1}
    \qquad \int \ud^p x\, \langle \tilt^{\tilde s}(x) \bulk \rangle =-\langle \delta_{\tilde s}\bulk \rangle\:. \qquad
       \end{equation}
\vspace{-0.3cm}
}}}
   \vspace{0.3cm} 

\noindent This is our first important identity. It has been known in the literature and appeared explicitly for the case of the displacement \cite{Billo:2016cpy} and the tilt \cite{Padayasi:2021sik}. Notice that the integral runs over the entire defect, i.e. the specific choice of time $T$ has decoupled. Since all insertions in $\bulk$ lie in the bulk the integral is perfectly convergent. Next, we consider 
\ba
    \langle \bulk_L [Q^{\tilde s}, Q^{\tilde t}]\bulk_R\rangle=f^{\tilde s \tilde t}_{\ \ u}\langle \bulk_L Q^u \bulk_R\rangle+f^{\tilde s \tilde t}_{\ \  \tilde u} \langle \bulk_L Q^{\tilde u} \bulk_R\rangle\:. \label{eq:id2}
\ea
Here the charges are inserted at time $T$, bulk $\bulk_L$ insertions have times to the future of $T$ and $\bulk_R$ insertions to the past. 
This gives \\

\setlength\fboxrule{1.2pt}
\setlength{\fboxsep}{8pt} 
{\hspace{-18pt}\noindent\fbox{\parbox{\textwidth-18pt}{
{\bf Second identity (defect double soft theorem):}
  \begin{multline}
    -\int_{ x^0>T}\int_{x'^0<T} \ud^{p} x\, \ud^p  x'\, \langle \bulk_L\, \tilt^{[\tilde s}( x) \tilt^{\tilde t]}( x') \bulk_R\rangle
    -\int_{ x^0>T}\ud^{p}  x \langle \bulk_L\, \tilt^{[\tilde s}( x)  \delta^{\tilde t]}\bulk_R\rangle\\
    -\int_{ x'^0<T} \ud^p  x'\, \langle \delta^{[\tilde s} \bulk_L\, \tilt^{\tilde t]}( x') \bulk_R\rangle-\langle \delta^{[\tilde s}\bulk_L\,  \delta^{\tilde t]}\bulk_R\rangle\\
    =f^{\tilde s \tilde t}_{\ \ u}\langle\bulk_L \delta_u \bulk_R\rangle+f^{\tilde s \tilde t}_{\ \  \tilde u} \left[\langle \bulk_L \delta_{\tilde u} \bulk_R\rangle+\int_{ x^0<T}\ud^p  x\, \langle\bulk_L\, \tilt^{\tilde u}( x)\, \bulk_R\rangle\right]\;.
\label{eq:inhomo}
    \end{multline}
\vspace{-0.3cm}
}}}
\vspace{0.3cm}

\noindent
This is our second basic identity. As it stands it looks somewhat cumbersome, but this is a consequence of its great generality. In applications several of the terms above actually drop out and we will get much simpler relations. Let us make now a couple of comments on this result.

Firstly, and unlike the previous identity, it is now not a priori guaranteed that the above is convergent, due to the integration region where the two $\mathbb T^{\tilde a}$ operators approach each other. However, in our derivation we merely used the definition of the charges as integrals of the currents as well as the implicit assumption that the algebra of the charges is well defined, i.e. that the surface operators $Q^{\tilde a}, Q^{\tilde b}$ have a finite limit when they approach other. In order to satisfy this regularity there is a constraint on the dimension of the first operator in the $\tilt^{[\tilde a} \tilt^{\tilde b]}$ OPE. With this aside, all integrals appearing above are actually finite. Divergences arise only as we take some of the bulk insertions to approach the defect. 

Another comment concerns the time dependence of this identity. To make expressions a bit simpler let us set $p=1$. Taking the derivative with respect to $T$ gives
  \begin{multline}
    \int\ud  x\, \langle \bulk_L \tilt^{[\tilde s}(T)\tilt^{\tilde t]}( x) \bulk_R\rangle
    + \langle \bulk_L\, \tilt^{[\tilde s}(T)  \delta^{\tilde t]}\bulk_R\rangle\\
    + \langle \delta^{[\tilde t} \bulk_L\,  \tilt^{\tilde s]}(T) \bulk_R\rangle=f^{\tilde s \tilde t}_{\ \  \tilde u}  \langle\bulk_L\, \tilt^{\tilde u}(T)\, \bulk_R\rangle\;.\label{dert}
    \end{multline}
It turns out that this is not a new constraint. Consider the first identity as applied to $\langle \bulk_L \mathcal J_\mu^{[\tilde s}( x_T) Q^{\tilde t]} \bulk_R\rangle$ with $ x_T^0=T$. It gives
  \begin{multline}
    \int\ud  x\, \langle \bulk_L  \mathcal J_\mu^{[\tilde s}( x_T)\mathcal \tilt^{\tilde t]}( x) \bulk_R\rangle
    + \langle \bulk_L\,  \mathcal J_\mu^{[\tilde s}( x_T)  \delta^{\tilde t]}\bulk_R\rangle\\
    + \langle \delta^{[\tilde t} \bulk_L\,  \mathcal J_\mu^{\tilde s]}( x_T)\bulk_R\rangle=f^{\tilde s \tilde t}_{\ \ \tilde u}  \langle\bulk_L\, \mathcal J_\mu^{\tilde u}( x_T)\, \bulk_R\rangle+f^{\tilde s \tilde t}_{\ \ u}  \langle\bulk_L\, \mathcal J_\mu^{u}( x_T)\, \bulk_R\rangle
      \;.
\end{multline}
We now push the current towards the defect and use \reef{eq:push}. Then we recover \reef{dert} on the nose as long as :
\ba
    \lim_{|y|\to 0} |y|^{d-p-2} y_a\, \mathcal J^{s,a}(y,{x})=0
\ea
which is indeed true according to our assumption of no relevant defect operators $\hcO^a$. Thus~our second identity is independent of $T$ as desired.

\section{Integrated constraints: tilts}
\label{sec:tiltSRApp}

\subsection{Setup}
In this section we will specialize the general formalism to the case where the defect partially breaks a continuous internal symmetry $G$ of the bulk CFT, preserving a subgroup $\hat{G}$. The set of all such defects is then described by the coset $G/\hat{G}$. This coset is then a conformal manifold spanned by all such defect CFTs. 

In this case the bulk current conservation gets modified as
\ba
    \partial_\mu J^{\tilde s,\mu}(y,{x})=\delta^{(d-p)}(y) t^{\tilde s}( x)\:. \label{eq:WIcurrent}
\ea
The operator $t^{\tilde a}$ is called the tilt and has protected dimension $p$. It is thus a marginal operator on the defect. Its goal in life is to allow us to move from one point on the conformal manifold to another. As the normalisation of the tilt has been fixed by the Ward identity, the tilt two-point function is part of the defect CFT data:
\ba
    \langle t^{\tilde s}(x) t^{\tilde t}(0)\rangle=\frac{\delta^{\tilde s\tilde t}\, \normt}{x^{2p}}\:.
\ea
In particular $\normt$ is a function on the conformal manifold which determines its metric. The Ward identity further implies that 
\ba
    \lim_{|y|\to 0} |y|^{d-p-2} y^a J^{\tilde s}_a(y,{x})=\, t^{\tilde s}({x})/V_{d-p-1}\:.\label{eq:DOEJ}
\ea
In particular this fixes the bulk-defect two point function:
\ba
\label{eq:Jt}
    \langle J_{\mu}^{\tilde s}(y,{x}_1) t^{\tilde t}({x}_2)   \rangle& = \delta^{\tilde s\tilde t}\,\mathbb T^{Jt}_{\mu}K_t^J(y,x_1,x_2) \,\normt/V_{d-p-1} \:,
\ea
with the structure\footnote{This is $Q_\text{BD}^2$ in the convention of~\cite{Billo:2016cpy}.}
\ba
\label{eq:structure-JO}
\mathbb T^{J\op}_m=\frac{2|y| x_{12,m}}{y^2+x_{12}^2}\ , \qquad \mathbb T^{J\op}_a=\frac{y_a}{|y|}\ \frac{y^2- x_{12}^2}{y^2 + x_{12}^2}\,.
\ea
As mentioned previously, this is the only current-scalar two point function which can be written down.

In the next few subsections we will apply the two fundamental identities to derive constraints on various defect four-point functions as well as two point form factors (i.e. bulk-defect-defect correlators). For simplicity we will consider the case where $G=O(N)$ and $\hat{G}=O(N-1)$ although it is straightforward to consider more general setups.
In this case the conformal manifold is the $N$ dimensional sphere. Since this is a maximally symmetric space the coefficient $\normt$ is independent of the point on the conformal manifold where we sit, and furthermore the curvature of the space is also entirely determined by this number. 

For the $O(N)$ group the generators can be denoted
\ba
Q^{A}\to Q^{[IJ]}\,, \qquad I,J=1,\ldots N
\ea
and tensors of $O(N)$ are chosen to transform as 
\begin{equation}
    [Q^{[IJ]}, V^{K_1\ldots K_P}]=\delta^{K_1 [I} V^{J] K_2\ldots}+\ldots +\delta^{K_P [I} V^{K_1\ldots J]}\,.
\end{equation}
We now consider that $O(N)$ symmetry is broken by a preferred direction $n^I=(1,0,\ldots 0)$. We set:
\ba
    Q^{s}\to Q^{[ij]}\,, \qquad Q^{\tilde s}\to Q^{[1i]}\,,
\ea
where $i = 2, \ldots, N$ are the directions perpendicular to $n^I$. Note that
\begin{equation}
    [Q^{[1i]},Q^{[1j]}]=Q^{[ij]} \Rightarrow 
    f^{\tilde s \tilde t}_{\ \  \tilde u}=0  \:.
\end{equation}
In the following we will denote
\ba
t^{\tilde s}=t^{[1i]}\quad& \longrightarrow \quad t^i\,, \\
\delta^{\tilde s}=\delta^{[1i]}
\quad  &\longrightarrow \quad  \delta^i\,.
\ea
With this notation and the above choices our two identities become:
\ba
\int \ud^p  x \, \langle t^i(x) \bulk\rangle=-\langle \delta^i \bulk\rangle \:.\label{eq:id1tilt}
\ea
and
\begin{multline}
    -\int_{x^0>T}\int_{ x'^0<T} \ud^p  x\, \ud^p  x'\, \langle \bulk_L\, t^{[i}( x) t^{j]}( x') \bulk_R\rangle
    -\int_{ x^0>T}\ud^p  x \langle \bulk_L\, t^{[i}( x)  \delta^{j]}\bulk_R\rangle\\
    -\int_{ x'^0<T} \ud^p  x'\, \langle \delta^{[i} \bulk_L\,  t^{j]}( x') \bulk_R\rangle-\langle \delta^{[i}\bulk_L\,  \delta^{j]}\bulk_R\rangle=\langle\bulk_L \delta^{[ij]} \bulk_R\rangle \:.\label{eq:id2tilt}
\end{multline}

\subsection{Form factor identities}
\label{ssec:formf-tilt}
We will now derive integrated constraints on form factors involving one bulk operator, one tilt and one defect operator. The strategy is to apply our first identity as applied to two bulk insertions. We then push one of the bulk insertions towards the defect where, using the DOE, it becomes a defect operator insertion. Schematically
\ba
\int \ud^p x \langle t (x) \Phi (x_1,y_1) \tilde \Phi (x_2,y_2) \rangle & =- \langle \delta \Phi (x_1,y_1) \tilde \Phi (x_2,y_2) \rangle - \langle \Phi(x_1,y_1) \delta \tilde \Phi (x_2,y_2)\rangle \\
\mbox{Push $\Phi$ to the defect}\quad &\Rightarrow\quad  \int \ud^p \langle t (x) \hcO (x_1) \tilde \Phi (x_2,y_2) \rangle=-\langle \hcO (x_1) \delta \tilde \Phi(x_2,y_2) \rangle\:.
\ea
On the righthand side we dropped a term. This could be either because it is zero by symmetry reasons, or because we assume the leading DOE operator in $\delta \tilde \Phi$ is of higher dimension than the one in $\Phi$ (once the constraint is obtained this assumption can be removed by analytic continuation). There are a great many possibilities for the choice of bulk operators depending on their $O(N)$ quantum numbers. Here we will restrict ourselves to a few representative cases. 

As a simple warmup application, let us consider correlator involving one bulk operator and one tilt. In particular consider a bulk operator $\Phi$ living in a tensor representation of $O(N)$ of rank $P+1$. In this case the two point function fixed by conformal invariance up to an overall constant:
\ba
\langle t^i( x_1) \Phi_{j J_1\ldots J_P}( x_2,y)\rangle= \sqrt{\normt}\, \mu^{\Phi}_{t}\,\frac{\delta^{i}_j\, n_{J_1}\ldots n_{J_P}}{|y|^{\Delta_\Phi-p}(x_{12}^2+y^2)^p}\,.
\ea
Applying the first identity gives:
\ba
-n^{J_1}\ldots n^{J_P}\int \ud^p \hat x' \,\langle t^{i}(\hat x') \Phi_{j J_1\ldots J_P}(\hat x,y) \rangle=\delta^{i}_j\, n^{J_0}n^{J_1}\ldots n^{J_P} \langle \Phi_{J_0 \ldots J_P}(\hat x,y)\rangle\:.
\ea
and thus we find the non-trivial result:
\ba
\mu^{\Phi}_{t}=-\frac{\mu^{\Phi }_{\mathds 1}2^{p-1} \Gamma \left( \frac{p+1}{2} \right)}{\pi^{\frac{p+1}{2}}\, \sqrt{\normt}}\:. \label{eq:vev-tilt}
\ea
where $\mu^{\Phi}_{\mathds 1}$ is simply $a_\Phi$, the vev of $\Phi$ in the presence of the defect. This result matches \cite{Padayasi:2021sik}, where this relation was derived for a boundary.

Now let us move on to more interesting applications which include an additional defect insertion on top of the tilt. Let us consider the first identity as applied to two bulk insertions, one of which is a vector under $O(N-1)$ and one of which is a scalar. We get
\ba
\int \ud^p  x\, \langle t^{i}( x) \Phi^j( x_1,y_1) \Psi( x_2,y_2)\rangle =\delta^{ij}\, \langle \Phi^1( x_1,y_1)\,\Psi( x_2,y_2)\rangle\:.
\ea
To obtain identities involving two defect insertions, we can push either of the bulk operators towards the defect. Concretely, with DOE:
\ba
\Psi_s( x,y)&\underset{|y|\to 0}=\mu^{\Psi}_{s} |y|^{\Delta_s-\Delta_{\Psi}}\, {\hat\op}_{s}+\ldots\,,\\
\Phi^{1}( x,y)&\underset{|y|\to 0}=\mu^{\Phi}_{s} |y|^{ \Delta_s-\Delta_{\Phi}} {\hat\op}_{s}+\ldots\,,\\
\Phi^j(x,y) &\underset{|y|\to 0}=\mu^{\Phi}_v\,|y|^{ \Delta_v-\Delta_{\Phi}} {\hat\op}^j_{v}+\ldots\,,
\ea
with $\hat\cO_s$, $\hat\cO_v$ operators singlets and vectors under $O(N-1)$ respectively. Let us first push~$\Psi$. The identity becomes:
\ba
\int \ud^p  x\, \langle t^{i}( x) \hat\cO_s( x_2) \Phi^j( x_1,y_1)\rangle=\delta^{ij} \langle \hat\cO_s( x_2) \Phi^{1}( x_1,y_1)\rangle \:.
\ea
Using \reef{eq:formf} we can write this as 
\ba
\boxed{ \int_0^1 \frac{\ud w}{w^{\frac{ \Delta_s}2+1}(1-w)^{1 - \frac{p}{2}}}\mathcal H^{\Phi}_{ts}(w)= -\mu^{\Phi}_{s} \frac{ \Gamma \left(\frac{p}{2}\right)}{\pi^{\frac{p}{2}} }\,.} \label{eq:tsPhi}
\ea
If we push $\Phi$ instead we are faced with a possible issue. Concretely the right hand side diverges if $\Delta_v> \Delta_s$. The idea then is to define the sum rule in the opposite regime (where we get zero on the RHS) and then everywhere else by analytic continuation. In this way we~get
\ba
\int \ud^p  x \, \langle t^i(x) \hcO_v^j( x_1) \Psi( x_2,y_2)\rangle=0
\ea
or equivalently 
\ba
\boxed{\int_0^1 %\frac{\ud w}{w^{\frac{ \Delta_v}2+1}\sqrt{1-w}} 
\frac{\ud w}{w^{\frac{ \Delta_v}2+1}(1-w)^{1 - \frac{p}{2}}} \mathcal H^{\Psi}_{tv}(w)=0\,.} \label{eq:psitv}
\ea
For the special case where $\hcO_v=t$ the exact same sum rule could have been derived by considering instead our identity applied to two bulk insertions of a broken current and an $O(N)$ scalar (this time without the need for doing the analytic continuation).

Next consider the case where one bulk operator is the broken current and the other a $O(N)$ vector. We get
\ba
\int \ud^p  x \, \langle t^i( x) J_\mu^{[1j]}( x_1,y_1) \Phi^{1}( x_2,y_2)\rangle=-\langle J_\mu^{[1j]}( x_1,y_1) \Phi^{i}( x_2,y_2)\rangle \\
\Rightarrow \int\, \ud^p  x \, \langle t^i( x) t^j( x_1) \Phi^{1}( x_2,y_2)\rangle=-\langle t^j( x_1)\Phi^{i}( x_2,y_2)\rangle \,,
\ea
where in the second line we pushed the broken current towards the defect and used \reef{eq:vev-tilt}. This can be rewritten as
\ba
\boxed{\int_0^1 \frac{\ud w}{w^{1 + \frac{p}{2}}(1-w)^{1 - \frac{p}{2}}} \mathcal H^{\Phi}_{tt}(w)= -\mu_{t}^{\Phi}\, \frac{\Gamma\left(\frac{p}{2}\right)}{\pi^{\frac{p}{2}}}\,.} \label{eq:phitt}
\ea

Moving on, consider now the case where the final bulk insertion is the current, broken or unbroken. Consider first the latter. We have:
\ba
\int\,\ud^p  x\, \langle t^i(x)\, \Psi( x_1,y_1) J_\mu^{[1j]}( x_2,y_2)\rangle&=0\\
\Rightarrow  
\int\,\ud^p  x\, \langle t^i(x)\, \hcO_s( x_1) J_\mu^{[1j]}( x_2,y_2)\rangle&=0\:.
\ea
Writing this in the form \reef{eq:formfv} it is possible to show that this translates into the constraints 
\begin{empheq}[box=\fbox]{align}
& \int_0^1 \frac{\ud w}{(1-w)^{1 - \frac{p}{2}}\,w^{\frac{ \Delta_s}2}}\, \mathcal H^{J,(-)}_{ts}(w)=0\:,\\
& \int_0^1 \frac{\ud w (1-w)^{\frac{p}{2}}}{w^{\frac{ \Delta_s}2+1}}\,\mathcal H^{J,(+)}_{ts}(w)=0\,.
\end{empheq}
Imposing conservation we find only one of these is an independent constraint.

Finally, consider
\ba
\int\,\ud^p  x\, \langle t^i(x)\, J_{\nu}^{[1j]}( x_1,y_1) J_\mu^{[kl]}( x_2,y_2)\rangle&=\langle J_\nu^{[1j]}( x_1,y_1) J_\mu^{[1l]}( x_2,y_2)\rangle\delta^{ik}-(k\leftrightarrow l)
\\
\Rightarrow  
\int\,\ud^p  x\, \langle t^i(x)\, t^j( x_1) J_\mu^{[kl]}( x_2,y_2)\rangle&=\langle t^j( x_1) J_\mu^{[1l]}( x_2,y_2)\rangle \delta^{ik}-(k\leftrightarrow l)\:.
\ea
The righthand side is fixed by the DOE of the broken current, cf. equation \reef{eq:DOEJ}. Writing out the overall tensor structure on the left as $\delta^{i[k}\delta^{l]j}$ we can rewrite this as 
\ba
\boxed{ \int_0^1 \frac{\ud w}{(1-w)^{1- \frac{p}{2}} w^{\frac{p}{2}}} \, \mathcal H^{J,(-)}_{tt}(w)= -\frac{1}{2} \frac{\normt}{V_{d-p-1}}\frac{\Gamma \left(\frac{p}{2}\right)}{\pi^{\frac{p}{2}} } \,.}
\ea

\subsection{Defect correlator identities}
In this subsection we work out integrated constraints on defect four-point correlators. The strategy for doing so is similar to what we did in the previous section, namely to start off with some bulk insertions which we later push towards the defect. We analyse the first and second identities in turn.

\subsubsection{First identity}
As our first application we consider bulk insertions corresponding to a broken current as well as two operators in some $O(N-1)$ irrep (sitting inside a larger $O(N)$ irrep). In this case we get simply
\ba
\label{eq:tilthom-x}
\int \ud^p  x\, \langle t^i(x) J_\mu^{[1j]}( x_1,y_1) \Phi_I ( x_2,y_2) \Phi_J( x_3,y_3)\rangle& =(\ldots) \\
\Rightarrow \int \ud^p  x\, \langle t^i(x)\, t^{j}( x_1) \hcO_I ( x_2) \hcO_J( x_3)\rangle=0\:,
\ea
where in the second step we pushed all operators to the defect. The point is that in the omitted terms on the RHS of the first line, at least one of the bulk operators gets transformed into a different $O(N-1)$ irrep (sitting inside the same $O(N)$ irrep). Since then all operators get pushed to the defect there will be a mismatch as we have seen in the previous subsection, and under suitable assumptions (which can be dropped after the fact) the outcome is that all terms drop out. The correlator on the left can be expressed as in \reef{eq:corr}, and depends on two cross-ratio $z$ and $\bar{z}$~\eqref{eq:zzbardef}. We use conformal invariance to go to the frame 
\begin{equation}
    x_1=0,\quad x_2=1,\quad x_3\rightarrow\infty 
\end{equation}
where the correlator takes the simple expression
\begin{equation}
\label{eq:4pt-01inf}
    \langle {\hcO}^J(0)t^{i}(x)t^{j}(1){\hcO}^I(x_3)\rangle ~\underset{x_3\rightarrow\infty}{\sim}~ \frac{1}{x_3^{2\Delta}} \ \frac{\mathcal G^{JijI}(z,\bar{z})}{|z|^{p-\Delta_\hcO}}\:.
\end{equation}

We now parametrize $x$ by $(r,\theta,\phi_i)$, with $\theta$ the angle between $x$ and the $(0,1)$ axis, and~$\phi^i$ the $p-2$ other angles that are trivially integrated, giving:
\begin{equation}
    \frac{V_{p-2}}{x_3^{2\Delta}} \int_0^\infty\ud r\int_0^\pi \ud\theta\ r^{p-1}(\sin\theta)^{p-2}\ \frac{\mathcal G^{JijI}(z,\bar{z})}{|z|^{p-\Delta_\hcO}} =0\:.
\end{equation}
where in the above the cross-ratios are simply $z=r e^{i \theta}$. Changing to these variables leads to

\setlength\fboxrule{1.2pt}
\setlength{\fboxsep}{8pt} 
{\hspace{-18pt}\noindent\fbox{\parbox{\textwidth-18pt}{
{\bf Tilt soft sum rules:}
\begin{equation}
\label{eq:tilthom}
    \iint_{\text{Im}\,z>0}\frac{\mathrm{d}z\mathrm{d}\bar{z}}{|z|^{p-\Delta_\hcO}}\left(\text{Im}\,z\right)^{p-2} \mathcal G^{JijI}(z,\bar{z}) = 0\ .
\end{equation}
\vspace{-0.3cm}
}}}
   \vspace{0.3cm} 
\noindent

More details on the derivation in the case of $p=1$ are given in appendix \ref{app:1d-SR}. 
The above sum rule has appeared before in \cite{Bashmakov:2017rko,Behan:2017mwi} in the context of marginal operators, but not specifically tailored to defect~CFTs.

\subsubsection{Second identity}
Let us now apply the second identity \eqref{eq:id2tilt} to derive constraints on defect four-point functions, and push the bulk insertions to the defect. In this case nearly all terms on the right-hand side drop out, except one
\begin{equation}
\label{eq:tiltinhom-x}
    \int_{x^0>T} \int _{x'^0<T} \ud^p x\ud^p x'\, \langle \hat{\op}^I(x_1) t^{[i}(x) t^{j]}(x') \hat{\op}^J(x_2)\rangle = -\langle \hcO^I Q^{[ij]}\hcO^J\rangle\:.
\end{equation}
We now wish to express the results in terms of cross-ratios. To do this it is convenient to evaluate the broken charges on a different codimension 1 surface, namely one that intersects the defect along a $p-1$ dimensional sphere, so that the regions below/above $T$ now become the regions inside/outside this sphere. By conformal symmetry we can set this sphere centered about the origin with unit radius and set $(x_1,x_2)=(0,\infty)$. As for the first identity case, we switch to radial coordinates and find:
\begin{equation}
    V_{p-1}V_{p-2}\int_0^1\ud r\int_1^\infty \ud r'\int_0^\pi \ud\theta\ \frac{r^{p-1}(\sin\theta)^{p-2}}{r'^{p+1}}\frac{\mathcal G^{I[ij]J}(z,\bar z)}{|z|^{p-\Delta_\hcO}}  = - Q_{IJ}^{[ij]}\:,
\end{equation}
where we defined 
\ba
\label{eq:tilt-RHS}
Q^{[ij]}_{IJ}:=\langle \hcO^I| Q^{[ij]}|\hcO^J\rangle\,.
\ea
and the cross-ratios are $z=\frac{r}{r'}e^{i\theta}$. Changing variables we are left with a single integral to perform, obtaining:

\setlength\fboxrule{1.2pt}
\setlength{\fboxsep}{8pt} 
{\hspace{-18pt}\noindent\fbox{\parbox{\textwidth-18pt}{
{\bf Tilt double soft sum rules:}
\begin{equation}
\label{eq:tiltinHom}
    \iint_{\text{Im}\,z>0,\,|z|<1} \frac{\ud z\ud\bar z}{|z|^{p-\Delta_\hcO}}(\text{Im}\,z)^{p-2}\log|z|\, \mathcal G^{I[ij]J}(z,\bar z) = Q_{IJ}^{[ij]}\,\frac{\Gamma(p-1)}{(2\pi)^{p-1}}\:.
\end{equation}
\vspace{-0.3cm}
}}}
   \vspace{0.3cm} 
   
\noindent
The computation for $p=1$ has to be treated separately and we do this in in appendix \ref{app:1d-SR}. 

\section{Integrated constraints: displacements}\label{sec:dispSRApp}

\subsection{Setup}
While not all CFTs may have an internal symmetry group, they are all invariant under bulk conformal symmetries. A conformal defect leads then to a particular protected defect operator called the displacement, associated to the breaking of the stress-tensor Ward identity:
\ba
\partial_\mu T^{\mu a} ({x},y)  = \delta^{d-p} (y) \disp^a ({x})\;, \label{eq:WIst}
\ea
where $a$ runs over the directions transverse to the defect, and $y$ is the coordinate perpendicular to the defect.
The displacement operator has protected dimension $\Delta_{\disp} = p+1$, and it is a vector under the $SO(d-p)$ rotational symmetry. Its normalization $\normD$ is a part of the defect CFT data:
\begin{equation}
        \langle \disp^{a}({x}) \disp^{b}(0)\rangle=\frac{\delta^{ab}\, \normD}{{x}^{2(p+1)}}\:.
\end{equation}
Note also that the Ward identity implies that 
\ba
    \lim_{|y|\to 0} |y|^{d-p-2} y_b\, T^{ab}(y, x)=\, \disp^a ( x)/V_{d-p-1} \:,
\ea
In order to formulate identities \reef{eq:id1} and \reef{eq:id2}, let us set up our conventions for the charges. The conformal generators form $SO(d+1,1)$ Euclidean conformal algebra 
\ba
    &[K^{\mu}, P^{\nu} ] = 2 \delta^{\mu \nu} D - 2 M^{\mu \nu}\:, \quad [D, K^{\mu}] = - K^{\mu}\:, \quad [D, P^{\mu}] = P^{\mu}\:, \\
    &[M^{\mu \nu} , P^{\rho}] = \delta^{\nu \rho} P^{\mu} - \delta^{\mu \rho} P^{\nu}\:, \quad [M^{\mu \nu} , K^{\rho}] = \delta^{\nu \rho} K^{\mu} - \delta^{\mu \rho} K^{\nu}\:, \\
    &[M^{\mu \nu}, M^{\rho \sigma}] = \delta^{\nu \rho} M^{\mu \sigma} - \delta^{\mu \rho} M^{\nu \sigma} +  \delta^{\nu \sigma} M^{\rho \mu} -  \delta^{\mu \sigma} M^{\rho \nu}\:. \label{eq:confAlg}
\ea
The introduction of the $p$-dimensional defect breaks the conformal group to $SO(p+1,1) \times SO(d-p)$. A basis of broken generators is given by the $P^a, K^a$ and $M^{ma}$, and satisfy the algebra 
\begin{equation}
\begin{aligned}
    [K^a,P^b]&=2\delta^{ab} D-2\,M^{ab}\,,& \qquad [M^{ma},M^{nb}]&=-\delta^{mn} M^{ab}-\delta^{ab} M^{mn}\:,\\
[M^{ma},P^b]&=\delta^{ab} P^m\,,&\qquad [M^{ma},K^b]&=\delta^{ab} K^m\:. \label{eq:brokenC}
\end{aligned}
\end{equation}
In particular note that broken generators always commute to unbroken ones as desired.
The generators are constructed by contracting the stress-tensor with a conformal Killing vector $\xi_\mu=\xi_\mu(x,y)$. We have
\begin{equation}
\begin{aligned}
    [Q^{\tilde s},|y|^{\Delta_\Phi} \Phi] &= (\xi^{\tilde s})^\mu \partial_\mu(|y|^{\Delta_\Phi} \Phi) \:, \\
    Q^{\tilde s}|0\rangle_{\mathcal D} &= \int_{ x_0<T} \mathrm{d}^p x\  \xi^{\tilde s}_a( x,0) D^{a}\ket{0}_\mathcal{D}\: . \label{eq:confcharge}
\end{aligned}
\end{equation}
Using
\ba
\xi_\nu^D = x_\nu\,,\quad \xi^{P_\mu}_\nu=\delta_{\mu\nu}\,, \quad \xi^{K_\mu}_\nu=2 x_\mu x_\nu-x^2 \delta_{\mu\nu}\,,\quad \xi^{M_{\rho\tau}}_{\nu}=x_\tau \delta_{\nu\rho}-x_\rho \delta_{\nu\tau}\:,
\ea
it follows that \eqref{eq:confcharge} contains three expressions for the broken generators:
\ba
    P^a \ket{0}_\mathcal{D} &=\int_{ x_0<T} \mathrm{d}^p x\, D^{a}\ket{0}_\mathcal{D}\:,\\
     M^{ma} |0\rangle_\mathcal{D} &=-\int_{ x_0<T} \mathrm{d}^p x\,  x^{m} D^{a}|0\rangle_\mathcal{D}\:,\\
    K^a \ket{0}_\mathcal{D} &= -\int_{ x_0<T} \mathrm{d}^p x\  x^2 D^{a}|0\rangle_\mathcal{D}\:.   
\ea    
After these preliminaries we are ready to state the version of our first identity \reef{eq:id1} as specialized to the displacements:
\ba
\label{eq:id1disp-3}
    \int \mathrm{d}^p x\ \langle D^{a}( x) \bulk \rangle &= -\langle \delta_{P^{a}} \bulk \rangle \:,\\
    \int \mathrm{d}^p x\  x^{m} \langle D^{a}( x) \bulk \rangle &= \langle \delta_{M^{ma}} \bulk \rangle, \\
    \int \mathrm{d}^p x\  x^2\langle D^{a}( x) \bulk \rangle &=\langle \delta_{K^a} \bulk\rangle\:. 
\ea
Before we move on to the statement of the second identity, let us make a brief comment. Since the displacement is not a marginal operator, it is not immediately obvious that these integrals are invariant under conformal transformations, and in fact taken separately they are not. However, taken together, they do form a closed set. To see this, recall that given an operator $\hcO$ in a CFT we can define a nonlocally related operator $\tilde \cO$ called its shadow \cite{Ferrara:1972uq,Ferrara:1972xe,Ferrara:1972kab}, by the definition:
\begin{equation}
    \tilde{\op} ( x_0) = \int \ud^p x \, \frac{1}{( x- x_0)^{2 (p- \Delta)}} \hat\op_{\Delta} ( x)\:.
\end{equation}
For the displacement $\Delta=p+1$, and hence we see that we can repackage the identities together as
\ba
\langle \tilde D^a(x_0) \bulk\rangle= \int \mathrm{d}^p x\ ( x -  x_0)^2\langle D^{a}( x) \bulk \rangle &= \langle \delta_{K^{a}_{ x_0}}\bulk\rangle\:, \label{eq:id1disp}
\ea
where $K^{a}_{ x_0}$ is a special conformal transformation about point $ x_0$, thus making conformal invariance manifest.

Let us now consider the second identity \eqref{eq:id2}. In this case we get
\begin{multline}
\label{eq:id2disp}
    -\int_{x^0>T}\int_{ x'^0<T} \mathrm{d}^{p} x\, \mathrm{d}^p  x'\, \langle \bulk_L\, \xi^{[\tilde s}_a(x)D^{a}(x) \xi^{\tilde t\,]}_b(x')D^{b}(x') \bulk_R\rangle -\int_{ x^0>T} \mathrm{d}^{p}  x\ \langle \bulk_L\, \xi^{[\tilde s}_a(x)D^{a}(x)  \delta^{\tilde t\,]}\bulk_R\rangle\\
    -\int_{ x'^0<T} \mathrm{d}^p  x\ \langle \delta^{[\tilde s}\bulk_L\,  \xi_b^{\tilde t]}(x') D^b(x') \bulk_R\rangle - \langle \delta^{[\tilde s}\bulk_L\,  \delta^{\tilde t]}\bulk_R\rangle
    =\langle\bulk_L \delta^{[\tilde s \tilde t]}\bulk_R\rangle \:.
\end{multline}
Generically there are four different commutators for the broken generators \reef{eq:brokenC}, this identity leads a priori to four different sum rules. However, as we will see in section \ref{sssec:disp-second-id}, when considering four-point functions of defect operators, only three of those are independent.

Once again it is not obvious that these identities are conformally invariant. One might wonder if we could again make use of the shadow formalism here as well to write the integrals in a conformally invariant way. However, due to the range of integration being only over a half-plane, this is not as obvious. Nevertheless we have checked that different choices of conformal frames map amount to reshufflings of the full set of identities.

\subsection{Form factor identities}  
Let us examine the consequences of \reef{eq:id1disp} on correlators with a single bulk operator insertion. As a warm up we consider constraints on the bulk-defect two point function involving such an operator and the displacement. We have
\ba
\langle D^a( x_1) \Psi( x_2,y)\rangle =\sqrt{\normD}\,\mu^{\Psi}_D\,\frac{y^a}{|y|^{\Delta_\Psi+1}}\frac{y^{p+1}}{(x_{12}^2+y^2)^{p+1}}\:,
\ea
and identity \reef{eq:id1disp} gives 
\ba
\label{eq:vev-disp}
\int \ud^p  x \, ( x- x_0)^2\langle D^a( x) \Psi(x',y')\rangle &= \delta_{K^a_{x_0}} \langle \Psi(x',y')\rangle \quad \Rightarrow \quad \mu^{\Psi}_D=\frac{2^p  \Gamma\left(\frac{p+1}{2}\right)}{\pi^{\frac{p+1}{2}}}\, \, \frac{\Delta_{\Psi} \mu^{\Psi}_{\mathds 1}}{\sqrt{\normD}}\:.
\ea
This is in perfect agreement with \cite{Billo:2016cpy}. 
Now let us move on to more interesting form factors. Starting with two scalar bulk insertions and pushing as usual we find
\ba
\int \ud^p  x \,  ( x-x_0)^2 \langle D^a( x) \hat\cO_s( x_1) \Psi( x_2,y)\rangle = \langle \hat\cO_s( x_1) \delta_{K^{a}_{ x_0}} \Psi( x_2,y)\rangle \:.
\ea
This can be rewritten in manifestly conformally invariant fashion as
\ba
\int_{0}^1 \frac{\ud w}{(1-w)^{1-\frac{p}{2}}\, w^{\frac{\Delta_s+1}2}} 
  \left(\frac{1-w}{w}+\frac{1-\tilde w}{\tilde w}\right)\, \mathcal H^{\Psi}_{Ds}(w)
=-\mu^{\Psi}_{s}\left[\frac{\Delta_\Psi-\Delta_{s}}{\tilde w}+2\Delta_{s}\right] \frac{\Gamma\left(\frac{p}{2}\right)}{\pi^\frac{p}{2} } \:,
\ea
with
\ba
\label{eq:wtilde}
\tilde w=\frac{ x_{01}^2 y^2}{(y^2+ x_{02}^2)(y^2+ x_{12}^2)}\:.
\ea
By matching the dependence in $\tilde w$ on both sides we recover two independent sum rules: 
\begin{empheq}[box=\fbox]{align}
 \int_{0}^1 \frac{\ud w}{(1-w)^{1-\frac{p}{2}} \, w^{\frac{\Delta_{s}+3}2}} 
  \, \mathcal H^{\Psi}_{Ds}(w)
&= - 2\mu^{\Psi}_{s}\Delta_{\Psi} \frac{\Gamma\left(\frac{p}{2}\right)}{\pi^\frac{p}{2} }\,, \\
 \int_{0}^1 \frac{\ud w}{(1-w)^{1-\frac{p}{2}} \, w^{\frac{\Delta_{s}+1}2}} 
   \, \mathcal H^{\Psi}_{Ds}(w)
&= - \mu^{\Psi}_{s}(\Delta_\Psi-\Delta_{s}) \frac{\Gamma\left(\frac{p}{2}\right)}{\pi^\frac{p}{2} } \,.
\end{empheq}
Next we consider the form factor:
\ba
\langle \disp^a( x_1) \disp^b( x_2) \Psi( x_3,y)\rangle=K^\Psi_{DD}(x_i) \left[\delta^{ab} \mathcal H^{\Psi,S}_{DD}(w)+\left(\frac{y^a y^b}{y^2}-\frac{\delta^{a b}}{d-1}\right)\, \mathcal H^{\Psi,T}_{DD}(w)\right] \:, \label{eq:ddy}
\ea
where $S,T$ stands for singlet/symmetric traceless tensor with respect to transverse rotations. We have the identity:
\ba
\int \ud^p x \, \langle \disp^a( x) \disp^b( x_1) \Psi( x_2,y)\rangle=\langle D^b( x_1) \delta_{K^{a}_{x_0}}\, \Psi( x_2,y)\rangle \:.
\ea
Computing the right hand side we find it takes a form similar to \reef{eq:ddy} but now for a form factor $\langle D \tilde D \Psi\rangle$, with $\tilde D$ the shadow of $D$, entirely analogous to our previous computation. We now find
\ba
\int_{0}^1 \frac{\ud w}{(1-w)^{1-\frac{p}{2}}\, w^{1 + \frac{p}{2}}} 
  &\left(\frac{1-w}{w}+\frac{1-\tilde w}{\tilde w} \right)\, \mathcal H^{\Psi,S}_{DD}(w)=-\frac{\mu^{\Psi}_{D}\sqrt{\normD}}{d-p}\,
\left[\frac{\Delta_\Psi-d}{\tilde w}+2(p+1)\right] \frac{\Gamma\left(\frac{p}{2}\right)}{\pi^\frac{p}{2} } \:,
\\
\int_{0}^1 \frac{\ud w}{(1-w)^{1-\frac{p}{2}}\, w^{1 + \frac{p}{2}}} 
  &\left(\frac{1-w}{w}+\frac{1-\tilde w}{\tilde w} \right)\, \mathcal H^{\Psi,T}_{DD}(w)= -\mu^{\Psi}_{D}\sqrt{\normD}\,\left[\frac{\Delta_\Psi-p}{\tilde w}+2(p+1)\right] \frac{\Gamma\left(\frac{p}{2}\right)}{\pi^\frac{p}{2}}  \:.
\ea

By matching the dependence in $\tilde w$ on both sides we again obtain two independent sum rules for each tensor structure:
\begin{empheq}[box=\fbox]{align}
 \int_{0}^1 \frac{\ud w}{(1-w)^{1-\frac{p}{2}} \, w^{2+\frac{p}{2}}}  \, \mathcal H^{\Psi,S}_{DD}(w)
&=-\frac{2\mu^{\Psi}_{D}\sqrt{\normD}}{d-p}
\left[\Delta_\Psi+p + 1-d\right] \frac{\Gamma\left(\frac{p}{2}\right)}{\pi^\frac{p}{2} }\:, \\
\int_{0}^1 \frac{\ud w}{(1-w)^{1-\frac{p}{2}} \, w^{1+\frac{p}2}}  \, \mathcal H^{\Psi,S}_{DD}(w)
&=-\frac{\mu^{\Psi}_{D}\sqrt{\normD}}{d-p}
\left[\Delta_\Psi-d\right]\frac{\Gamma\left(\frac{p}{2}\right)}{\pi^\frac{p}{2} } \:,
\end{empheq}
and 
\begin{empheq}[box=\fbox]{align}
 \int_{0}^1 \frac{\ud w}{(1-w)^{1 - \frac{p}{2}} \, w^{2+\frac{p}2}} \, \mathcal H^{\Psi,T}_{DD}(w)
&=-2\mu^{\Psi}_{D}\sqrt{\normD}
\left[\Delta_\Psi+1\right]\frac{\Gamma\left(\frac{p}{2}\right)}{\pi^\frac{p}{2} } \:, \\
 \int_{0}^1 \frac{\ud w}{(1-w)^{1 - \frac{p}{2}} \, w^{1+\frac{p}2}} \, \mathcal H^{\Psi,T}_{DD}(w)
&=-\mu^{\Psi}_{D}\sqrt{\normD}
\left[\Delta_\Psi-p\right] \frac{\Gamma\left(\frac{p}{2}\right)}{\pi^\frac{p}{2} }\:.
\end{empheq}

\subsection{Defect correlator identities}
We now move on to examine the consequences of our first and second identities for correlators involving displacement operators \eqref{eq:id1disp}, \eqref{eq:id2disp}, for specific four-point functions. The derivation is very similar to the one section~\ref{sec:tiltSRApp}.  

\subsubsection{First identity}
We specialize the first identity \eqref{eq:id1disp} to a correlator containing a bulk stress tensor and two operators transforming in some irreducible representation of transverse rotations. After pushing to the defect as usual we get
\ba
\label{eq:disphom-x}
    \int \mathrm{d}^p x\, ( x- x_0)^2 \langle D^{a}( x)D^{b}( x_1)\hat \op^I( x_2)\hat \op^J( x_3)\rangle = 0\: .
\ea
The computation is similar as for the tilt, and only changes in the $z,\bar{z}$ powers. The result~is:

\setlength\fboxrule{1.2pt}
\setlength{\fboxsep}{8pt} 
{\hspace{-18pt}\noindent\fbox{\parbox{\textwidth-18pt}{
{\bf Displacement soft sum rules:}
\begin{align}
    &\iint_{\text{Im}\,z>0}\frac{\mathrm{d}z\mathrm{d}\bar{z}}{|z|^{p+1-\Delta_\hcO}}\left(\text{Im}\,z\right)^{p-2} \mathcal G^{JabI}(z,\bar{z}) = 0\ , \\
    & \iint_{\text{Im}\,z>0}\frac{\mathrm{d}z\mathrm{d}\bar{z}}{|z|^{p+1-\Delta_\hcO}}z\left(\text{Im}\,z\right)^{p-2} \mathcal G^{JabI}(z,\bar{z}) = 0\ , \\
    & \iint_{\text{Im}\,z>0}\frac{\mathrm{d}z\mathrm{d}\bar{z}}{|z|^{p+1-\Delta_\hcO}}\bar{z}\left(\text{Im}\,z\right)^{p-2} \mathcal G^{JabI}(z,\bar{z}) = 0\ , \\
    & \iint_{\text{Im}\,z>0}\frac{\mathrm{d}z\mathrm{d}\bar{z}}{|z|^{p+1-\Delta_\hcO}}z\bar{z}\left(\text{Im}\,z\right)^{p-2} \mathcal G^{JabI}(z,\bar{z}) = 0\ .
\label{eq:disphom}
\end{align}
\vspace{-0.3cm}
}}}
   \vspace{0.3cm} 

\noindent
More details on the derivation in the case $p=1$ are given in appendix \ref{app:1d-SR}, where we show they agree with the ones found by \cite{Gabai:2025zcs}.

\subsubsection{Second identity}
\label{sssec:disp-second-id}

We now apply the second identity \reef{eq:id2disp} again for two bulk insertions transforming in some irrep of transverse rotations, and push the bullk insertions to the defect. As for the tilt, nearly all terms on the right-hand side drop out, except one
\begin{multline} 
\label{eq:dispinhom-x}
   \int_{x^0>T}\int_{x'^0<T} \mathrm{d} x \mathrm{d} x^\prime\ (x- x_0)^2 ( x^\prime- x_0')^2\langle \hat\op^I( x_1)D^{a}( x)D^{b}( x^\prime)\hat\op^J( x_2)\rangle\\
   - (a\leftrightarrow b, ~  x_0 \leftrightarrow  x_0')= -\langle \hat\op^I( x_1)\delta^{[ab]}\hat\op^J( x_2)\rangle\:,
\end{multline}
where
\ba
\delta^{[ab]}&:=\delta_{K^a_{x_0}}\delta_{K^b_{x_0'}}- (a\leftrightarrow b, ~  x_0 \leftrightarrow  x_0')\:.
\ea
We now proceed as for the tilt: we integrate over the ball $r<1$ and the space $r'>1$, and push one operator to 0 and one to $\infty$. Even though the sum rules can be obtained conformally with this setup, it is easier to consider each one of them separately. There are (a~priori) four different sum rules coming from the four commutators of broken generators~\eqref{eq:brokenC}.

The~one coming from the $[K^a,P^b]$ commutator is  straightforward, as only $r^2$ and $r'^2$ appear
\begin{equation}
    \frac{V_{p-1}V_{p-2}}{2} 
    \iint_{\text{Im}\,z>0,\,|z|<1} \frac{\ud z\ud\bar z}{|z|^{p+1-\Delta_\hcO}}(\text{Im}\,z)^{p-2}(-\log|z|) \big(|z|^2 \mathcal G^{IabJ}-\mathcal G^{IbaJ}\big) = -2(\Delta_\hcO\delta^{ab}\delta^{IJ}+M_{IJ}^{ab})\:,
\end{equation}
leading to two independent sum rules when (anti-)symmetrizing, and where we defined 
\ba
\label{eq:disp-RHS}
M^{ab}_{IJ}:=\langle \hcO^I| M^{ab}|\hcO^J\rangle\,.
\ea
For $[M^{ma},M^{nb}]$, the integration over the angle has to be performed more carefully because of the additional $x^m x'^n$. Projecting orthogonally $x'^n$ on $x^m$ gives a $\cos(\theta-\theta')$ and the integral over the angles
\begin{equation}
    \int \ud^{p-1}\hat x\, \hat x^m \hat x^n = \frac{V_{p-1}}{p} \delta^{mn}\:,
\end{equation}
leading to the sum rule
\begin{equation}
    \frac{V_{p-1}V_{p-2}}{2 p} 
    \iint_{\text{Im}\,z>0,\,|z|<1} \frac{\ud z\ud\bar z}{|z|^{p+1-\Delta_\hcO}}(\text{Im}\,z)^{p-2}\left(\frac{z+\bar z}{2}\right)(-\log|z|) \mathcal G^{I[ab]J} = -M_{IJ}^{ab}\:.
\end{equation}
Finally, the $[M^{ma},P^{b}]$ and $[M^{ma},K^{b}]$ do not give any new constraint in this choice of frame. All in all we have thus three independent sum rules

\setlength\fboxrule{1.2pt}
\setlength{\fboxsep}{8pt} 
{\hspace{-18pt}\noindent\fbox{\parbox{\textwidth-18pt}{
{\bf Displacement double soft sum rules:}
\begin{align}
    \iint_{\text{Im}\,z>0,\,|z|<1} \frac{\ud z\ud\bar z}{|z|^{p+1-\Delta_\hcO}}(\text{Im}\,z)^{p-2}\big(|z|^2-1\big)\log|z|\, \mathcal G^{I(ab)J} &= 4\Delta_\hcO\delta^{ab}\delta^{IJ}\,\frac{\Gamma(p-1)}{(2\pi)^{p-1}}\:, \\
    \iint_{\text{Im}\,z>0,\,|z|<1} \frac{\ud z\ud\bar z}{|z|^{p+1-\Delta_\hcO}}(\text{Im}\,z)^{p-2}\big(|z|^2+1\big) \log|z|\, \mathcal G^{I[ab]J} &=4M_{IJ}^{ab}\,\frac{\Gamma(p-1)}{(2\pi)^{p-1}}\:, \\
    \iint_{\text{Im}\,z>0,\,|z|<1} \frac{\ud z\ud\bar z}{|z|^{p+1-\Delta_\hcO}}(\text{Im}\,z)^{p-2}\left(\frac{z+\bar z}{2}\right)\log|z|\, \mathcal G^{I[ab]J} &=2 M_{IJ}^{ab}\,\frac{p\Gamma(p-1)}{(2\pi)^{p-1}}\:.
\label{eq:dispinhom}
\end{align}
\vspace{-0.3cm}
}}}
   \vspace{0.3cm} 

\noindent
More details on the derivation in the case $p=1$ are given in appendix \ref{app:1d-SR}, and again we find they agree with the ones found in \cite{Gabai:2025zcs}.

\subsubsection{New 2 tilts 2 displacements sum rule}

A special case appears when the correlation function consists of two displacements and two tilts. One can derive an additional double soft sum rule, on top of the ones already derived for the tilt and displacement with two generic defect operators. We apply the second identity~\eqref{eq:id2} again but now take the commutator between a conformal charge and the charge of $O(N)$ global symmetry. This commutator results in a RHS which is zero, but the LHS is nontrivial. Using a particular conformal frame, we get
\begin{equation}
\begin{aligned}
\label{eq:tilt/dispinhom-x}
    \int_{r<1} \ud^p x \int_{r'>1}\ud^p x^\prime\ \Big(x'^n \langle t^j(0) t^i(x)D^a(x')D^b(\infty)\rangle -x^n \langle t^j(0) D^a(x)t^i(x')D^b(\infty)\rangle \Big)=0\:,
\end{aligned}
\end{equation}
where $n=0,1,2$ depending on the chosen conformal Killing vector $\xi$ associated to the integrated displacement. Evaluating these integrals we obtain three sum rules. However, upon using the first identity sum rules~\eqref{eq:4disphom} with two tilts and two displacements, they reduce to only one independent sum rule:

\setlength\fboxrule{1.2pt}
\setlength{\fboxsep}{8pt} 
{\hspace{-18pt}\noindent\fbox{\parbox{\textwidth-18pt}{
{\bf Mixed tilt/displacement double soft sum rules:}
\begin{equation}
    \iint_{\text{Im}\,z>0,\,|z|<1} \ud z\ud\bar z\ (\text{Im}\,z)^{p-2}\log|z|\,\big(\mathcal G^{ibja}-|z|\mathcal G^{abji}\big)  = 0\:.
\label{eq:disptiltinhom}
\end{equation}
\vspace{-0.3cm}
}}}
   \vspace{0.3cm} 

\noindent

\section{An {\em avant-go\^ut} of displacinos}

\label{sec:dispinoSRApp}

\subsection{Setup}
In this section we will consider consequences of broken supersymmetries in the bulk. For lack of space and time, and as the logic is quite similar to the previous sections, we will be brief and consider only a few examples, leaving a more careful analysis and further results for future work.

In this case the broken Ward identity for a bulk super-current takes the form:
\ba
P^+_{\alpha\beta} \partial_\mu J^{\mu,\beta}({x},y)  = \delta^{d-p} (y) \dino_\alpha ({x})\;, \qquad P^+_{\alpha\beta}:=\frac{(1+n_a\Gamma^{a})_{\alpha \beta}}{2} \label{eq:WIsc}\:,
\ea
where we recall that $\Gamma^\mu_{\alpha\beta} J_\mu^{\beta}=0$.
The displacino operator $\dino$ has protected dimension $\Delta_{\dino} = p+\frac 12$ and it transforms as a spinor under the $SO(d-p)$ rotational symmetry. Here we will focus on the simplest case where we have a line defect embedded in a CFT in three spacetime dimensions. In this case the
rotation group is $U(1)$, under which vector operators such as the displacement become a doublet $(D,\bar D)$ of charge $q=(1,-1)$ and the displacino a doublet $(\psi,\bar \psi)$ of charge $q=(\frac 12,-\frac 12)$. The normalization of the displacino two point function is a part of the defect data,
\ba
\langle \psi(x) \bar \psi(0)\rangle=\frac{\normd}{x^{3}} \:,
\ea
and note that the Ward identity implies that 
\ba
    \lim_{|y|\to 0}  y_a\, P^+_{\alpha\beta} J^{a,\beta}(y, x)=\, \dino_\alpha (x)/(2\pi) \:,
\ea
In order to formulate identities \reef{eq:id1} and \reef{eq:id2}, let us set up our conventions for the charges. We consider a three-dimensional bulk $\mathfrak{o}\mathfrak{s}\mathfrak{p} (\mathcal{N}|4;\mathbb{R})$ algebra and consider a one-dimensional defect that might or might not preserve part of the supersymmetry. The \emph{broken} superconformal generators then satisfy the algebra 
\ba
\{Q,Q\}&=2 P\,,& \qquad \{S,S\}&=2 K\,,& \qquad
\{Q,\bar S\}&=-D- R+\ldots\\
\{Q,\bar Q\}&=-P^0\,,& \qquad \{S,\bar S\}&=-K^0\,,& \{Q,S\}&=M\,,& \quad \{\bar Q,\bar S\}&=\bar M\:.
\ea
Here $M,\bar M$ stand for broken rotation generators built from $M^{0a}$ and where the precise form of $R$ and the dots represent contributions of broken generators which depend on the details of the full superconformal algebra. A simple example is where in the bulk we have minimal supersymmetry ($\mathcal{N}=1$) in which case the line defect does not preserve any supersymmetry and we have simply $R=-i M_{12}$, and $[R,Q]=\frac 12 Q, [R,P]=P$, etc. More generally it may mix with broken $R$-symmetry generators.
When acting on the defect we have 
\ba
    Q \ket{0}_\mathcal{D} &=\int_{ x_0<T} \mathrm{d} x\, \dino(x) \ket{0}_\mathcal{D}\:,\\
     S |0\rangle_\mathcal{D} &=\int_{ x_0<T} \mathrm{d} x\,  x\, \psi(x)|0\rangle_\mathcal{D}\:,
\ea    
After these preliminaries we are ready to consider the specialization of identities \reef{eq:id1} and \reef{eq:id2}. We will consider only consequences of the identities for defect correlators

\subsection{Defect correlator identities}
We begin with the first identity. As for the displacement, we can repackage the identities corresponding to $S,Q$ in a single formula, namely
\ba
\int_{-\infty}^{\infty} \ud x \, (x-x_0) \langle \psi(x)\hcO_1(x_1) \hcO_2(x_1)\hcO_3(x_1)\,.
\ea
manifesting their conformal invariance.
As usual, this identity can be derived by first starting with bulk insertions and then pushing them towards the defect. We assume as usual that terms on the righthand side arising from the BOE of rotated bulk operators may be dropped. Let us now specialize to the case where one of the insertions is a displacino, and the other two are identical bosonic or fermionic defect operators. Then a simple calculation yields

\setlength\fboxrule{1.2pt}
\setlength{\fboxsep}{8pt} 
{\hspace{-18pt}\noindent\fbox{\parbox{\textwidth-18pt}{
{\bf Displacino soft sum rules:}
\ba
\label{eq:dinohom}
\int_0^1\, \ud z\, \left[2 z^{\hdO-\frac{1}{2}}\mathcal G^{\bar \cO\bar \cO \psi\psi}(z)+\eta^{\hcO}\,\mathcal G^{\bar \cO\psi \bar \cO \psi}(z) \right]&=0 \:, \\
\int_0^1\, \ud z\, \left[ z^{\hdO-\frac 12}\mathcal G^{\cO\bar \cO (\psi\bar\psi)}(z)+\eta^{\cO}\,\mathcal G^{\cO\psi \bar \cO \bar \psi}(z) \right]&=0 \:, \\
\int_0^1\, \ud z\, \Big[ z^{\hdO-\frac 32}\left[\mathcal G^{\cO\bar \cO \psi\bar\psi}(z)+(z-1)\mathcal G^{\cO\bar \cO \bar \psi\psi}(z)\right]
+\,z\,\eta^{\cO} \,\mathcal G^{\cO\psi \bar \cO \bar \psi}(z) \Big]&=0 \:,
\ea
\vspace{-0.3cm}
}}}
 \vspace{0.3cm} 
 
\noindent
with $\eta^\cO=\pm 1$ for bosonic/fermionic operators.
Specializing to $\cO=\dino, \bar \cO=\bar \dino$ they all reduce~ to
\ba
\int_0^1\, \ud z\, \Big[ \mathcal G^{\psi\bar \psi \psi\bar\psi}(z)+(z-1)\mathcal G^{\psi\bar \psi \bar \psi\psi}(z)-z
\,\mathcal G^{\psi\psi \bar \psi \bar \psi}(z) \Big]&=0\:.
\ea
Moving on, now let us consider the double soft sum rules. We find that there are three independent constraints: 

\setlength\fboxrule{1.2pt}
\setlength{\fboxsep}{8pt} 
{\hspace{-18pt}\noindent\fbox{\parbox{\textwidth-18pt}{
{\bf Displacino double soft sum rules:}
\ba
\label{eq:dinoinhom}
\int_0^1 \ud z\,\big[z^{\hdO-\frac 32}(1+z)\,\mathcal G^{\cO(\psi \bar\psi)\bar\cO}(z)\, \log(z)-\eta^{\hcO} (1-2z) \mathcal G^{\cO\psi \bar \cO \bar \psi}(z)\log\left(\tfrac{1-z}z\right)\Big]&=2\hdO\langle \cO|\bar \cO\rangle\\
\int_0^1 \ud z\,\big[z^{\hdO-\frac 32}\, (1-z) \,\mathcal G^{\cO[\psi \bar \psi]\bar \cO}(z)\, \log(z)- \eta^{\hcO}  \mathcal G^{\cO\psi \bar \cO \bar \psi}(z)\log\left(\tfrac{1-z}z\right)\Big]&=-2 R_{\cO}\langle \cO|\bar \cO\rangle\\
\int_0^1\ud z\,\Big[ 2z ^{\hdO-\frac 32}(1+z) \mathcal G^{\bar \cO \psi\psi \bar \cO}(z)\, \log(z)-\eta^{\hcO} (1-2z) \, \mathcal G^{\bar \cO \psi \bar \cO \psi}(z)\, \log\left(\tfrac{1-z}z\right)\Big]&=0\:.
\ea
\vspace{-0.3cm}
}}}
   \vspace{0.3cm} 
   
\noindent
In the special case where we take $\cO,\bar \cO$ to be the displacino itself, these reduce to two independent sum rules which may be chosen as the first and third ones above.

It is possible that the defect still preserves some amount of supersymmetry, the minimal amount being two supercharges \footnote{See \cite{Agmon:2020pde} for a full classification of allowed supersymmetric line defects.}. In this case the displacino and displacement operators are related by supersymmetry and so are their correlation functions. Using this we have succeeded in verifying that at least some of the displacement sum rules then reduce to the above.

\section{Functionals and sum rules }\label{sec:funcSR}
In the previous two sections we worked out a number of integrated constraints on defect four-point correlators as well as form factors. Plugging the DOE and OPE these integrated constraints directly translate into sum rules on defect CFT data. However it can be advantageous to make suitable combinations of such sum rules with those arising from crossing of four-point functions and locality of form factors. In this way we can arrive at sum rules which are essentially diagonalized by sparse correlators such as those arising in free theory limits, allowing us to obtain analytic results in perturbation theory. As we will see, it is also possible to obtain non-perturbative, tight and rigorous bounds on certain pieces of CFT data.

In this section we will focus on the case $p=1$. The soft and double-soft sum rules for $p=1$ are given explicitly in appendix~\ref{app:1d-SR}, while the form factor sum rules can be found by simply setting $p=1$. For the form factor sum rules, a generalization of this section to $p>1$ is straightforward. As for defect correlators it would be interesting to combine the sum rules derived in previous sections with the dispersive formulae derived in \cite{Caron-Huot:2017vep,Caron-Huot:2020adz} or perhaps using a Mellin space approach \cite{Penedones:2019tng,Gopakumar:2021dvg}.
\subsection{Form factors}
\label{ssec:SRcases}
We begin by considering form factor sum rules. 
Consider for instance the constraint on form factors $\langle t \hcO_s \Phi\rangle$ given in equation \reef{eq:tsPhi}. We can use the DOE to write this as
\ba
\int_0^1 \frac{\ud w}{w^{\frac{\Delta_s}2+1}\sqrt{1-w}} \left(\sum_{\Delta} \mu^{\Phi}_{\Delta}\lambda_{ts\Delta} H_{\Delta}(w)\right) =\mu^{\Phi}_{s}\,. \label{eq:naive}
\ea
We could now proceed by performing the integral term by term. Instead, let us use a dispersion relation for the form factor:\footnote{We have used the definition of the discontinuity 
\ba
    \mathcal{I}_z [f(z)] = \lim_{\epsilon \to 0^{+}} \frac{f(z + i \epsilon) - f(z - i \epsilon)}{2 \pi i}\:.
\ea
}
\ba
\mathcal H^{\Phi}_{ts}(w)=\int_{-\infty}^0 \frac{\ud w'}{\pi} \frac{w^{\alpha+1}}{w'(w'-w)}\, \mathcal I_w[(w')^{-\alpha} \mathcal H^{\Phi}_{ts}(w')]\:.
\ea
This dispersion relation always holds for sufficiently large parameter $\alpha$. The advantage of doing this is that now plugging in the DOE now automatically kill blocks with dimension $\Delta=2\alpha+2n$. Plugging this formula in the previous one we can perform both integrals to arrive at
\ba
\sum_{\Delta} \mu^{\Phi}_{\Delta}\lambda_{ts\Delta} \, \kappa^{\Phi}_{ts} (\Delta)=\mu^{\Phi}_s\:,
\ea
where the precise $\kappa$ functional depends on the choice of $\alpha$ but satisfies
\ba
\kappa^{\Phi}_{ts}(\Delta) \underset{\Delta\to \Delta_s}=\frac{2}{\Delta-\Delta_s}\,, \qquad \kappa^{\Phi}_{ts}(2\alpha+2n)=0\,, \quad n\geq 0\,. 
\ea

In practice, our trick of introducing a dispersion relation is equivalent to replacing the conformal blocks in \reef{eq:naive} by so-called local blocks \cite{Levine:2023ywq}, which manifest locality of the form factor (namely smoothness for $w\geq 1$) . The advantage of using the local blocks is that the above sum rule can be chosen to have zeros on the free theory result by setting $\alpha=1+\hat \Delta_s$.
This makes it particularly useful to extract new defect CFT data in perturbation theory as we will see later on.

Let us now provide an explicity formula for the functional. We actually do a slightly more general computation which can be used to find sum rules for general form factor constraints. Concretely: 
\ba
~&I(\Delta|\Delta_1,\Delta_2,\alpha,\beta):=\int_0^1 \frac{\ud w}{\sqrt{w(1-w)} w^{\frac{\Delta_1+\Delta_2}2}} w^{1+\beta+\alpha} \, \int_{-\infty}^0 \frac{\ud w'}{\pi}\, \frac{\mathcal I_{w'}[w'^{-\alpha} H_{\Delta}^{\Delta_{12}}(w')]}{w'(w'-w)}\\
~ & =
\sqrt{\pi } \Gamma \left(\Delta +\tfrac{1}{2}\right) \left(\Gamma \left(\tfrac{2 \beta +\Delta -\Delta _1-\Delta _2+1}{2}\right) \,
\pFqt{3}{2}{
\tfrac{\Delta +\Delta
   _{12}}2,
   \tfrac{\Delta +\Delta _{12}}2,
   \tfrac{2 \beta +\Delta -\Delta _1-\Delta _2+1}2
   }
{
\tfrac{2 \beta +\Delta -\Delta _1-\Delta
   _2+2}2,
\Delta +\frac{1}{2}}{1} \right.
\\
  ~ & \left.-\frac{\Gamma \left(\tfrac{2\alpha +\Delta _{12}+2}2\right)^2 \Gamma
   \left(\tfrac{2\alpha +2\beta -\Delta _1-\Delta _2+3}{2}\right)}{\Gamma \left(\tfrac{\Delta +\Delta _{12}}2\right)^2}\,
   \pFqt{4}{3}{
     1,\tfrac{2\alpha +\Delta _{12}+2}2,\tfrac{2\alpha +\Delta_{12}+2}2,\tfrac{2\alpha +2\beta -\Delta _1-\Delta _2+3}{2}}
     {\tfrac{2\alpha +2\beta -\Delta _1-\Delta _2+4}{2},\tfrac{2\alpha -\Delta+4}2,\tfrac{2 \alpha
   +\Delta +3}2}{1} \right)\:.\label{eq:formSRgen}
\ea
We now have 
\ba
\kappa^{\Phi}_{ts}(\Delta)=I(\Delta|1,\Delta_s,\alpha,0)\:.
\ea
In special cases it is possible to provide simpler formulae. Consider the constraints \reef{eq:psitv} and \reef{eq:phitt} as applied to two tilt defect insertions, setting $\alpha=1$. Then we get
\ba
\sum_{\Delta} \lambda_{tt\Delta} \mu^{\Psi}_{\Delta}\, \kappa^{\Psi}_{tt}(\Delta)&=0 \:,\\
\sum_{\Delta} \lambda_{tt\Delta} \mu^{\Phi}_{\Delta}\, \kappa^{\Phi}_{tt}(\Delta)&=\frac{\mu_{\mathds 1}^{\Phi}}{\sqrt{\normt}}\:,
\ea
with
\ba
\kappa^{\Psi}_{tt}(\Delta)=\kappa^{\Phi}_{tt}(\Delta)=\frac{2^{\Delta +2}}{\sqrt{\pi}}\,\frac{ \sin \left(\frac{\pi  \Delta }{2}\right)
   \Gamma \left(\Delta +\frac{1}{2}\right)}{(2-\Delta) (\Delta -1) \Gamma (\Delta +2)}\:.\label{eq:PhittFunc}
\ea

The form factor sum rules including the displacement are also of the form \eqref{eq:formSRgen}. We have for example
\ba
    \kappa^{\Psi,1}_{D s} &= I(\Delta|2,\Delta_s,\alpha,0)\:, &&\quad \kappa^{\Psi,2}_{D s} = I(\Delta|2,\Delta_s,\alpha,1)\:, \\
    \kappa^{\Psi,1}_{DD,S,T} &= I(\Delta|2,2,\alpha,0)\:, &&\quad \kappa^{\Psi,2}_{DD,S,T} = I(\Delta|2,2,\alpha,1)\:.
\ea

\subsection{Defect correlators}
Now let us move on to discuss defect correlators, and in particular integrated constraints on four tilts or four displacements. We will comment on what can be done for more general four-point functions later. 

For both tilts and displacements there are $O(N-1)$ indices to worry about, but let us first ignore this. An integrated constraint then takes the schematic form
\ba
\int_0^1 \ud z\, h(z) \mathcal G(z)=C\:,
\ea
where $h(z)$ is some relatively simple function and $\mathcal G$ is the correlator for identifical fields of dimension $\Df$ (in practice $\Df=1$ or $2$). When plugging in the OPE this gives the sum rule
\ba
\sum_{\Delta} \lambda^2_{\Delta} \omega(\Delta)=C\,, \qquad \omega(\Delta)=\int_{0}^1 \ud z\, h(z)\, G_{\Delta}(z)\:,
\ea
Strictly speaking the functional action $\omega(\Delta)$ as written above is initially defined only for sufficiently large $\Delta$ and everywhere else by analytic continuation. As discussed previously, we would like to obtain an improved version of this sum rule which is trivialized by nearly free solutions. The idea for doing so is similar to what we did for the form factors, that is, to use a dispersive formula for the correlator. However here it is more useful to consider such a formula involving the double discontinuity rather than the single:
\ba
\mathcal G(z)=-\int_0^1 \ud z' g_\Df(z',z) \ud^2\,\mathcal G(z')\,, \qquad \ud^2 \mathcal G(z):=\mathcal G(z)-(1-z)^{-2\Df}\,\mbox{Re}\, \mathcal G(\tfrac{z}{z-1})\:.
\ea
This is because this gives
\ba
\ud^2 G_{\Delta}(z)=2\sin^2\left[\tfrac{\pi}2(\Delta-2\Df)\right]\, G_{\Delta}(z)\:,
\ea
and this double-zero property will translate into that of the functional. In previous work \cite{Paulos:2020zxx}, it was demonstrated how suitable $g_{\Df}(z',z)$ can be found. Although generically their construction is involved, it does simplify significantly for integer $\Df$. We can then hope to construct a new functional by setting
\ba
\rho(\Delta):=\int_0^1 \ud z\, m(z) \ud^2 G_{\Delta}(z)\,, \qquad m(z):=-\int_{0}^1 \ud z'h(z) g_{\Df}(z',z)\:.
\ea
As the expressions for the $g_{\Df}$ are quite involved, it is better to use a shortcut. In particular it is possible to directly construct $m(z)$. Let us see how this can be done. As explained in \cite{Paulos:2020zxx}, the kernel $g_{\Df}(z',z)$ is found by solving the following functional equation for $z\in (0,1)$:
\ba
z^{2\Df-2}\mbox{Re}_z\, g_{\Df}(z',\tfrac{z-1}z)-g_\Df(z',z)-g_{\Df}(z',1-z)=\delta(z'-z)+\delta(1-z'-z)\label{eq:funceqn}
\ea
for analytic $g$ away from $z<0$ and $z>1$. Thus to find $m(z)$ we can solve instead
\ba
z^{2\Df-2}\mbox{Re}_z\, m(\tfrac{z-1}z)+m(z)+m(1-z)=-h(z)-h(1-z)\,\quad z\in(0,1)\:.
\ea
For simple $h(z)$ this can be found by using an ansatz. Consider for instance the four tilts soft sum rule \reef{eq:tilthom-p=1}. In this case $\Df=1$ and $h(z)=1$, and hence we have to solve
\ba
\mbox{Re}_z\, m(\tfrac{z-1}z)+m(z)+m(1-z)=-2
\ea
which has the solution
\ba
m(z)=\frac 23\, \frac{(1-z)^2}{z}\,.
\ea
It is now straightforward to compute $\rho(\Delta)$. Let us define
\ba
\rho(\Delta)&=\frac{4 \sin^2(\tfrac{\pi}2\Delta)}3\, \int_0^1 \frac{\ud z}{z} (1-z)^2 G_{\Delta}(z)\\
&=\frac 43 \frac{\sin^2(\tfrac{\pi}2\Delta)}{a^{\tt gff}_{\Delta}}\,
\frac{(\Delta -\tfrac 12) \big(6-2 c_\Delta +(\Delta -2)_4 \Psi(\Delta) \big)}{(\Delta -2) (\Delta +1)}\:,
\ea
where for convenience we defined:
\ba
\Psi(\Delta):=\psi ^{(1)}\left(\frac{\Delta
   }{2}\right)-\psi ^{(1)}\left(\frac{\Delta +1}{2}\right)\,, \qquad c_\Delta:=\Delta(\Delta-1)\:,
\ea
and denoted with $a_{\Delta}^{\tt gff}$ the GFF OPE density, evaluated for $\Df=1$:
\ba
a_{\Delta}^{\tt gff}=\frac{\sqrt{\pi } 2^{3-2 \Delta } \Gamma (\Delta ) \Gamma \left(\Delta +2 \Delta _{\phi }-1\right)}{\Gamma \left(\Delta -\frac{1}{2}\right) \Gamma \left(\Delta -2 \Delta _{\phi }+1\right) \Gamma
   \left(2 \Delta _{\phi }\right)^2}\:.
\ea
From this result we get the four tilts soft sum rule:
\ba
    \sum_{I,\hat{\Delta}} (\lambda^{I}_{tt\hat{\op}})^2 f_{I}^{t,\text{soft}} (\hat{\Delta}) = 0\:, \quad I = S, T, A\:,
\ea
where
\ba
    f_{S}^{t,\text{soft}} (\Delta) &= \rho(\Delta)\,, \quad    f_{T}^{t,\text{soft}} (\Delta) = \left(\frac{N-2}{N-1}\right) \rho(\Delta) \:, \quad 
    f_{A}^{t,\text{soft}} (\Delta) &= 0\:.\label{eq:tiltHomFunc}
\ea
A very similar logic holds for the first displacement soft sum rule \eqref{eq:DH1}. In this case $\Df=2$ and $h(z)=z(z-1)+1$ and we find
\ba
m(z)=\frac 25 \, \frac{(1-z)^4}{z}\:.
\ea
The sum rule looks similar,
\ba
    \sum_{I,\hat{\Delta}} (\lambda^{I}_{tt\hat{\op}})^2 f_{I}^{D,\text{soft}1} (\hat{\Delta}) = 0\:, \quad I = S, T, A\:,
\ea
where the channels now refer to the transverse rotation group $O(d-p)$
Again we get a relatively simple functional action with double zeros: 
\ba
f_{T}^{D,\text{soft}1}(\Delta)=\left(\frac{d-2}{d-1}\right)\,f_{S}^{D,\text{soft}1}(\Delta)\,, \quad f_A^{D,\text{soft}1}(\Delta)=0\:,
\ea
\ba
    f_{S}^{D,\text{soft}1}= 
\frac{\sin^2(\tfrac{\pi}2\Delta)}{a^{\tt gff}_{\Delta}}
\frac{2 \Delta -1}{(\Delta-4)(\Delta+3)}\,\Big(600-2 c_\Delta[130+c_\Delta(c_\Delta-21)]+
(\Delta-4)_8\, \Psi(\Delta)\Big)\:.
\ea
Before proceeding, let us explain how we may arrive at similar sum rules for those integrated constraints where the $O(N-1)$ structure is important. These take the form
\ba
\int_0^1 \ud z\, \sum_I h^I(z) \mathcal G^I(z)=C\:,
\ea
where the sum runs over $S,T,A$. The logic is a reasonably straightforward generalization of what we have discussed above, see \cite{Ghosh:2021ruh} for details. Now we express
\ba
\mathcal G_I(z)= \int \ud z' \sum_{J} g_\Df^{I,J}(z,z')\, \ud^2_J \mathcal G^J(z')\:,
\ea
where $\ud^2_I \mathcal G$ is defined as before for $I=S,T$ but 
\ba
\ud^2_A \mathcal G(z):=\mathcal G(z)+(1-z)^{-2\Df}\,\mbox{Re}\, \mathcal G(\tfrac{z}{z-1}) \quad \Rightarrow\quad \ud^2_A G_{\Delta}(z)=2\cos^2[\tfrac{\pi}2(\Delta-2\Df)]\, G_{\Delta}(z) \:.
\ea
The kernel $g_{\Df}^{I,J}$ satisfies a set of functional equations very similar to \reef{eq:funceqn}, but it is simpler to directly solve them for
\ba 
m^J(z):= \int_0^1 \ud z'\,\sum_I h^I(z') g_{\Df}^{I,J}(z',z)\:.
\ea
Let us show one concrete example of this procedure. For the tilt double-soft sum rule \reef{eq:4t-inhomogeneous} we find
\begin{multline}
    \int_0^{1} \mathrm{d}z \bigg[
    \Big(
    \mathcal G^{(1)}_+(z)-3\,\mathcal G^{(2)}_{ +}(z)\Big)
    \log[z(1-z)]-3\,\mathcal G_-(z) \log\left(\tfrac{1-z}z\right)
    \bigg]\\
    =
    \int_0^1 \ud z \sum_I m^I(z) \ud^2_I \mathcal G^I(z)
\end{multline}
with
\ba
m^S(z)&=-\frac{4 \left(z^2+z-3 z \log (z)-2\right)}{3 z}\,, \\
m^T(z)&=-\frac{N+1}{2(N-1)} m^S(z)\,, \\
m^A(z)&=-2(z-1)+2 \log (z)\,.
\ea
Plugging in the OPE and integrating block by block we find the sum rule
\ba
    \sum_{I,\hat{\Delta}} (\lambda^{I}_{tt\hat{\op}})^2 f_{I}^{t,\text{dsoft}} [\hat{\Delta}] = \frac{1}{\normt}\:, \quad I = S, T, A\:,
\ea
where 
\ba
    f_{S}^{t,\text{dsoft}} [\Delta] &= \frac{4}{3}\,\frac{\sin^2[\tfrac{\pi}2(\Delta)]}{a_{\Delta}^{\tt gff}}\, \frac{2\Delta-1}{\Delta(\Delta-1)}\, \left[
 c_\Delta^2\, \Psi (\Delta )-\frac{2 c_\Delta  (c_\Delta -3)+12}{(\Delta -2) (\Delta
   +1)}
\right]\,, \\
   f_{T}^{t,\text{dsoft}} [\Delta] &= - \frac{N+1}{2(N-1)}\, f_{S}^{t,\text{dsoft}}[\Delta]\:,  \\
   f_{A}^{t,\text{dsoft}} [\Delta] &=4\,\frac{\sin^2[\tfrac{\pi}2(\Delta)]}{a_{\Delta}^{\tt gff}}\, \frac{2\Delta-1}{\Delta(\Delta-1)}\left[2-2 c_\Delta\,+ c_\Delta^2  \Psi (\Delta )\right]\,.\label{eq:tiltInhom}
\ea
We have also applied this procedure to the remaining displacement soft and double soft constraints. The results are provided in the attached notebook.

Let us briefly comment on how this procedure can be generalized to more general correlators which do not involve identical operators. In this case it has not yet been worked out how to write down dispersive sum rules for such correlators. However, dispersive functional bases have been constructed for general mixed systems of correlators \cite{Ghosh:2023wjn}. 
Thus we can improve the integral constraints derived in this work  `by hand', i.e. by combining said bases with appropriately chosen coefficients so as to produce (double) zeros at desired locations.

\subsection{Simple bounds}
\begin{figure}
    \centering
    \begin{subfigure}{0.5\linewidth}
    \includegraphics[width=0.9\linewidth]{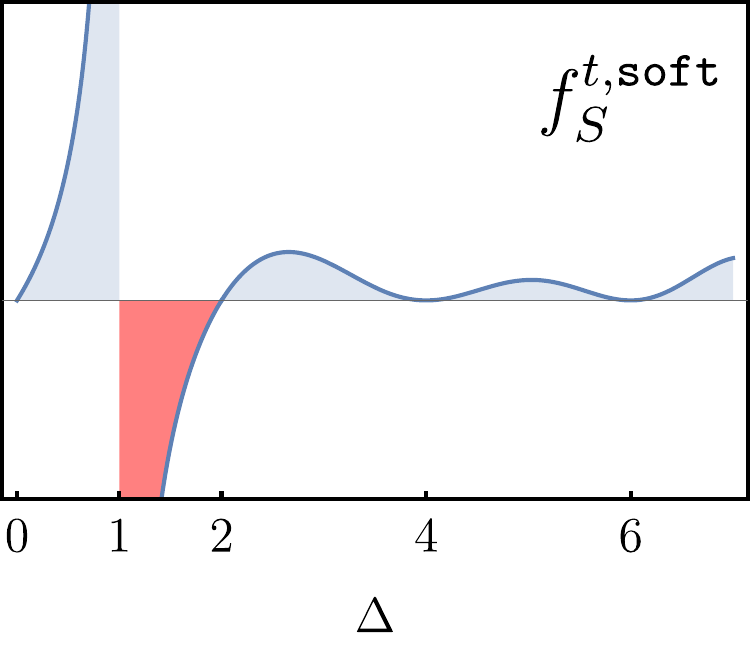}
    \end{subfigure}\begin{subfigure}{0.5\linewidth}
        \includegraphics[width=0.9\linewidth]{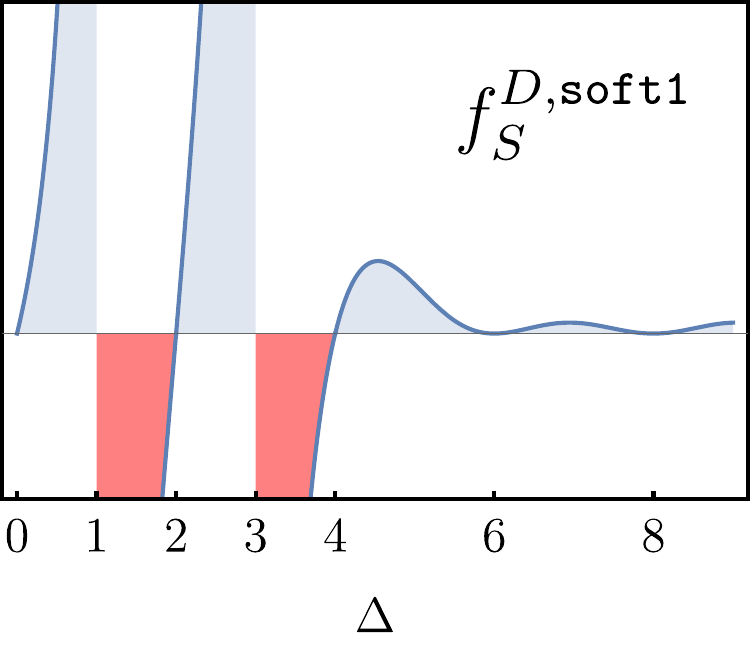}
    \end{subfigure}
    \caption{Soft sum rules and bounds. In this figure we show the S-channel action of soft sum rules on tilts and displacements, which have simple poles at $\Delta=1$ for the first and $\Delta=1,3$ for the second. Negative contributions to the sum rule from the negative regions in the vicinity of the poles are required to cancel positive contributions from the blue regions.}
    \label{fig:funcSR}
\end{figure}

The sum rules derived above can be combined with those for crossing and locality in a full numerical bootstrap setup, and we will take this up in future work. Nevertheless, we will now show that even without doing any further work it is already possible to obtain non-trivial rigorous bounds on defect CFT data. Concretely, let us examine the soft sum rules derived above for both tilts and displacements. In both cases the $A$ channel does not participate and the functional action in the $T$ channel is proportional (with a positive constant) to that in the $S$ channel. They can be written as
\ba
\sum_{\Delta\in t\times t} a_{tt,\Delta}f_S^{t,\tt soft}(\Delta)=0\,, \qquad \sum_{\Delta\in D\times D} a_{DD,\Delta}f_S^{D,\tt soft1}(\Delta)=0
\ea
with
\ba
a_{tt,\Delta}:=(\lambda^S_{tt\Delta})^2+\frac{N-2}{N-1}(\lambda^T_{tt\Delta_S})^2\,, \qquad a_{DD,\Delta}:=(\lambda^S_{DD\Delta_S})^2+\frac{d-2}{d-1}(\lambda^T_{DD\Delta})^2\,.
\ea
In figure \ref{fig:funcSR} we plot the two relevant functional actions. By positivity we can immediately deduce the following:

\setlength\fboxrule{1.2pt}
\setlength{\fboxsep}{8pt} 
{\hspace{-18pt}\noindent\fbox{\parbox{\textwidth-18pt}{
{\bf Spectral bounds on line defects}
\begin{itemize}
\item Conformal line defects must have an operator of dimension $\Delta$, which is either a singlet or traceless symmetric tensor under transverse rotations satisfying
\ba
\Delta\in [1,2] \cup [3,4]\:.
\ea
\item If the defect breaks a global symmetry then there must exist a parity-even operator with dimension $\Delta$ satisfying
\ba
\Delta\in [1,2]\:.
\ea
\end{itemize}
\vspace{-0.3cm}
}}}
   \vspace{0.3cm} 
\noindent

The second item above requires some explanation. Although we focused on a particular pattern of symmetry breaking, under the stated assumption we can obtain similar tilt sum rules much more generally. In particular it is always true that
\ba
\int \ud  x \, \langle t^{i}(x)t^{i}(x_1)t^{i}(x_2)t^{i}(x_3)\rangle=0\:.
\ea
Then the sum rule, initially derived for $O(N)$ still applies for more general patterns of global symmetry breaking, when restricted to those operators appearing in the $t^{i}\times t^{i}$ OPE, which are necessarily parity even.\footnote{Recall we are focusing on parity and inversion preserving defects in this work.}

These sum rules significantly strengthen the 1d CFT bootstrap bounds. For instance, when maximizing the gap in the OPE of equal external operators, the result is a correlator describing generalized free fermions with the gap $\Delta_g^{\tt max}=1+2\Df$ \cite{Mazac:2016qev,Mazac:2018mdx}. In the tilt case this would give $\Delta_g^{\tt max}=3$, but thanks to the sum rule above this is now lowered to $\Delta_g^{\tt max}=2$. Similarly for displacements we have $\Delta_g^{\tt max}=5 \to 4$. For tilts our bound is tight, as exemplified for instance by the 1/2 BPS line defect in $\mathcal N$=4 SYM. In that context our bound translates into a constraint on the dimension of the first long operator, which indeed has  dimension interpolating between 1 and 2.

\section{The pinning line defect in the \texorpdfstring{$O(N)$}{O(N)} model revisited}\label{sec:pinning}

In order to test the validity of the sum rules above we apply them in section \ref{sec:bounds} to a well-studied perturbative model: the pinning line defect in the $O(N)$ CFTs. The bulk CFT is the O(N) Wilson-Fisher model in $d = 4 - \varepsilon$ dimensions, described by the Lagrangian
\begin{equation}
    \mathcal{L}_{\text{bulk}} = 
    \frac{1}{2} (\partial \Phi_I)^2 + \frac{\lambda_0}{4!} (\Phi_I\Phi^I)^2 \:,
\end{equation}
where $\Phi_a,\, a = 1, \ldots, N$ is a vector under the global $O(N)$ symmetry and the mass term is tuned to zero. The coupling $\lambda_0$ gets renormalized:
\begin{align}
\blambda_0 &= M^\veps \blambda\left(1 + \frac{N+8}{3\veps}\, \blambda - \frac{3N+14}{6\veps}\, \blambda^2 + \frac{(N+8)^2}{9\veps^2}\, \blambda^2  + \op(\lambda^3)\right). 
\end{align}
For ease of notation, we have defined the reduced coupling $\bar{\lambda}_0=\frac{\lambda_0}{(4\pi)^2}$.
Its $\beta$-function and the Wilson-Fisher fixed point are given by 
\begin{equation}
    \blambda_* = \frac{3\veps}{N+8} + \frac{9(3N+14)\veps^2}{(N+8)^3} + \op(\veps^3).
\end{equation}
The pinning defect is constructed by integrating one component of the vector field $\Phi^I$ over an infinite line. Here we will pick $\Phi^1$ such that the defect CFT is given by
\begin{align}
    S = S_{\text{bulk}} + h_0 \int^{\infty}_{-\infty} \ud x\, \Phi^1 (x,y=0)\:, \label{eq:pinningaction}
\end{align}
with $x$ the direction along the defect, and $y$ the direction transverse to the defect. The operator $\Phi(x,y=0)$ that is integrated is a defect operator, it will be denoted in the following as $\hat{\phi}(x)$. The bare coupling $h_0$ gets renormalized as \cite{Allais:2014fqa}
\begin{align}
h_0 &= M^\frac{\veps}{2}h \left(1 + \frac{\blambda h^2}{12\veps} + \frac{\blambda^2}{12\veps}\Big(\frac{N+2}{6}-\frac{N+8}{9}h^2-\frac{h^4}{4}\Big) + \frac{\blambda^2}{12\veps^2}\Big(\frac{N+8}{9}h^2-\frac{h^4}{8}\Big) + \op(\lambda^2) \right)\:.
\end{align}
leading to the $\beta$-function and fixed point
\ba
    \blambda_* &= \frac{3\veps}{N+8} + \frac{9(3N+14)\veps^2}{(N+8)^3} + \op(\veps^3),\\
    \qquad h^2_* &= (N+8) + \frac{4N^2+45N+170}{2(N+8)}\veps+ \op(\veps^2)\:. \label{eq:pinningFP}
\ea
We will give results for this model perturbative in the bulk coupling $\lambda_0$. From \eqref{eq:pinningFP} one can see that the defect coupling $h_0$ is not small, but of order $O(1)$, so at every order in $\lambda_0$, an arbitrary number of defect insertions are allowed. However, in practice this number is finite.

The defect CFT data has been studied perturbatively in the $d=4-\varepsilon$ expansion in the literature to first and second order in $\varepsilon$ \cite{Cuomo:2021kfm,Gimenez-Grau:2022czc,Gimenez-Grau:2022ebb,Bianchi:2022sbz, Gabai:2025zcs, Belton:2025hbu}. However, in order to test our newly found sum rules, we need additional data, especially for bulk-to-defect OPE coefficients. In particular, we want to compute the correlator $\langle \Phi_I \hat{\phi}_j \hat{\phi}_k \rangle$ up to $O(\varepsilon)$, and a handful of bulk-to-defect OPE coefficients that will only appear at higher order in this correlator. We report more results, e.g. for $\langle (\Phi_I \Phi_J) \hat{\phi}_k \hat{\phi}_l \rangle$, in appendix \ref{app:morepinning}.

\subsection{Conventions and summary of the results}

We follow the conventions of~\cite{Cuomo:2021kfm,Gimenez-Grau:2022czc}. From the action~\eqref{eq:pinningaction}, we find the scalar propagator in free theory
\begin{align}
\begin{tikzpicture}[baseline,valign]
  \draw[dashed] (0, 0) -- (0.8, 0);
 \end{tikzpicture} 
 \;\, \equiv \;
 \langle \Phi_I(x_1,y_1) \Phi_J(x_2,y_2) \rangle_{\lambda_0=h_0=0}
 \, = \;
 \frac{C_{\Phi,\text{free}} \, \delta_{IJ}}{\left(y_{12}^2+x_{12}^2\right)^{\Delta_{\Phi,\text{free}}}} \, ,
 \label{eq:free-prop}
\end{align}
with the free theory values
\begin{equation}
\Delta_{\Phi,\text{free}} = \frac{d-2}{2}\, , \qquad \kappa \equiv C_{\Phi,\text{free}} = \frac{\Gamma \! \left(\frac{d}{2}\right)}{2 \pi ^{d/2} (d-2)}\, .   
\end{equation}
From the action, we can also extract the vertices
\begin{equation}
\begin{tikzpicture}[baseline,valign]
\draw[dashed] ( 0.2, 0.2) -- (-0.2, -0.2);
\draw[dashed] (-0.2, 0.2) -- ( 0.2, -0.2);
\node at (0,0) [bcirc] {};
\end{tikzpicture} 
\;\; \equiv \; 
- \lambda_0 \int \mathrm{d}^d X \ldots \, , \qquad
\begin{tikzpicture}[baseline,valign]
\draw[thick] (-0.1, 0) -- (0.7, 0);
\draw[dashed](0.3, 0) -- (0.3, 0.6);
\node at (0.3, 0) [dcirc] {};
\end{tikzpicture}
\;\; \equiv \;
- h_0 \int_{-\infty}^\infty \mathrm{d} x \ldots \, .  
\end{equation}
The scaling dimensions and normalizations of the bulk and defect operators acquire corrections in $\veps$. Defining the renormalized fundamental fields $[\Phi_I]$ as
\be
    \Phi_I = Z_{\Phi} [\Phi_I]\:,
\ee
with $Z_{\Phi}$ the wavefunction renormalization, the anomalous dimensions are given by
\begin{equation}
    \gamma = \beta_\lambda \frac{\partial \log Z}{\partial \lambda} + \beta_h \frac{\partial \log Z}{\partial h}\:.
\end{equation}
Tuning $\lambda, h$ to the fixed points $\lambda_{\star},h_{\star}$ then gives the corrections to the scaling dimensions $\Delta_{\op}$.
The corrections to the operator normalizations can be found from the two-point functions and have been computed e.g. in \cite{Cuomo:2021kfm,Gimenez-Grau:2022ebb}. 
The results for the displacement and the tilt, which are important for our checks of the defect soft sum rules are summarized in table~\ref{tab:CFT-data}.
For the bulk operators, we consider $\Phi_I$ as well as products of $\Phi_I$ in various representations of $O(N)$:
\be
    \Phi^2 = \Phi_I \Phi_I\:, \quad T_{IJ} = \Phi_{I} \Phi_{J} - \frac{\delta_{IJ}}{N} \Phi_K \Phi_L\:.
\ee
For the defect operators, the lowest-dimensional ones are defined in terms of $\hat{\phi}_1, \hat{\phi}_{i}$ as
\ba
    t_{i} = h_{\star} \hat{\phi}_{i}\:, \quad \hat{V}_{i} = \hat{\phi}_1 \hat{\phi}_{i}\:, \quad \hat{T}_{ij} = \frac{1}{2}\hat{\phi}_{(i}\hat{\phi}_{j)} - \frac{\delta_{ij}}{N-1} \hat{\phi}_{k} \hat{\phi}_{k}\:.
\ea
The operators $\hat{s}_{\pm}$ are defined in \cite{Cuomo:2021kfm} and given as a linear combination of $\hat{\phi}_{1}^2$ and $\hat{\phi}_{i}^2$:
\be
    \hat{s}_{\pm} = \frac{N + 16 \mp \sqrt{N(N+40) + 320}}{4(N+8)} \hat{\phi}_i^2 - \hat{\phi}_1^2 + \op(\veps)\:.
\ee
The linear combinations will receive corrections at $\op(\veps)$.
We also consider defect operators with transverse derivatives
\begin{equation}
    D_{a}= \hat{\op}^S_{1,0} = h_\star\partial_{a}\hat{\phi}_1, \quad \hat{\op}^V_{1,0} = \partial_a\hat{\phi}_{i}\:, \quad \hat{\op}^S_{2,0} = \partial_{a} \partial_{b} \hat{\phi}_1\:, \quad \hat{\op}^V_{2,0} = \partial_{a} \partial_{b} \hat{\phi}_{i}\:,
\end{equation}
which have transverse spin $s \neq 0$. The notation is borrowed from \cite{Gimenez-Grau:2022ebb}, where
\begin{equation}
    \hat{\op}^{S}_{s,0} \sim \partial_{a_1} \ldots \partial_{a_s} \hat{\phi}_1\:, \quad \hat{\op}^{V}_{s,0} \sim \partial_{a_1} \ldots \partial_{a_s} \hat\phi_{i}\:,
\end{equation}
and $S,V$ indicate an $O(N-1)$ scalar or vector respectively.

\begin{table}
    \centering
    \begin{tabular}{c|c|c}
         Operator & $\Delta$ & Reference\\ 
                  \hline \\[-1em]
         $\Phi$ & $1-\frac{\veps}{2}+\frac{\veps^2(N+2)}{4(N+8)^2}+ \op(\veps^3)$ & \cite{Cuomo:2021kfm}\\
         $\Phi^2$ & $2-\frac{6\veps}{N+8} + \op(\veps^2)$ & \cite{Cuomo:2021kfm}\\
         $T^{\{IJ\}}$ & $2-\frac{\veps(N+6)}{N+8} + \op(\veps^2)$ & \cite{Cuomo:2021kfm}\\
          \hline \\[-1em]
         $\hat{\phi}_1$ & $1+\veps - \veps^2 \frac{3 N^2 + 49 N+ 194}{2 (N+8)^2} + \op(\veps^3)$ & \cite{Cuomo:2021kfm}\\
         $t_{i}$ & 1 \text{ (exact)} & $\--$\\
         $D_{a}$ & 2 \text{ (exact)} & $\--$\\
         $\hat{\op}^{V}_{1,0} $ & $2-\frac{\veps}{3} + \op(\veps^2)$ & \cite{Cuomo:2021kfm}\\
         $\hat{V}_{i}$ & $2-\frac{\veps(N+10)}{N+8} + \op(\veps^2)$ & \cite{Cuomo:2021kfm}\\
         $\hat{T}_{\{ij\}}$ & $2-\frac{2\veps}{N+8} + \op(\veps^2)$ & \cite{Cuomo:2021kfm}\\
         $\hat{s}_\pm$ & $2+ \veps \frac{3N+20\pm\sqrt{N^2+40N+320}}{2(N+8)} + \op(\veps^2)$ & \cite{Cuomo:2021kfm}\\ 
         $\hat{\op}^{S}_{2,0}$ & $3 - \frac{\veps}{5} + \op(\veps^2)$ & \cite{Gimenez-Grau:2022ebb,Bianchi:2022sbz}\\
         $\hat{\op}^V_{2,0}$ & $3 - \frac{2 \veps}{5} + \op(\veps^2)$ & \cite{Gimenez-Grau:2022ebb,Bianchi:2022sbz}\\
    \end{tabular}
    \caption{Scaling dimensions of bulk and defect operators.} 
    \label{tab:defect-op}
\end{table} 

Besides scaling dimensions, the defect CFT data consists of defect OPE coefficients ${\lambda}_{\hat{\op}_1\hat{\op}_2\hat{\op}_3}$, and bulk-to-defect coefficients $\mu^{\op_1}_{\hat{\op}_2}$. All the defect OPE coefficients relevant for us were computed in~\cite{Gimenez-Grau:2022czc,Bianchi:2022sbz}. For the case of the bulk-to-defect coefficients, the expectation values of the bulk operators $a_{\op_1} \equiv \mu^{\op_1}_{\mathds 1}$ were already considered in~\cite{Cuomo:2021kfm} and will not be repeated here. Bulk-to-defect coefficients involving other defect operators besides the defect identity were computed in~\cite{Gimenez-Grau:2022ebb}\footnote{Bulk-to-defect coefficients were also obtained using defect conformal bootstrap in \cite{Gimenez-Grau:2022ebb,Bianchi:2022sbz}.}. Below, we compute additional ones, that have to our knowledge not been computed yet. The results are summarized in table~\ref{tab:CFT-data}.

Finally, all the operators in this section are not canonically normalized: their normalization can be found in the literature~\cite{Cuomo:2021kfm,Gimenez-Grau:2022czc,Gimenez-Grau:2022ebb} and we do not list it here. One then has to be careful when computing the OPE and bulk-to-defect coefficients to properly rescale the results by the normalizations, e.g
\begin{equation}
\label{eq:general-2pt-Bb}
\langle \op (y,x_1)\hat{\op}(x_2)\rangle = \sqrt{C_{\op}C_{\hat{\op}}}\, \mu^{\op}_{\hat{\op}}\, K^{\op}_{\hat \op}(y,x_1,x_2)\:.
\end{equation}

\begin{table}
    \centering
    \begin{tabular}{c|c|c}
        Observable & Value (to order $O(\veps)$) & Reference \\ \hline & \\[-1em]
        $\lambda_{tt\hat{\phi}_1}$ & $\frac{\pi\veps}{\sqrt{N+8}}$ & \cite{Gimenez-Grau:2022czc} \\
        $\lambda_{DD\hat{\phi}_1}$ & $\frac{\pi\veps}{\sqrt{N+8}}$ & \cite{Gimenez-Grau:2022czc} \\
        $\lambda_{D\hat{\op}^{V}_{1,0}t}$ & $\frac{\pi\veps}{3\sqrt{N+8}}$ & \cite{Gimenez-Grau:2022czc} \\
        $\lambda_{\hat{\phi}_1 t\hat{V}}$ & $1-\frac{\veps}{N+8}$ & \cite{Gimenez-Grau:2022czc} \\
        $\lambda_{tt\hat{T}}$ & $\sqrt{2}\big(1-\frac{\veps}{N+8}\big)$ & \cite{Gimenez-Grau:2022czc} \\ 
        $\lambda_{tt\hat{s}_\pm}$ & $\frac{1}{\sqrt{N-1}}\sqrt{1\mp\frac{N+18}{\sqrt{S_N}}}\left(1-\veps \frac{3N+20\pm\sqrt{S_N}}{4(N+8)} \right)$ & \cite{Belton:2025hbu} \\
        $\lambda_{D D \hat{\op}^{S}_{2,0}}$ & $\frac{2 \pi  \veps }{5 \sqrt{N+8}}$ & new \\
        $\lambda_{D \hat{\op}^{V}_{1,0}\hat{\op}^{V}_{2,0}}$ & $\frac{2 \pi  \veps }{15 \sqrt{N+8}}$ & new
        \\ \hline & \\[-1em]
        $\mu^{\Phi}_{\hat{\phi}_1}$ & $1 + 3\veps\frac{\log2-1}{2}$ & \cite{Gimenez-Grau:2022ebb} \\
        $\mu^{\Phi}_{t}$ & $1 + \veps\frac{\log2-1}{2}$ & \cite{Gimenez-Grau:2022ebb} \\
        $\mu^{\Phi}_{D}$ & $\sqrt{2}\left(1+ \veps\frac{3\log2-4}{6}\right)$ & \cite{Gimenez-Grau:2022ebb} \\
        $\mu^{\Phi^2}_{\hat{\phi}_1}$ & $-\frac{\sqrt{N+8}}{\sqrt{2N}}\left(1 + \veps\big(\frac{(N+14)\log2}{N+8}-\frac{7N^2+119N+438}{4(N+8)^2}\big)\right)$ & \cite{Gimenez-Grau:2022ebb} \\
         $\mu^{\Phi^2}_{D}$ & $ -\frac{\sqrt{N+8}}{\sqrt{N}}\left(1 + \veps\big(\frac{6\log2}{N+8}-\frac{5N^2+137N+578}{12(N+8)^2}\big)\right)$ & \cite{Gimenez-Grau:2022ebb} \\
        $\mu^{T}_{\hat{\phi}_1}$  & $-\frac{\sqrt{N+8}}{\sqrt{2}}\left(1 + \veps\big(\frac{2(N+7)\log2}{N+8}-\frac{5N^2+103N+438}{4(N+8)^2}\big)\right)$ & new \\
        $\mu^{T}_{t}$  & $-\frac{\sqrt{N+8}}{2\sqrt{2}}\left(1 + \veps\big(\frac{(N+6)\log2}{N+8}-\frac{N^2+39N+182}{4(N+8)^2}\big)\right)$ & new \\
        $\mu^\Phi_{\hat{V}}$ & $-\frac{\veps}{\sqrt{N+8}}$ & new \\
        $\mu^{\Phi}_{\hat{s}_\pm}$ & $\frac{\veps\sqrt{N-1}}{2 \sqrt{N+8}(S_N)^\frac{1}{4}} \frac{\pm(N+12)-\sqrt{S_N}}{\sqrt{\mp (N+18) +\sqrt{S_N}}}$ & new \\
        $\mu^{T}_{\hat{V}}$ & $\frac{1}{\sqrt{2}}\Big(1+\veps\frac{4\log 2-3}{2}\Big)$ & new \\ 
        $\mu^{T}_{\hat{T}}$ & $1+\veps(\log 2-1)$ & new \\
        $\mu^{\Phi^2}_{\hat{s}_\pm}$ & $\frac{\mp(N+16)+\sqrt{S_N}}{2\sqrt{2N}}\sqrt{1\pm\frac{N+18}{\sqrt{S_N}}}\left(1-\veps \frac{3N+32\pm\sqrt{S_N}}{4(N+8)}(1-2\log2) \right)$& new \\
        $\mu^{T}_{\hat{s}_\pm}$ & $\frac{\pm(3N+16)-\sqrt{S_N}}{2\sqrt{2}(N-1)}\sqrt{1\pm\frac{N+18}{\sqrt{S_N}}}\left(1-\veps \frac{N^2+26N+64\mp(N-2)\sqrt{S_N}}{8(N+8)}+\veps \frac{5N+32\pm\sqrt{S_N}}{2(N+8)}\log2 \right)$ & new \\
        \hline & \\[-1em]
        $C_t$ & $\frac{N+8}{4 \pi ^2} + \frac{\left(3 N^2+29 N+106\right) \varepsilon }{8 \pi ^2 (N+8)}$ & \cite{Gimenez-Grau:2022czc}\\
        $C_D$ & $\frac{N+8}{2 \pi ^2} + \frac{\left(5 N^2+23 N+62\right) \varepsilon }{12 \pi ^2 (N+8)}$ & \cite{Gimenez-Grau:2022czc}   \\
    \end{tabular}
    \caption{Conformal data - defect OPE and bulk-to-defect coefficients - to order $\op(\veps)$. 
    We~defined $S_N=N^2+40N+320$. The normalizations $C_t$ and $C_D$ differ from those in \cite{Gimenez-Grau:2022czc} by a factor of $h_{\star}$.}
    \label{tab:CFT-data}
\end{table}

\subsection{New defect three-point function} 

The three-point function between one $\hat\op_{2,0}$ and two~$\hat\op_{1,0}$ can be computed from the diagram
\begin{equation}
\langle\partial_{\{ab\}}\hat\phi_I(x_1)\, \partial_c\hat \phi_J(x_2)\, \partial_d\hat \phi_K(x_3)\rangle \quad=\quad 
  \begin{tikzpicture}[baseline,valign]
  \draw[thick] (-0.15, 0) -- (1.1, 0);
  \draw[dashed] (0.9,0) arc (20:160:0.45);
  \draw[dashed](0.3, 0) -- (0.5, 0.3);
  \draw[dashed](0.7, 0) -- (0.5, 0.3);
  \node at (0.7, 0) [dcirc] {};
  \node at (0.5, 0.3) [bcirc] {};
  \end{tikzpicture} \quad+\quad \dots
\end{equation}

This diagram is already at one-loop and we can thus compute it at $d=4$
\ba
&\lim _{|y| \rightarrow 0} \frac{\partial^2}{\partial y_{1, \{ab\}}} \frac{\partial}{\partial y_{2, c}} \frac{\partial}{\partial y_{3, d}} \int \frac{\ud x_4 \ud x_5\ud^3 y_5}{(y_{15}^2+x_{15}^2)(y_{25}^2+x_{25}^2)(y_{35}^2+x_{35}^2)(y_{45}^2+x_{45}^2)}\\
=\ &-8\pi \int \frac{\ud x_5\ud^3 y_5\ y_{5a}y_{5b}\big((y_5^2+x_{15}^2)\delta_{cd}-4y_{5c}y_{5d}\big)}{|y_5|(y_{15}^2+x_{15}^2)^3(y_{25}^2+x_{25}^2)^2(y_{35}^2+x_{35}^2)^2}\\
=\ &\frac12\big(\delta_{a(c}\delta_{d)b}-\tfrac2{N-1}\delta_{ab}\delta_{cd}\big)\frac{16 \pi^4 / 15}{\sqrt{x_{12}^6 x_{13}^6 x_{23}^2}}\:.
\ea

This leads directly to the OPE coefficients
\begin{equation}
    \lambda_{DD\hat\op_{2,0}^S} = 3 \lambda_{D\hat\op_{1,0}^V\hat\op_{2,0}^V} = 3 \lambda_{\hat\op_{1,0}^V\hat\op_{1,0}^V\hat\op_{2,0}^S} = \frac{2\pi\veps}{5\sqrt{N+8}}+\op(\veps^2)\:.
\end{equation}

\subsection{New bulk-defect two-point function}

The bulk-to-defect coefficients obtained from the correlator $\langle\Phi\hat{\phi}\rangle$ were computed in~\cite{Gimenez-Grau:2022ebb}. The new bulk-to-defect coefficients showcased in table~\ref{tab:CFT-data} appear in the bulk-defect two-point function $\langle\Phi(\hat{\phi}\hat{\phi})\rangle$ that we compute below at order $O(\veps)$, and in $\langle(\Phi_I\Phi_J)\hat{\phi}_K\rangle$ and $\langle(\Phi\Phi)(\hat{\phi}\hat{\phi})\rangle$ that we leave to the appendix~\ref{ssec:Phi2phi} and \ref{ssec:Phi2phi2}

At $O(\veps)$, the correlator $\langle\Phi_I(\hat{\phi}_J\hat{\phi}_K)\rangle$ is given by a single Feynman diagram:
\begin{equation}
\langle\Phi_I(y_1,x_1)(\hat{\phi}_J\hat{\phi}_K)(x_2)\rangle \quad=\quad 
  \begin{tikzpicture}[baseline,valign]
  \draw[thick] (0, 0) -- (1.2, 0);
  \draw[dashed](0.6, 0.5) -- (0.6, 1);
  \draw[dashed] (0.3,0) arc (190:110:0.45);
  \draw[dashed] (0.3,0) arc (310:350:0.8);
  \draw[dashed](0.9, 0) -- (0.6, 0.5);
  \node at (0.9, 0) [dcirc] {};
  \node at (0.6, 0.5) [bcirc] {};
  \end{tikzpicture} \quad+\quad \dots   
\end{equation}

The integral appearing in this diagram reduces again to the divergence of the Bloch-Wigner function with two coincident external points~\eqref{eq:X1233} integrated over the defect. The result is shown in appendix~\eqref{eq:Phi-phi2-integral}. The integral does not diverge in $\frac1\veps$, coherent with the fact that the correlator starts at $O(\veps)$. All in all, we find
\begin{equation}
\begin{tikzpicture}[baseline,valign]
  \draw[thick] (0, 0) -- (1.2, 0);
  \draw[dashed](0.6, 0.5) -- (0.6, 1);
  \draw[dashed] (0.3,0) arc (190:110:0.45);
  \draw[dashed] (0.3,0) arc (310:350:0.8);
  \draw[dashed](0.9, 0) -- (0.6, 0.5);
  \node at (0.9, 0) [dcirc] {};
  \node at (0.6, 0.5) [bcirc] {};
  \end{tikzpicture} \quad=\quad -\ \frac{h_0 \kappa^\frac{3}{2} (\delta_{IJ}\delta_{K1}+\delta_{IK}\delta_{J1}+\delta_{JK}\delta_{I1})}{|y_1|^{-1} (y_1^2+x_{12}^2)^{2}}\left(\frac{\blambda_0}{3}\right)\:,
\end{equation}
from which we obtain the non-zero bulk-to-defect coefficients
\begin{equation}
\mu^{\Phi}_{\hat{V}} = -\frac{\veps}{\sqrt{N+8}}, \qquad  \mu^{\Phi}_{\hat{s}_\pm} = \veps\frac{\sqrt{N-1}}{2 \sqrt{N+8}(S_N)^\frac{1}{4}} \frac{\pm(N+12)-\sqrt{S_N}}{\sqrt{\mp (N+18) +\sqrt{S_N}}}\, . 
\end{equation}

\subsection{Bulk-defect-defect three-point functions}

We now consider bulk-defect-defect correlators~\eqref{eq:formf}, which admit an (infinite) expansion in conformal blocks multiplied by bulk-to-defect and defect OPE coefficients~\eqref{eq:formf-blockexp}. The correlators are not entirely fixed by conformal symmetry anymore, but depend on a cross ratio $w$~\eqref{eq:cross-ratio}. We consider below the correlator $\langle \Phi \hat{\phi}\hat{\phi}\rangle$ to order $\op(\veps)$, and leave the computation of $\langle (\Phi\Phi) \hat{\phi}\hat{\phi}\rangle$ to the appendix~\ref{ssec:Phi2phiphi}.

At $O(\veps)$, the correlator $\langle\Phi_I\hat{\phi}_J\hat{\phi}_K\rangle$ is given by four diagrams:
\begin{equation}
\langle \Phi_I(y_1,x_1)\hat{\phi}_J(x_2)\hat{\phi}_K(x_3)\rangle ~~~=~~~ 
  \begin{tikzpicture}[baseline,valign]
  \draw[thick] (0, 0) -- (1.5, 0);
  \draw[dashed](0.3, 0) -- (0.3, 1);
  \draw[dashed] (1.3,0) arc (20:160:0.4);
  \node at (0.3, 0) [dcirc] {};
  \end{tikzpicture} ~~~+~~~ \begin{tikzpicture}[baseline,valign]
  \draw[thick] (0, 0) -- (1.7, 0);
  \draw[dashed](0.3, 0) -- (0.3, 1);
  \draw[dashed] (1.5,0) arc (20:160:0.45);
  \draw[dashed](0.9, 0) -- (1.1, 0.3);
  \draw[dashed](1.3, 0) -- (1.1, 0.3);
  \node at (0.9, 0) [dcirc] {};
  \node at (1.3, 0) [dcirc] {};
  \node at (1.1, 0.3) [bcirc] {};
  \node at (0.3, 0) [dcirc] {};
  \end{tikzpicture} ~~~+~~~ \begin{tikzpicture}[baseline,valign]
  \draw[thick] (0, 0) -- (2, 0);
  \draw[dashed](0.5, 0) -- (0.5, 1);
  \draw[dashed](0.2, 0) -- (0.5, 0.5);
  \draw[dashed](0.8, 0) -- (0.5, 0.5);
  \node at (0.2, 0) [dcirc] {};
  \node at (0.5, 0) [dcirc] {};
  \node at (0.8, 0) [dcirc] {};
  \node at (0.5, 0.5) [bcirc] {};
  \draw[dashed] (1.8,0) arc (20:160:0.4);
  \end{tikzpicture} ~~~+~~~  \begin{tikzpicture}[baseline,valign]
  \draw[thick] (0, 0) -- (1.2, 0);
  \draw[dashed](0.6, 0) -- (0.6, 1);
  \draw[dashed](0.3, 0) -- (0.6, 0.5);
  \draw[dashed](0.9, 0) -- (0.6, 0.5);
  \node at (0.6, 0) [dcirc] {};
  \node at (0.6, 0.5) [bcirc] {};
  \end{tikzpicture} ~~~+~~~ \dots
\end{equation}
The first three disconnected diagrams are combinations of one-point function diagrams computed in \cite{Allais:2014fqa}, and diagrams contributing to the wavefunction renormalization of $\hat{\phi}$. The last diagram involves a non-trivial integral that can be solved with Schwinger parametrization in $d = 4$, using the conformal frame where $|y_1|=1$, $x_1=0$, $x_3\rightarrow\infty$:
\begin{equation}
\label{eq:Int1Bu2Bo1De}
\int \frac{\ud x_4\ud^3 y_4}{y_4(y_{14}^2+x_{14}^2)(y_{4}^2+x_{24}^2)(y_{4}^2+x_{34}^2)} = \frac{\pi^2}{|y_1|x_{23}^2}\frac{\pi^2+4\big[\cosh^{-1}(\sqrt{w})\big]^2}{2}\:.   
\end{equation}
Interestingly, the integral can also be computed by first performing the angular part of the bulk integration. The result can then be recast into a Witten diagram that was shown in~\cite{Levine:2023ywq} to be proportional to a local block~$\mathcal{L}_1^0$: 
\ba
\label{eq:Int1Bu2Bo1De-Witten}
\int \frac{\ud x_4 \ud^3 y_4}{y_4(y_{14}^2+x_{14}^2)(y_{4}^2+x_{24}^2)(y_{4}^2+x_{34}^2)}\ &=\ \frac{4\pi^2}{|y_1|x_{23}^2}
\times \begin{tikzpicture}[baseline,valign]
    \filldraw [fill=gray!25, thick] (0,0) circle (1);
    \node (P1) at (150:1){{\tiny$\bullet$}};
    \node (P2) at (210:1){{\tiny$\bullet$}};
    \node (X) at (0:0){{\tiny$\bullet$}};
    \draw [thick] (150:1) -- (180:1/2);
    \draw [thick] (210:1) -- (180:1/2);
    \draw [thick] (180:1/2) -- (0:0);
    \draw (P1) node [left]{$\hat{\phi}$};        
    \draw (P2) node [left]{$\hat{\phi}$};        
    \draw (X) node [right]{$\Phi$};  
\end{tikzpicture} \\
&\hspace{-6cm} = \frac{4\pi^2}{|y_1|x_{23}^2}\left(\frac{\pi}2 G_1(w) + \sum_{n=1}^\infty \left( \frac{-1}4\right)^{n} \frac{\sqrt{\pi} \Gamma(2n-1) }{n\Gamma(2n-\frac12)} G_{2n}(w) \right) \:,
\ea
which matches the closed-form result~\eqref{eq:Int1Bu2Bo1De} order by order in $w$ . All in all, the last diagram gives a contribution
\begin{equation}
\begin{tikzpicture}[baseline,valign]
  \draw[thick] (0, 0) -- (1.2, 0);
  \draw[dashed](0.6, 0) -- (0.6, 1);
  \draw[dashed](0.3, 0) -- (0.6, 0.5);
  \draw[dashed](0.9, 0) -- (0.6, 0.5);
  \node at (0.6, 0) [dcirc] {};
  \node at (0.6, 0.5) [bcirc] {};
  \end{tikzpicture} = 
\frac{\kappa^{3/2}(\delta_{IJ}\delta_{K1}+\delta_{IK}\delta_{J1}+\delta_{JK}\delta_{I1})}{|y_1|x_{23}^2} \left(\frac{\bar{\lambda}h}6\right)\frac{\pi^2+4\big[\cosh^{-1}(\sqrt{w})\big]^2}{2}.    
\end{equation}
Using~\eqref{eq:formf}, we get
\ba
    \mathcal H^{\Phi}_{tt}(w) &= -\frac{\sqrt{N+8}}{2}\left(1+ \veps \frac{\pi^2+4\big[\cosh^{-1}(\sqrt{w})\big]^2}{2(N+8)} + \frac{\veps\log2}2 + \veps \frac{N^2-3N-22}{4(N+8)^2} \right)\:, \\
    \mathcal H^{\Phi}_{\hat{\phi}_1 t}(w) &= \veps \frac{\pi^2+4\big[\cosh^{-1}(\sqrt{w})\big]^2}{4(N+8)}\:, \\
    \mathcal H^{\Phi}_{\hat{\phi}_1\hat{\phi}_1}(w) &= -\frac{\sqrt{N+8}}{2}\left(1+ 3\veps \frac{\pi^2+4\big[\cosh^{-1}(\sqrt{w})\big]^2}{2(N+8)} + \frac{\veps\log2}2 + \veps \frac{N^2-3N-22}{4(N+8)^2}\right)\:.
\ea
Expanding the above expressions in conformal blocks provides us with a check of part of the CFT data listed in Table~\ref{tab:CFT-data}.
We also get information on a tower of operators $\hat{\phi}^{2n}$ of tree-level conformal dimension $\Delta = 2n$, that is contained in the Witten diagram expansion~\eqref{eq:Int1Bu2Bo1De-Witten}:
\begin{equation}
    \llangle \mu^{\Phi}_{\hat{\phi}^{2n}}\lambda_{tt\hat{\phi}^{2n}}\rrangle = \llangle \mu^{\Phi}_{\hat{\phi}^{2n}}\lambda_{\hat{\phi}_1 t\hat{\phi}^{2n}}\rrangle = \frac13 \llangle \mu^{\Phi}_{\hat{\phi}^{2n}}\lambda_{\hat{\phi}_1\hat{\phi}_1\hat{\phi}^{2n}}\rrangle = \frac{2\veps}{\sqrt{N+8}}\left( \frac{-1}4\right)^{n} \frac{\sqrt{\pi} \Gamma(2n-1)}{n\Gamma(2n-\frac12)}\:.
\end{equation}
The notation $\llangle \cdot \rrangle$ is used to indicate a sum over degenerate states of similar conformal dimension. Note that the $n=1$ state matches the CFT data of the operators $\hat{V}$ and $\hat{s}_\pm$, and gives us non-trivial identities between the $\hat{s}_+$ and $\hat{s}_-$ contributions:
\ba
\llangle \mu^{\Phi}_{\hat{\phi}^{2}}\lambda_{tt\hat{\phi}^{2}}\rrangle &= \mu^{\Phi}_{\hat{s}_+}\lambda_{tt \hat{s}_+} + \mu^{\Phi}_{\hat{s}_-}\lambda_{tt \hat{s}_-} = -\frac{\veps}{\sqrt{N+8}}\,, \\
\llangle \mu^{\Phi}_{\hat{\phi}^{2}}\lambda_{\hat{\phi}_1\hat{\phi}_1\hat{\phi}^{2}}\rrangle &= \mu^{\Phi}_{\hat{s}_+}\lambda_{\hat{\phi}_1\hat{\phi}_1 \hat{s}_+} + \mu^{\Phi}_{\hat{s}_-}\lambda_{\hat{\phi}_1\hat{\phi}_1 \hat{s}_-} = -\frac{3\veps}{\sqrt{N+8}}\, .
\ea

\subsection{Bulk current operator}
Besides sum rules for correlators involving the tilt operator, we also derived sum rules for correlators involving the $O(N)$ symmetry current $J_{\mu}^{[IJ]}$. In order to test the power of these sum rules, we repeat the calculations above, now taking $J_\mu^{[IJ]}\propto \Phi^{I}\overset{\leftrightarrow}{\partial}_\mu\Phi^{J}$ as one of the external operators. It has a non-vanishing bulk-to-defect coefficient with the tilt, and with the defect operators $\hat{A}^{ij}=\hat\phi^{i}\overset{\leftrightarrow}{\partial}_x\hat\phi^{j}$ and~$\hat{B}^{i}=\hat\phi^{i}\overset{\leftrightarrow}{\partial}_x\hat\phi^{1}$ that have not been considered before. Since $J_{\mu}^{[IJ]}$ has bulk spin $\ell = 1$, the tensor structures appearing in the correlation functions will be more complicated than before. It is also antisymmetric in its $O(N)$ indices, which causes some Feynman diagrams to vanish, in particular disconnected ones. This is equivalent to the statement that $J_{\mu}^{[IJ]}$ does not have a one-point function. 
A consistency check on our calculations is given by the fact that the bulk current is conserved,
so all correlators involving a bulk current have to satisfy a conservation equation
\begin{equation}
    \partial^\mu \langle J_\mu \op_1 \cdots \op_n \rangle =(\text{contact terms})\:.
\end{equation}

\subsubsection{New CFT data}

\paragraph*{Conformal dimensions} As before, the first step is to renormalize the two-point function, and find the conformal dimensions of the operators. The bulk current has protected dimension $d-1$, and the new defect operators have conformal dimensions
\begin{equation}
    \Delta_{\hat A} = 3+\op(\veps^2)\ , \qquad \Delta_{\hat B} = 3+\veps+\op(\veps^2)\:.
\end{equation}
More details on the computation can be found in appendix~\ref{sssec:current-renorm}.

\paragraph*{OPE coefficients} A new OPE coefficient we will need is the one between $\hat A$ or $\hat B$, and two defect operators $\hat \phi_I$ 
\begin{equation}
    \lambda_{tt\hat A}=1\ , \qquad \lambda_{\hat{B}t\hat{\phi}_1}=1+\frac\veps4\:.
\end{equation}
The computation of the diagrams can be found in appendix~\ref{sssec:A/Bphiphi}. Note that the OPE coefficient $\lambda_{tt\hat{A}}$ had already been found in~\cite{Gimenez-Grau:2022czc} by expanding the four-point function~$\langle tttt\rangle$.

\paragraph*{DOE coefficients} Due to the $O(N-1)$ symmetry preserved by the defect, the only bulk-defect two-point functions between $J_{\mu IJ}$ and low-lying defect operators $(\Delta_0 <4)$, are those with the tilt, and the defect operators $\hat{A}$ and $\hat{B}$
\begin{gather}
        \mu^{J}_{t} = -\frac{\sqrt{N+8}}{4}\left[1+ \veps\Big(\log2 - \frac{19N+86}{4(N+8)^2}\Big) + \op(\veps^2) \right] \\
    \mu^{J}_{\hat{A}} = 1+\veps \Big(\log2 - \frac34\Big), \qquad \mu^{J}_{\hat{B}} = 1+ 2\veps \Big(\log2 - \frac34\Big).
\end{gather}
More details on the computations can be found in appendix~\ref{sssec:Jt} and~\ref{sssec:JA/B}.

\subsubsection{Bulk-defect-defect three-point functions}

Finally, we can compute the form factors with a bulk current, the easiest case being with two $\hat\phi_I$. It can be obtained from the correlator $\langle J_{\mu IJ}\, \hat{\phi}_K \hat{\phi}_L  \rangle$, which has the form~\eqref{eq:formfv} with the tensor structures~\eqref{eq:structure-JOO}.
At $O(\veps)$ it receives contributions from three diagrams:
\begin{equation}
\langle J_{\mu IJ}(y_1,x_1)\hat\phi_K(x_2)\hat\phi_L(x_3)\rangle \quad =\quad 
\begin{tikzpicture}[baseline,valign]
  \draw[thick] (0, 0) -- (1, 0);
  \draw[dashed](0.2, 0) -- (0.5, 1);
  \draw[dashed](0.8, 0) -- (0.5, 1);
\end{tikzpicture} \quad+\quad \begin{tikzpicture}[baseline,valign]
  \draw[thick] (0, 0) -- (1.5, 0);
  \draw[dashed](0.3, 0) -- (0.7, 1);
  \draw[dashed](0.6, 0) -- (0.9, 0.6);
  \draw[dashed](0.9, 0) -- (0.9, 0.6);
  \draw[dashed](1.2, 0) -- (0.9, 0.6);
  \draw[dashed](0.7, 1) -- (0.9, 0.6);
  \node at (0.3, 0) [dcirc] {};
  \node at (0.9, 0) [dcirc] {};
  \node at (0.9, 0.6) [bcirc] {};
\end{tikzpicture} \quad+\quad \begin{tikzpicture}[baseline,valign]
  \draw[thick] (0, 0) -- (1.5, 0);
  \draw[dashed](0.3, 0) -- (0.7, 1);
  \draw[dashed](0.6, 0) -- (0.9, 0.6);
  \draw[dashed](0.9, 0) -- (0.9, 0.6);
  \draw[dashed](1.2, 0) -- (0.9, 0.6);
  \draw[dashed](0.7, 1) -- (0.9, 0.6);
  \node at (0.6, 0) [dcirc] {};
  \node at (1.2, 0) [dcirc] {};
  \node at (0.9, 0.6) [bcirc] {};
\end{tikzpicture} \quad+\quad \dots \ \:. 
\end{equation}
The expressions for each diagram to $O(\veps)$ can be found in the appendix~\ref{sssec:JOO}.
Putting everything together, we get the following form factors that can be expanded in conformal blocks. This provides a check on the OPE data previously computed
\begin{align}
    \mathcal H_{tt}^{J(-)}(w) &= w\Big(1+\veps \big(\log2-\tfrac34 \big) \Big) = \mu^{J}_{\hat{A}}\lambda_{tt\hat{A}} H^{(-)}_{\Delta_{\hat{A}}}(w)\:, \\
    \mathcal H_{t\hat{\phi}_1}^{J(-)}(w) &= w \Big(1+\veps\big(2\log2-\tfrac54+\tfrac12\log w\big) \Big) = \mu^{J}_{\hat{B}}\lambda_{t\hat{\phi}_1\hat{B}}H^{(-)}_{\Delta_{\hat{B}}}(w)\:, \\
    \mathcal H_{t\hat{\phi}_1}^{J(+)}(w) &= -\frac{\veps w}2 \left(1+\frac{\cosh^{-1}\sqrt{w}}{\sqrt{w(w-1)}}\right) = \mu^{J}_{t}\lambda_{tt\hat{\phi}_1}H_{\Delta_t}(w)\:. 
\end{align}

\section{Higher-order perturbative predictions}
\label{sec:bounds}

The functional sum rules derived above can also be used to extract new predictions in perturbation theory. For this, we will take the example of the pinning field line defect in the $O(N)$ CFT in $d = 4 - \veps$ dimensions discussed in section \ref{sec:pinning}. The input we will use is the spectrum of lowest-lying defect operators, given in table \ref{tab:defect-op}, and OPE coefficients as well as bulk-to-defect OPE coefficients given in table \ref{tab:CFT-data}. In particular, we will assume that there are no relevant operators with $\Delta < p = 1$, that the lowest scalar operator in the symmetric traceless representation of $O(N-1)$ has dimension $\Delta_{\hat{T}} = 2 + \op(\veps)$, and the lowest scalar operator in the antisymmetric representation of $O(N-1)$ has dimension $\Delta_{\hat{A}} = 3 + \op(\veps)$. Furthermore, we assume that the four-point functions of tilts and displacements are equal to a GFF correlator in $d=4$. The CFT data is given perturbatively as
\ba
    \Delta &= \Delta^{(0)} + \veps \Delta^{(1)} + \veps^2 \Delta^{(2)} + \ldots\,,\\
    \lambda &= \lambda^{(0)} + \veps \lambda^{(1)} + \veps^2 \lambda^{(2)} + \ldots\,,\\
    C_{\op} &= C_{\op}^{(0)} + \veps C_{\op}^{(1)} + \veps^2 C_{\op}^{(2)} + \ldots\,.
\ea
The sum rules are satisfied order by order in $\veps$.

\subsection{Tilt sum rules in perturbation theory}

Let us start with the sum rules for tilt operators derived in section \ref{sec:funcSR}. We use the functional sum rules of \eqref{eq:tiltHomFunc} and \eqref{eq:tiltInhom} for the correlator of four tilt operators to compare with the perturbative results for defect OPE coefficients and anomalous dimensions for the pinning line defect computed in section \ref{sec:pinning}. We perform the same check for the bulk-defect-defect correlators. 

\subsubsection{Four tilts}

The sum rules involving four tilts can be checked by inserting the anomalous dimensions and defect OPE coefficients of the lowest operators appearing in the $t^{\hat{i}} \times t^{\hat{j}}$ OPE. Using the sumrule $f^{t, \text{soft}}_I [\Delta]$ given in \eqref{eq:tiltHomFunc}, we see that there is a pole at $\Delta = 1$, a single zero at $\Delta = 2$, and double zeros for $\Delta = 2+ 2n, n\in \mathbb{Z}_{+}$ in the singlet and traceless symmetric channels, while the antisymmetric channel does not contribute at all. This ensures that at order $\op(\veps)$, the only exchanged operators which contribute are $\hat{\phi}_1$ and $\hat{s}_\pm$ in the singlet channel, and $\hat{T}$ in the symmetric traceless channel. We thus have
\begin{equation}
    \underbrace{\big(\lambda^{(1)}_{tt\hat{\phi}_1}\big)^2}_{\op({\veps^2)}}~ \underbrace{f_{S}^{t,\text{soft}}[\Delta_{\hat{\phi}_1}]}_{\op(\frac1\veps)} ~~+~~ \underbrace{\left(\big(\lambda^{(0)}_{tt\hat{T}}\big)^2 f_{T}^{t,\text{soft}} [\Delta_{\hat{T}}] + \sum_{i = \pm} \big(\lambda^{(0)}_{tt\hat{s}_{i}}\big)^2 f_{S}^{t,\text{soft}}[\Delta_{\hat{s}_i}] \right)}_{\op(\veps)}  ~=~ 0\:.
\end{equation}
This equation is satisfied by the CFT data of tables~\ref{tab:defect-op} and \ref{tab:CFT-data}. 

A similar check can be done for the double-soft sum rule \eqref{eq:tiltInhom}. Here an additional piece of dCFT data enters: the normalization of the tilt two-point function $C_t$ given in table \ref{tab:CFT-data}. From \eqref{eq:tiltInhom} we find that for $f^{t,\text{dsoft}}_{I}[\Delta]$ there is a single zero at $\Delta = 2$, and double zeros for $\Delta = 2 + 2n, n \in \mathbb{Z}_{+}$ in the singlet and traceless symmetric channels, as before. The difference with respect to the soft sum rules is that there is now a double pole for $\Delta = 1$. The antisymmetric channel does contribute to the double-soft sum rule, and has double zeros for odd $\Delta = 1 + 2n, n \in \mathbb{Z}_{+}$. At $\mathcal{O}(1)$ the sum rules satisfy
\begin{align}
    \underbrace{\big(\lambda^{(1)}_{tt\hat{\phi}_1}\big)^2}_{\op({\veps^2)}}~ \underbrace{f_{S}^{t,\text{dsoft}}[\Delta_{\hat{\phi}_1}]}_{\op(\frac{1}{\veps^2})} ~=~ \underbrace{\frac{2}{C_t}}_{\op(1)} \:.
\end{align}
At $\op(\veps)$ we obtain our first new prediction: $\lambda^{(2)}_{tt\hat{\phi}_1}$. The sum rule is given by
\begin{align}
     \underbrace{\big(\lambda^{(1)}_{tt\hat{\phi}_1} + \lambda^{(2)}_{tt\hat{\phi}_1}\big)^2}_{\op({\veps^3)}}~ \underbrace{f_{S}^{t,\text{dsoft}}[\Delta_{\hat{\phi}_1}]}_{\op(\frac1{\veps^2})} ~+~ \underbrace{\left(\big(\lambda^{(0)}_{tt\hat{T}}\big)^2 f_{T}^{t,\text{dsoft}} [\Delta_{\hat{T}}] + \sum_{i = \pm} \big(\lambda^{(0)}_{tt\hat{s}_{i}}\big)^2 f_{S}^{t,\text{dsoft}}[\Delta_{\hat{s}_i}] \right)}_{\op(\veps)}  ~=~ \underbrace{\frac{2}{C_t}}_{\op(\veps)}\:,
\end{align}
where in order to extract $\lambda^{(2)}_{tt\hat{\phi}_1}$ we have to input $\Delta_{\hat{\phi}_1}$ to $\op(\veps^2)$, which has been computed in \cite{Cuomo:2021kfm}. We obtain
\begin{equation}
    \lambda^{(2)}_{tt\hat{\phi}_1} = -\frac{\pi  \left(9 N^2+127 N+494\right)}{4 (N+8)^{5/2}}\:.\label{eq:OPE2ttphi1}
\end{equation}

\subsubsection{\texorpdfstring{$\langle \Phi^I \, t_{\hat{j}} \, \hat\phi_{k} \rangle$}{<Phi t phi>}}

To check the bulk-defect-defect sum rules, we focus on one particular bulk operator, namely $\Phi^I$. There are two different three-point functions: $\langle \Phi^1 t^i t^j \rangle$ and $\langle \Phi^i t^j \phi_1 \rangle$. As can be seen from \eqref{eq:tsPhi} and \eqref{eq:phitt}, these sum rules have a nontrivial RHS which depends on the normalization of the tilt and a bulk-to-defect two-point function. 

\paragraph{$\langle \Phi^1 \, t_{\hat{j}} \, t_{\hat{k}} \rangle$.} The simplest case is that of $\langle \Phi^1 t^i t^j \rangle$, where the two defect operators have the same dimension. The functional sum rules depend on a parameter $\alpha$, which dictates the position of the zeros. For $\alpha=1$, given in \eqref{eq:PhittFunc}, there is a simple pole at $\Delta = 1$, and simple zeros for $\Delta = 2 + 2n, n \in \mathbb{Z}_+$. At $\op(\veps)$ we find
\begin{equation}
    \underbrace{\big(\lambda^{(1)}_{tt\hat{\phi}_1} + \lambda^{(2)}_{tt\hat{\phi}_1}\big) \big(\mu^{\Phi(0)}_{\hat\phi_1} + \mu^{\Phi(1)}_{\hat\phi_1}\big)}_{\op({\veps^2)}}~ \underbrace{\kappa_{tt}^{\Phi} [\Delta_{\hat{\phi}_1}]}_{\op(\frac1\veps)} + \sum_{i=\pm} \underbrace{\big(\lambda^{(1)}_{tt\hat{s}_i} + \lambda^{(2)}_{tt\hat{s}_i}\big) \big(\mu^{\Phi(1)}_{\hat{s}_i}\big)}_{\op({\veps^2)}}~ \underbrace{\kappa_{tt}^{\Phi} [\Delta_{\hat{s}_i}]}_{\op(\frac1\veps)} ~=~ \underbrace{\frac{\mu^{\Phi}_{t}}{\sqrt{\normt}}}_{\op(\veps)}\:.
\end{equation}
We can again extract the value of $\lambda^{(2)}_{tt\hat{\phi}_1}$ from this sum rule, and find that it agrees with~\eqref{eq:OPE2ttphi1}, an important consistency check on this result. 

\paragraph{$\langle \Phi^{\hat{i}} \, t_{\hat{j}} \, \hat\phi_{1} \rangle$.} Let us now move to the second correlator $\langle \Phi^{\hat{i}} \, t_{\hat{j}} \, \hat\phi_{1} \rangle$. The two defect operators now have a difference in scaling dimension and the sum rule depens on $\Delta_{\hat{\phi}_1}$. As before, we set $\alpha = 1$. At $\op(\veps)$ we find
\begin{align}
     \underbrace{\big(\lambda^{(1)}_{tt\hat{\phi}_1} + \lambda^{(2)}_{tt\hat{\phi}_1}\big) \big(\mu^{\Phi(0)}_{t} + \mu^{\Phi(1)}_{t}\big)}_{\op({\veps^2)}}~ \underbrace{\kappa^{\Phi}_{t\hat{\phi_1}} [1]}_{\op(\frac1\veps)} +  \underbrace{\lambda^{(0)}_{\hat{\phi}_1 t \hat{V}} \,  \mu^{(1)}_{\Phi \hat{V}}}_{\op({\veps)}}~ \underbrace{\kappa^{\Phi}_{t\hat{\phi_1}} [\Delta_{\hat{V}}]}_{\op(1)} ~=~ \underbrace{\frac{\mu^{\Phi}_{\hat{\phi}_1}}{\sqrt{C_t}}}_{\op(\veps)}\:.
\end{align}
This is satisfied by the data in tables~\ref{tab:defect-op} and~\ref{tab:CFT-data} and~\eqref{eq:OPE2ttphi1}.

One can consider different representations of $O(N)$ for the bulk operators and see that they are also in agreement with the perturbative results of section~\ref{sec:pinning}. We will stop here, however, and turn our attention to the sum rules involving the displacement.

\subsection{Displacement sum rules in perturbation theory}

The same checks can be performed for the displacement sum rules. 
Since now we have access to multiple sum rules, coming from the different conformal generators, we can combine them in such a way that poles in certain channels get canceled, while poles in other channels remain, allowing us to find higher order perturbative data from a minimal set of lower order data.

\subsubsection{Four displacements}

One such combination is the following:
\begin{align}
     f^{D, \text{Com}1}_{I} [\Delta]  &= - \frac{1}{3} f^{D, \text{soft}1}_{I} [\Delta] + \frac{1}{12} f^{D, \text{soft}2}_{I} [\Delta] + f^{D, \text{dsoft}1}_{I} [\Delta] + \frac{1}{16} \Big( f^{D, \text{dsoft}2}_{I} [\Delta] + f^{D, \text{dsoft}3}_{I} [\Delta] \Big)\nonumber \\
     & = \frac{1}{C_D}\:.
\end{align}
This particular combination results in a sum rule for which the singlet channel is zero, and which has a double pole in the symmetric traceless channel for $\Delta = 3$, single zeros for $\Delta = 2,4$, and double zeros for $\Delta = 4 + 2n, n \in \mathbb{Z}_{+}$. In the antisymmetric channel, there is a double pole at $\Delta = 2$ and double zeros for $\Delta = 3 + 2n, n \in \mathbb{Z}_{+}$. With this, we are able to obtain defect CFT data for the lightest operator $\partial_i \partial_j \hat{\phi}_1$ in the symmetric traceless channel, which at $\op(1)$ gives the only contribution:
\begin{align}
     \underbrace{\big(\lambda^{(1)}_{DD \partial_i \partial_j \hat{\phi}_1} \big)^2 }_{\op({\veps^2)}}~ \underbrace{f_{T}^{D,\text{Com}1} [\Delta_{\partial_i \partial_j \hat\phi_1}]}_{\op(\frac{1}{\veps^2})}  ~=~ \underbrace{\frac{1}{C_D}}_{\op(1)}\:.
\end{align}
At $\op(\veps)$ there are two new unknowns: the OPE coefficient $\lambda^{(2)}_{DD \partial_i \partial_j \hat{\phi}_1} $ and the dimension $\Delta^{(2)}_{\partial_i \partial_j \hat{\phi}_1}$. At this point we can only constrain one in terms of the other: 
\begin{equation}
    \lambda^{(2)}_{DD \partial_i \partial_j \hat{\phi}_1}  = -\frac{\pi  \left(300 \Delta^{(2)}_{\partial_i \partial_j \hat{\phi}_1} (N+8)^2+N (37 N+307)+1078\right)}{150 (N+8)^{5/2}}\:.\label{eq:lDDdidjphi}
\end{equation}
A different combination of soft and double-soft sum rules leads to the symmetric traceless channel being zero, while having double poles at $\Delta = 1$ and $\Delta = 3$ in the singlet channel. This combination is given by
\begin{align}
     f^{D, \text{Com}2}_{I} [\Delta]  &= - \frac{16}{3} f^{D, \text{soft}1}_{I} [\Delta] - \frac{8 (N-2)}{3 (N+1)} f^{D, \text{soft}2}_{I} [\Delta] - \frac{16 (N-3)}{(N+1)} f^{D, \text{dsoft}1}_{I} [\Delta] \nonumber \\
     &+  f^{D, \text{dsoft}2}_{I} [\Delta] - \frac{N-3}{N+1} f^{D, \text{dsoft}3}_{I} [\Delta]  = - \frac{16 (N-3)}{(N+1)} \frac{1}{C_D}\:.
\end{align}
At $\op(1)$ we find
\begin{align}
     \underbrace{\big(\lambda^{(1)}_{DD \hat{\phi}_1} \big)^2 }_{\op({\veps^2)}}~ \underbrace{f_{S}^{D,\text{Com}2} [\Delta_{\hat\phi_1}]}_{\op(\frac{1}{\veps^2})}  + \underbrace{\big(\lambda^{(1)}_{DD \hat{\phi}^3_{\pm}} \big)^2 }_{\op({\veps^2)}}~ \underbrace{f_{S}^{D,\text{Com}2} [\Delta_{\hat\phi^3_{\pm}}]}_{\op(\frac{1}{\veps^2})}  ~=~ - \frac{16 (N-3)}{(N+1)} \underbrace{\frac{1}{C_d}}_{\op(1)}\:.
\end{align}
The data for the defect operators $\hat{\phi}^3_{\pm}$ is new and when computed in perturbation theory would require some unmixing as was the case for $s_{\pm}$. Its dimension was computed in \cite{Belton:2025hbu}, where they called the operator $s^{'}_{\pm}$:
\begin{equation}
    \Delta_{\hat{\phi}^3_{\pm}} = \Delta_{s^{'}_{\pm}} = 3 + \frac{5 N+ 46 \pm \sqrt{N^2+ 52 N+ 388}}{2(N+8)} \veps + \op(\veps^2)\:.
\end{equation}
We can immediately extract a relation between the OPE coefficients, and find
\begin{equation}
    \frac{\lambda^{(1)}_{DD \hat{\phi}^3_{+}} }{\left(5 N+ 46+\sqrt{N(N+52)+388}\right)^2}+\frac{\lambda^{(1)}_{DD \hat{\phi}^3_{-}} }{\left(5 N + 46-\sqrt{N (N+52)+388}\right)^2} = \frac{\pi^2 (4 - N)}{3 (N-1) (N+8)^3} \:.
\end{equation}

\subsubsection{\texorpdfstring{$\langle \Phi D_{i} D_{j} \rangle$}{Phi Di Dj}}

We move on to the bulk-defect-defect sum rules. Let us look at the simplest case of $\langle \Phi D_{i} D_{j} \rangle$. We have two sum rules for the singlet channel, and two for the symmetric traceless channel. Setting $\alpha=1$, for the singlet channel, the first sum rule has a pole for $\Delta = 1,3$ and single zeros at $\Delta = 2 n, n \in \mathbb{Z}_{+}$. The second sum rule has only a pole at $\Delta = 1$ and single zeros at $\Delta = 2 + 2 n, n \in \mathbb{Z}_{+}$. Let us look at the second sum rule first. At $\op(\veps)$ we find 
\begin{align}
     \underbrace{\big(\lambda^{(1)}_{DD\hat{\phi}_1} + \lambda^{(2)}_{DD\hat{\phi}_1}\big) \big(\mu^{\Phi(0)}_{\hat{\phi}_1} + \mu^{\Phi(1)}_{\hat{\phi}_1}\big)}_{\op({\veps^2)}}~ \underbrace{\kappa^{\Phi,2}_{DD,S} [\Delta_{\hat{\phi}_1}]}_{\op(\frac1\veps)} ~=~ \underbrace{- \frac{\mu^{\Phi}_{D}}{\sqrt{C_t}} \frac{(\Delta_{\Phi} - 4 + \veps)}{3 - \veps}}_{\op(\veps)}\:.
\end{align}
From this we can extract $\lambda_{DD\hat{\phi}_1}^{(2)}$:
\begin{equation}\label{eq:lamDDphi1}
    \lambda_{DD\hat{\phi}_1}^{(2)} = -\frac{\pi  (29N^2 + 413N +1610)}{12 (N+8)^{5/2}}\:,
\end{equation}
which agrees with the value found in \cite{Gabai:2025zcs}. When we now look at the first singlet sum rule, we find a contribution from $\hat{\phi}_{\pm}^3$:
\begin{align}
     \underbrace{\big(\lambda^{(1)}_{DD\hat{\phi}_1} + \lambda^{(2)}_{DD\hat{\phi}_1}\big) \big(\mu^{\Phi(0)}_{\hat{\phi}_1} + \mu^{\Phi(1)}_{\hat{\phi}_1}\big)}_{\op({\veps^2)}}~ \underbrace{\kappa^{\Phi,1}_{DD,S} [\Delta_{\hat{\phi}_1}]}_{\op(\frac1\veps)} +  \sum_{i = \pm} \underbrace{\lambda^{(1)}_{DD\hat{\phi}^3_i} \mu^{\Phi(1)}_{\hat{\phi}^3_i}}_{\op({\veps^2)}}~ \underbrace{\kappa^{\Phi,1}_{DD,S} [\Delta_{\hat{\phi}^3_i}]}_{\op(\frac1\veps)} ~=~ \underbrace{-2 \frac{\mu^{\Phi}_{D}}{\sqrt{C_t}} \frac{(\Delta_{\Phi} - 2 + \veps)}{3 - \veps}}_{\op(\veps)}\:.
\end{align}
Using \eqref{eq:lamDDphi1}, we find the relation
\begin{equation}
    \frac{\lambda^{(1)}_{DD\hat{\phi}^3_{+}} \mu^{\Phi(1)}_{\hat{\phi}^3_{+}}}{5 N + 46 + \sqrt{N(N+52) + 388}} + \frac{\lambda^{(1)}_{DD\hat{\phi}^3_{-}} \mu^{\Phi(1)}_{\hat{\phi}^3_{-}}}{5 N + 46 - \sqrt{N(N+52) + 388}} = 0\:.
\end{equation}
A similar analysis for the symmetric traceless channel leads to a prediction for $\lambda_{DD \partial_i \partial_j \hat{\phi}_1}^{(2)}$ which fully agrees with \eqref{eq:lDDdidjphi}.

\subsection*{Acknowledgements}
It is a pleasure to acknowledge discussions with Ant\'onio Antunes, Costas Bachas, Connor Behan, Barak Gabai, Julius Julius, and No\'e Suchel. This work was co-funded by the European Union (ERC, FUNBOOTS, project number 101043588). Views and opinions expressed are however those of the author(s) only and do not necessarily reflect those of the European Union or the European Research Council. Neither the European Union nor the granting authority can be held responsible for them.

%%%%%%%%%%%%%%%%%%%%%%%%%%%%%%%%%%%%%
%%%%%%%%%%%%%%%%%%%%%%%%%%%%%%%%%%%%%
\newpage

\appendix

\section{Further tilt form factors}\label{app:srtilt}

Below we list additional sum rules for form factors of a bulk and two tilt operators, where the bulk operator is in a different $O(N)$ representation than the ones presented in section~\ref{ssec:formf-tilt}.

\subsection{\texorpdfstring{$\langle \Phi^{\{IJ\}} \, t^k t^l \rangle$}{<Phi(IJ) t t>}}
We consider the case where the bulk operator transforms as a symmetric traceless tensor of $O(N)$ which we will call $T$. Choosing the other insertion to be the broken current and following the same logic as in previous examples in section \ref{ssec:formf-tilt}, it is easy to find:
\ba
  &-\int \ud^p  x \,\langle t^i( x) t^j( x_1) T^{11}(y_2, x_2)\rangle =2\langle t^j( x_1)T^{i1}(y_2, x_2)\rangle\,,
    \\
    &-\int \ud^p x \,\langle t^i( x) t^j(\hat x_1) T^{\{kl\}}(y_2, x_2)\rangle=-\delta^{ki}\langle t^j( x_1)T^{1l}(y_2, x_2)\rangle-\delta^{il}\langle t^j( x_1)T^{k1}(y_2, x_2)\rangle \,,\\
    & \ \ \ +\frac{2 \delta^{kl}}{N-1}\,\langle t^j( x_1)T^{i1}(y_2, x_2)\rangle \,,
\ea
where $\{\}$ stands for the projection operation $P$ which symmetrizes indices and removes the trace with $P^2=P$, such that $T^{\{kl\}}$ is a symmetric traceless tensor of $O(N-1)$. Now we can set 
 \ba
 \langle t^i(x) t^j( x_1) T^{11}(y,x_2)\rangle&=\delta^{ij}\,K^{T}_{tt}(x_i,y)\, \mathcal H^{T}_{tt}(w)\,,\\
 \langle t^i(x) t^j(x_1) T^{\{kl\}}(y_2,x_2)\rangle&=\delta^{i\{k}\delta^{l\}j}\,K^{T}_{tt}(x_i,y)\, \mathcal H^{T}_{tt}(w)\,,\\
 \langle t^j(x_1)T^{i1}(y_2,x_2)\rangle&=\delta^{ij}\sqrt{\normt}\, \mu^{T}_t\,K^{T}_{t}(y,x_i)\,,
 \ea
and change integration variables to $w$ to arrive at the final sum rules:
\begin{equation}
\begin{aligned}
     &\int \frac{\ud w}{w^{1 + \frac{p}{2}} (1-w)^{1 - \frac{p}{2}}} \, \delta^{ij} \mathcal{H}^{T}_{tt}(w) = - 2 \delta^{ij} \sqrt{\normt} \, \mu^{T}_t \frac{\Gamma\left(\frac{p}{2}\right)}{\pi^{\frac{p}{2}}} \:, \\    
  &\int \frac{\ud w}{w^{1 + \frac{p}{2}} (1-w)^{1 - \frac{p}{2}}} \,\delta^{i\{k}\delta^{l\}j} \, \mathcal{H}^{T}_{tt}(w) = \Big(\delta^{ki} \delta^{jl} + \delta^{il} \delta^{jk}  -\frac{2 \delta^{ij} \delta^{kl}}{N-1} \Big)\,\sqrt{\normt}\, \mu^{T}_t \frac{\Gamma\left(\frac{p}{2}\right)}{\pi^{\frac{p}{2}}}\,.
\end{aligned}
\end{equation}

\subsection{\texorpdfstring{$\langle \Phi^{[IJ]} \, t^k t^l \rangle$}{<Phi[IJ] t t>}}

Finally let us consider the case where $\Psi^{KL}$ now transforms in an antisymmetric representation. 
We now get
\begin{multline}
    -\int \ud^p x \langle t^i(x) J^{[1j]}(y_1,x_1) \Psi^{[KL]}(y_2,x_2)\rangle=\langle J^{[ij]}(y_1,x_1) \Psi^{[KL]}(y_2,x_2)\rangle \\
     -\delta^{iK}\langle J^{[1j]}(y_1,x_1) \Psi^{1L}(y_2,x_2)\rangle+\delta^{iL}\langle J^{[1j]}(y_1,x_1) \Psi^{1K}(y_2,x_2)\rangle\,.
\end{multline}
To proceed let us assume   
\ba
    y^{\mu}J_\mu^{[ij]}(y,x)\underset{|y|\to 0}=|y|^{p+1-d+\Delta_j} j^{[ij]}(x)\,, \qquad \Delta_j>p\,,
\ea
i.e. there are no marginal or relevant defect operators in the antisymmetric representation of $O(N-1)$. Under these circumstances we can push $y_1$ to zero to obtain
\ba
    \int \ud w  \frac{\ud w}{w^{1 + \frac{p}{2}} (1-w)^{1 - \frac{p}{2}}} \,  \mathcal{H}^{\Psi^{[KL]}}_{tt}(w)=
     - \Big(\delta^{iK}\delta^{jL} - \delta^{iL} \delta^{jK}\Big) \, \sqrt{C_t} \mu^{\Psi^{[IJ]}}_{t} \frac{\Gamma\left(\frac{p}{2}\right)}{\pi^{\frac{p}{2}}}\,.
\ea

\section{More details on the \texorpdfstring{$p=1$}{p=1} defect sum rules}
\label{app:1d-SR}

\subsection{First identity}

Similarly as for the general case, we use conformal invariance to go to the frame $(x_1,x_2,x_3)=(0,1,\infty)$. However, here one has to be careful with the ordering of the operators: the domain of integration splits into three different regions.

\subsubsection{Tilt} 

We start from~\eqref{eq:tilthom-x}. The correlator has the expression~\eqref{eq:4pt-01inf}, with here $p=1$. The sum rule becomes
\begin{equation}
    \int_{-\infty}^0 \frac{\ud x}{(1-x)^2} \, z^{\hdO-1}\, \mathcal G_{ijIJ}(z)\bigg|_{z=\frac{x}{x-1}}+
\int_0^1 \ud x \, z^{\hdO-1}\, \mathcal G_{jiIJ}(z)\bigg|_{z=x}+\int_1^\infty \frac{\ud x}{x^2} \, \mathcal G_{jIiJ}(z)\bigg|_{z=\frac 1x} =0\:,
\end{equation}
which becomes after change of variable
\begin{flalign}
\text{\bf Tilt soft sum rule:}\qquad\int_0^{1} \mathrm{d} z\left[z^{\Delta_\hcO-1}{\mathcal{G}}_{(ij)IJ}(z)+{\mathcal{G}}_{iJjI}(z)\right]=0\,. &&\label{eq:tilthom-p=1}
\end{flalign}

A special case correspond to taking $\hcO_I$ to be the tilt. In this case we have the correlator of four operators in the fundamental representation of $O(N-1)$ and we can use~\eqref{eq:4pt-channel-split} to rewrite this as
\begin{flalign}
\text{\bf 4 tilts soft sum rule:}\qquad\int_0^{1} \mathrm{d}z \,\mathcal G^{(1)}_{+}(z) = 0\:. &&\label{eq:srtiltHom}
\end{flalign}

\subsubsection{Displacement} 

The derivation starts from~\eqref{eq:disphom-x}. It would be possible to obtain the sum rules in this conformal setup, but it will be easier to expand directly in $x_0$ instead of doing in the end.
The three sum rules become
\begin{equation}
\begin{aligned} 
    \int_{-\infty}^0 \frac{\ud x\, x^n}{(1-x)^4} \, z^{\hdO-2}\, \mathcal G_{abIJ}(z)\bigg|_{z=\frac{x}{x-1}} &+
\int_0^1 \ud x \, x^n\, z^{\hdO-2}\, \mathcal G_{baIJ}(z)\bigg|_{z=x} \\
&+\int_1^\infty \frac{\ud x \,x^n}{x^4}\, \mathcal G_{bIaJ}(z)\bigg|_{z=\frac 1x} =0\:,
\end{aligned}
\end{equation}
with $n\in\{0,1,2\}$. After change of variable, we obtain
\begin{flalign}
\label{eq:disphom-p=1}
& \text{\bf Displacement soft sum rules:}  && \nonumber \\
&\int_0^{1} \mathrm{d} z\left[z^{\Delta_{\hat\op}-2}\Big(z^n(z-1)^{2-n} {\mathcal{G}}_{abIJ}(z)+z^n {\mathcal{G}}_{baIJ}(z)\Big)+z^{2-n}{\mathcal{G}}_{aJbI}(z)\right]=0 \:, \quad n\in\{0,1,2\}\:. \hspace{-1cm}&& 
\end{flalign}

When all four operators are displacements only two such sum rules are independent. In the $O(n)$ invariant notation of \reef{eq:4pt-channel-split} we find: 
\label{eq:4disphom}
\begin{flalign}
&\text{\bf 4 displacements soft sum rules:} && \nonumber\\
    &\int_0^{1} \mathrm{d}z(z^2-z+1) \mathcal G^{(1)}_{+}(z) = 0 \, ,\label{eq:DH1} &&\\
    &\int_0^{1} \mathrm{d}z\bigg[(2z^2-2z-1)\Big(\mathcal G_+^{(1)}(z)-3\mathcal G_+^{(2)}(z)\Big)+3(2z-1)\mathcal G_-(z)\bigg] = 0 \,.
\end{flalign}

\subsection{Second identity}

For the second identity, we use conformal invariance to go to the frame $(T,x_1,x_2)=(0,-1,1)$. Here also the domain of integration is split due to the different orderings of the operators. Naively there are four different regions but they are pairwise identical because of the relations $\langle \hcO \tilt \tilt \hcO\rangle=\langle \tilt \hcO \hcO \tilt\rangle$ and $\langle \tilt \hcO \tilt\hcO\rangle=\langle \hcO \tilt \hcO \tilt\rangle$. 

\subsubsection{Tilt} 

For the tilt, we start from~\eqref{eq:tiltinhom-x}. For $p=1$, the left-hand side becomes
\begin{multline}
\int_0^1 \ud x \int_{-1}^0  \ud x'\, \frac{2(z-1)^2}{(x-x')^2}\, z^{\hdO-1}\mathcal G_{I[ij]J}(z)\bigg|_{z=\frac{(1-x)(1+x')}{(1+x)(1-x')}}\\
+\int_1^\infty \ud x \int_{-1}^0  \ud x'\, \frac{2}{(x'-x)^2}\, \mathcal G_{[i|I|j]J}(z)\bigg|_{z=\frac{(1-x)(1+x')}{2(x'-x)}}
\end{multline}
which can be recast as
\ba
\int_0^1 \ud z \int_{\frac{z-1}{z+1}}^0 \ud x'\, \frac{4}{1-x'^2}\, z^{\hdO-1}\mathcal G_{I[ij]J}(z)+\int_0^{\frac 12} \ud z \int_{2z-1}^0 \ud x'\, \frac{4}{1-x'^2}\, \mathcal G_{[i|I|j]J}(z)\:.
\ea
The right-hand side is the same as in the general case: \eqref{eq:tilt-RHS}. Performing the $x'$ integrals and using crossing symmetry we find:
\begin{flalign}
&\text{\bf Tilt double soft sum rule:}&&\nonumber \\
&\int_0^{1} \mathrm{d} z\left[2\,z^{\Delta_\hcO-1}\,\log (z)\, {\mathcal{G}}_{I[ij]J}(z)+{\mathcal{G}}_{[i|I|j]J}(z)\, \log \big(\tfrac{z}{1-z}\big)\right]=Q^{[ij]}_{IJ}\: .
\end{flalign}

Specializing to the case where the two defect operators are also tilts, one can again further simplify the sum rule:
\begin{flalign}
\label{eq:4t-inhomogeneous}
& \text{\bf 4 tilts double soft sum rule:} &&\nonumber \\
    &\int_0^{1} \mathrm{d}z \bigg[
    \Big(
    \mathcal G^{(1)}_+(z)-3\,\mathcal G^{(2)}_{ +}(z)\Big)
    \log[z(1-z)]+3\,\mathcal G_-(z) \log\left(\tfrac z{1-z}\right)
    \bigg]  =\normt \: .
\end{flalign}

\subsubsection{Displacement} 

For the displacement, we start from~\eqref{eq:dispinhom-x}. We proceed as in the general case by expanding in $x_0$ and $x'_0$. For each commutator of~\eqref{eq:brokenC}, there is a couple $(m,n)\in\{(2,0),(1,1),(1,0),(1,2)\}$ for which the left-hand side of the sum rule is
\begin{multline}
\int_0^1 \ud x \int_{-1}^0  \ud x'\, \frac{2x^mx'^n(z-1)^4}{(x-x')^4}\, z^{\hdO-2}\mathcal G_{IabJ}(z)\bigg|_{z=\frac{(1-x)(1+x')}{(1+x)(1-x')}}\\
+\int_1^\infty \ud x \int_{-1}^0  \ud x'\, \frac{2x^mx'^n}{(x'-x)^4}\, \mathcal G_{aIbJ}(z)\bigg|_{z=\frac{(1-x)(1+x')}{2(x'-x)}}-(a\leftrightarrow b, m\leftrightarrow n)\:.
\end{multline}

This can be recast as
\begin{multline}
    \int_0^1 \ud z \int_{\frac{z-1}{z+1}}^0 \ud x'\, \frac{4x^n(1+x+z-xz)^{2-m}(1+x-z+xz)^{m}}{(1-x'^2)^3}z^{\hdO-2}\, \mathcal G_{IabJ}(z) \\ 
    + \int_0^{\frac 12} \ud z \int_{2z-1}^0 \ud x'\, \frac{4x^n(1+x-2z)^{2-m}(1+x-2xz)^{m}}{(1-x'^2)^3}\, \mathcal G_{aIbJ}(z)-(a\leftrightarrow b, m\leftrightarrow n)\:.
\end{multline}

These integrals can be performed and produce some logs similarly as in the tilt case. However, they also produce some polynomial terms. They can be canceled upon using the first identity sum rules~\eqref{eq:disphom-p=1}. Note that this did not happen in the general $p$ case because of the choice of the frame. The right-hand side is the same as in the general case:~\eqref{eq:disp-RHS}. In the end, the sum rules combine into the three independent constraints
\begin{flalign}
& \text{\bf Displacement double soft sum rules} && \nonumber \\
&\int_0^{1} \mathrm{d} z\,\left[2z^{\Delta_\hcO-2}\,z\, {\mathcal{G}}_{I[ab]J}(z)\log z
+z(z-1)\, {\mathcal{G}}_{[a|I|b]J}(z)\log\big(\tfrac{z}{1-z}\big) \right] = 2M_{IJ}^{[ab]}\:, && \\ %\BG{3}
&\int_0^{1} \mathrm{d} z\,\left[2z^{\Delta_\hcO-2}\,(z^2+1)\,{\mathcal{G}}_{I[ab]J}(z)\log z
+(2z^2-2z+1)\, {\mathcal{G}}_{[a|I|b]J}(z)\log\big(\tfrac{z}{1-z}\big) \right] = 4M_{IJ}^{[ab]}\:, && \\
&\int_0^{1} \mathrm{d} z\,\left[2z^{\Delta_\hcO-2}\,(z^2-1)\,{\mathcal{G}}_{I(ab)J}(z)\log z
+(2z-1)\, {\mathcal{G}}_{(a|I|b)J}(z)\log\big(\tfrac{z}{1-z}\big) \right] = 4\Delta_\mathcal{O}C_\op\delta_{ab}\delta_{IJ}\:. &&
\end{flalign}

We have checked these agree with the ones obtained in \cite{Gabai:2025zcs}. In the special case where all operators are displacements we get: 
\begin{flalign}
&\text{\bf 4 displacements double soft sum rules} &&\nonumber \\
&\int_0^{1}\mathrm{d}z\ \Big[{\mathcal G_+^{(1)}(z)-3\mathcal G_+^{(2)}(z)}+\big(2z-1\big)\mathcal G_-(z)\Big]\big[z \log z+(1-z)\log(1-z)\big] = -2C_D \,,&& \\
&\int_0^1\mathrm{d}z\ \bigg[\Big[(z^2+4z-3)\mathcal G_+^{(1)}(z)+(z^2-4z+1)\mathcal G_+^{(2)}(z)\Big]\log z &&\nonumber\\
&\qquad+\ \Big[(z^2-6z+2)\mathcal G_+^{(1)}(z)+(z^2+2z-2)\mathcal G_+^{(2)}(z)\Big]\log (1-z) &&\nonumber\\
&\qquad +\ \mathcal G_-(z)\big[(z^2-1)\log z -(z^2-2z)\log(1-z)\big]\bigg] = 0\,, &&\\
&\int_0^1\mathrm{d}z\ \bigg[\Big[\mathcal G_+^{(1)}(z)-3\mathcal G_+^{(2)}(z)\Big]\big[(z^2+4z+1)\log z + (z^2-6z+6)\log(1-z)\big] &&\nonumber \\
&\qquad+\ \mathcal G_-(z)\big[(13z^2-8z+3)\log z - (13z^2-18z+8)\log(1-z)\big]\bigg]= 0 \, .&&
\end{flalign}

\subsubsection{Mixed tilt/displacement} 

We start from~\eqref{eq:tilt/dispinhom-x}
\ba
    0&= 
    \int_1^\infty \ud x \int_{-\infty}^1 \ud x' \Big(x'^n \langle t^j(1) t^i(x)D^a(x')D^b(-1)\rangle -x^n \langle t^j(1) D^a(x)t^i(x')D^b(-1)\rangle \Big)\:,
\ea
where $n=0,1,2$ depending on the chosen conformal Killing vector $\xi$ associated to the integrated displacement. Evaluating these integrals we obtain three sum rules. However, upon using the first identity sum rules~\eqref{eq:4disphom} with two tilts and two displacements, they reduce to only one independent sum rule:
\begin{flalign}
&\text{\bf Tilt and displacement double soft sum rule} && \nonumber \\
&\int_{0}^1 \, \ud z \Big[ \Big(\mathcal{G}^{ibja} (z) - z\, \mathcal G^{abji} (z) \Big)\log z + z\, \mathcal{G}^{abij}(z) \log \big(\tfrac{1-z}{z}\big)  \Big] = 0\:. &&
\end{flalign}

\section{Integrals}

\subsection{Bulk integrals}

\subsubsection{Chain integral}

A first generic integral that is useful is the \emph{chain integral}~\cite{Kleinert2001}
\begin{equation}
\label{eq:chain-int}
\int\frac{\mathrm{d}^{D} x_3}{(x_{13}^2)^a(x_{23}^2)^b} = \frac{\Gamma\left(\frac D 2 - a\right)\Gamma\left(\frac D 2 - b\right)\Gamma\left(a+b-\frac D 2\right)}{\Gamma\left(a\right)\Gamma\left(b\right)\Gamma\left(D-a-b\right)}\frac {\pi^{\frac D 2}}{(x_{12}^2)^{a+b-\frac{D}{2}}}\:.
\end{equation}

It can be computed by introducing Schwinger parameters:
\begin{equation}
\begin{aligned}
    \int\frac{\mathrm{d}^{D} x_3}{(x_{13}^2)^a(x_{23}^2)^b} &= \frac{1}{\Gamma(a)\Gamma(b)}\iint_0^\infty\mathrm{d}s_1\mathrm{d}s_2\,s_1^{a-1}s_2^{b-1}\int\mathrm{d}^{D} x_3\ e^{-(s_1 x_{13}^2+s_2 x_{23}^2)} \\
    &= \frac{\pi^\frac{D}{2}}{\Gamma(a)\Gamma(b)}\iint_0^\infty\mathrm{d}s_1\mathrm{d}s_2\,\frac{s_1^{a-1}s_2^{b-1}}{(s_1+s_2)^\frac{D}{2}}\,e^{-\frac{s_1s_2}{s_1+s_2}x_{12}^2} \\
    &=\frac{\pi^\frac{D}{2}}{\Gamma(a)\Gamma(b)}\int_0^1\mathrm{d}\alpha\,\alpha^{a-1}(1-\alpha)^{b-1}\int_0^\infty\mathrm{d}s\ s^{1+(a-1)+(b-1)-\frac{D}{2}}\ e^{-s\alpha(1-\alpha)x_{12}^2} \\
    &= \frac{\pi^\frac{D}{2}}{(x_{12}^2)^{a+b-\frac{D}{2}}}\frac{\Gamma\big(a+b-\frac{D}{2}\big)}{\Gamma(a)\Gamma(b)}\int_0^1\mathrm{d}\alpha\,\frac{\alpha^{a-1}(1-\alpha)^{b-1}}{\big(\alpha(1-\alpha)\big)^{a+b-\frac{D}{2}}} \\
    &= \frac{\pi^\frac{D}{2}}{(x_{12}^2)^{a+b-\frac{D}{2}}}\frac{\Gamma\big(a+b-\frac{D}{2}\big)}{\Gamma(a)\Gamma(b)}\ B\left(\frac{D}{2}-b,\,\frac{D}{2}-a\right)\:.
\end{aligned}
\end{equation}

\subsubsection{Standard integrals}

Two other general integrals that appear in particular are
\begin{align}
Y_{123}&:=\quad\begin{tikzpicture}[baseline,valign]
    \draw[dashed](0, 0) -- (0.5, 0);
    \draw[dashed](0.5, 0) -- (1, 0.3);
    \draw[dashed](0.5, 0) -- (1, -0.3);
    \node at (0.5, 0) [bcirc] {};
\end{tikzpicture} \quad = \int\mathrm{d}^d x_4\ I_{14}I_{24}I_{34}\: ,\\
\vspace{0.2cm}
X_{1234}&:=\quad\begin{tikzpicture}[baseline,valign]
    \draw[dashed](0, 0.2) -- (1, 0.8);
    \draw[dashed](1, 0.2) -- (0, 0.8);
    \node at (0.5, 0.5) [bcirc] {};
\end{tikzpicture}\quad = \int\mathrm{d}^d x_5\ I_{15}I_{25}I_{35}I_{45}\: .
\end{align}
with $I_{ij}$ the scalar propagator~\eqref{eq:free-prop}. These integrals can be solved analytically (see~\cite{Drukker_2009} for a modern notation), but the relevant results that we need are their one-dimensional limit, and their pinching limit $Y_{122}$ and $X_{1233}$. The latter are log-divergent: they are usually computed in point-splitting regularization~\cite{Drukker_2009}, but we will need it in dimensional regularization in $d=4-\veps$. Note that the $Y$-integral can be expressed as the limit of $X_{1234}$ where one of the external points is sent to $\infty$
\begin{equation}
    Y_{123}=\lim _{x_4 \rightarrow \infty}  \frac 1 {I_{n4}}X_{1234}\: . 
\end{equation}

For the one-dimensional limit, a useful solution of the $X$-integral is given in term of the Bloch-Wigner function
\begin{equation}
\begin{aligned}
    X_{1234}&=\frac{I_{12} I_{34}}{16 \pi^2} \chi \bar{\chi} D(\chi, \bar{\chi})\:, \\
    D(\chi, \bar{\chi})&:=\frac{1}{\chi-\bar{\chi}}\left(2 \operatorname{Li}_2(\chi)-2 \operatorname{Li}_2(\bar{\chi})+\log \chi \bar{\chi} \log \frac{1-\chi}{1-\bar{\chi}}\right)\:, \\
    \chi \bar{\chi} &=\frac{I_{13} I_{24}}{I_{12} I_{34}}, \quad(1-\chi)(1-\bar{\chi})=\frac{I_{13} I_{24}}{I_{14} I_{23}}\:.
\end{aligned}
\end{equation}

It is then straightforward to obtain its one-dimensional limit
\begin{equation}
\begin{aligned}
X_{1234} & =\frac{I_{12} I_{34}}{16 \pi^2} \chi^2 D(\chi, \chi) \\
& =-\frac{I_{12} I_{34}}{8 \pi^2} \frac{\chi}{1-\chi}(\chi \log \chi+(1-\chi) \log (1-\chi))\: .
\end{aligned} 
\end{equation}

This gives for the $Y$-integral
\begin{equation}
    Y_{123}=-\frac{1}{2} \left(\frac{1}{\tau_{13}\tau_{23}} \log \frac{\tau_{12}}{\tau_{13}}+\frac{1}{\tau_{12}\tau_{13}} \log \frac{\tau_{23}}{\tau_{13}}\right)\: .
\end{equation}

For the pinching limit, a useful identity~\cite{Artico_2025}
\begin{equation}
\label{eq:X1233-Y122}
    X_{1233}=\frac{1}{2}\left(I_{13} Y_{223}+I_{23} Y_{113}\right)-\frac{I_{13} I_{23}}{32 \pi^2} \log \frac{I_{13} I_{23}}{I_{12}^2}\:,
\end{equation}
can be obtained (via the point-splitting regularization) with the solution of the $X$-integral in term of the $\Phi$ function~\cite{Usyukina:1992jd}
\begin{equation}
\begin{aligned}
    X_{1234}&=\frac{1}{16 \pi^2} I_{13} I_{24} \Phi\left(\frac{I_{13} I_{24}}{I_{12} I_{34}}, \frac{I_{13} I_{24}}{I_{14} I_{23}}\right)\:, \\
    \Phi(r, s)&:=\frac{1}{A} \operatorname{Im}\left\{\operatorname{Li}_2 e^{i \varphi} \sqrt{\frac{r}{s}}+\log \sqrt{\frac{r}{s}} \log \left(1-e^{i \varphi} \sqrt{\frac{r}{s}}\right)\right\} \:,\\
e^{i \varphi}&:=i \sqrt{-\frac{1-r-s-4 i A}{1-r-s+4 i A}}, \quad A:=\frac{1}{4} \sqrt{4 r s-(1-r-s)^2}\:.
\end{aligned}
\end{equation}

A direct computation with the chain integral~\eqref{eq:chain-int} gives us the result for $Y_{122}$ in dimensional regularization
\begin{equation}
    Y_{122} = \frac{2\pi^2}{(x_{12}^2)^{1-\veps}}\left(\frac1\veps +\frac{3-\aleph}2+\mathcal{O}(\veps)\right)\:.
\end{equation}

Inserting it into~\eqref{eq:X1233-Y122}, we obtain the pinching limit of the $X$-integral in dimensional regularization
\begin{equation}
\label{eq:X1233}
    X_{1233} = \frac{2\pi^2}{(x_{13}^2 x_{23}^2)^{1-\veps}(x_{12}^2)^\frac{\veps}{2}}\left(\frac1\veps +\frac{3-\aleph}2+\mathcal{O}(\veps)\right)\:.
\end{equation}

\subsection{Defect integrals}

In the presence of a defect, other class of integrals appear in the computations. We first have integrals over the defect, the simplest case being
\begin{equation}
    \int_{-\infty}^{+\infty} \frac{\ud x_2}{\left(y_1^2+x_{12}^2\right)^\Delta}= \frac{\sqrt{\pi}\,\Gamma\left(\Delta-\frac{1}{2}\right)}{\Gamma\left(\Delta\right)}\frac{1}{(y_1^2)^\frac{2\Delta-1}2}\:,
\end{equation}
which can also be proven with Schwinger parameters.

\subsubsection{Chain integral with additional bulk-to-defect propagators}

Another useful set of integrals are coupled bulk and defect integrals. The simplest case is the generalization of the chain integral with additional bulk-to-defect propagators
\begin{equation}
\int\frac{\ud x_3\mathrm{d}^{D-1} y_3}{(y_{13}^2+x_{13}^2)^a(y_{23}^2+x_{23}^2)^b(y_3^2)^\frac c2}\:.
\end{equation}

\begin{itemize}
\item When the two operators are on the defect $y_1=y_2=0$, the generalization is straightforward
\begin{equation}
\begin{aligned}
    &\int\frac{\ud x_3\mathrm{d}^{D-1} y_3}{(y_{3}^2+x_{13}^2)^a(y_{3}^2+x_{23}^2)^b(y_3^2)^\frac c2}\\
    =\ &\frac{\Gamma\left(\frac{D-c-1}{2}\right)}{\Gamma\left(\frac{D-1}{2}\right)}\,\frac{\Gamma\left(\frac {D-c} 2 - a\right)\Gamma\left(\frac {D-c} 2 - b\right)\Gamma\left(a+b-\frac {D-c} 2\right)}{\Gamma\left(a\right)\Gamma\left(b\right)\Gamma\left(D-c-a-b\right)}\frac{\pi^{\frac D2}}{(x_{12}^2)^{a+b-\frac{D-c}{2}}}\ \:.
\end{aligned}
\end{equation}

The proof relies again on Schwinger parameters, with a split of the integral into a defect, radial and angular part (the latter being trivial)
\begin{equation}
\begin{aligned}
    &\int\frac{\ud x_3\mathrm{d}^{D-1} y_3}{(y_{3}^2+x_{13}^2)^a(y_{3}^2+x_{23}^2)^b(y_3^2)^\frac c2} \\
    =\ &\frac{1}{\Gamma(a)\Gamma(b)}\iint_0^\infty\mathrm{d}s_1\mathrm{d}s_2\,s_1^{a-1}s_2^{b-1}\int\frac{\ud x_3\mathrm{d}^{D-1} y_3}{(y_3^2)^\frac c2}\ e^{-s_1(y_{13}^2+ x_{13}^2)-s_2 (y_{23}^2+x_{23}^2)} \\
    =\ &\frac{2\pi^\frac{D-1}{2}}{\Gamma(a)\Gamma(b)\Gamma\big(\frac{D-1}{2}\big)}\iint_0^\infty\mathrm{d}s_1\mathrm{d}s_2\,s_1^{a-1}s_2^{b-1}\int_{-\infty}^{+\infty}\mathrm{d}x_3\,e^{-(s_1x_{13}^2+s_2x_{23}^2)}\int_0^\infty\mathrm{d} r_3\, r_3^{D-2-c}\ e^{-(s_1+s_2)r_3^2} \\
    =\ &\frac{\pi^\frac{D}{2}\Gamma\big(\frac{D-c-1}{2}\big)}{\Gamma(a)\Gamma(b)\Gamma\big(\frac{D-1}{2}\big)}\iint_0^\infty\mathrm{d}s_1\mathrm{d}s_2\,\frac{s_1^{a-1}s_2^{b-1}}{(s_1+s_2)^\frac{D-c-1+1}{2}}\,e^{-\frac{s_1s_2}{s_1+s_2}x_{12}^2}\:.
\end{aligned}
\end{equation}

\item When only one operator is on the defect $y_2=0$, and the other operator is in the bulk, the integral becomes non-trivial. Some results can be obtained by using the $X_{1233}$ expansion~\eqref{eq:X1233}, pushing $y_1,y_2\rightarrow0$, and integrating over one of them, for example for $(a,b,c)=(2-\veps,1-\frac{\veps}{2},1-\veps)$:
\begin{equation}
\label{eq:Phi2-phi-integral}
   \int \frac{\ud x_3 \ud^{3-\veps}y_3}{\big(y_{13}^2+x_{13}^2\big)^{2-\veps}\big(y_3^2+x_{23}^2\big)^{1-\frac \veps 2}y_3^{1-\veps}}
    = \ \frac{\pi^2}{y_1^{1-2\veps}(y_1^2+x_{12}^2)^{1-\frac{\veps}{2}}} \left(\frac2\veps + 3-\aleph + 2\log 2 + \op(\veps) \right)\:,
\end{equation}
and for $(a,b,c)=(1-\frac{\veps}{2},2-\veps,1-\veps)$:
\begin{equation}
\label{eq:Phi-phi2-integral}
   \int \frac{\ud x_3 \ud^{3-\veps}y_3}{\big(y_{13}^2+x_{13}^2\big)^{1-\frac{\veps}2}\big(y_3^2+x_{23}^2\big)^{2-\veps}y_3^{1-\veps}}
    = \ \frac{-2\pi^2}{y_1^{-1+\veps}(y_1^2+x_{12}^2)^{2-2\veps}}\left(1 + \frac\veps2\big(7-\aleph - 4\log 2\big) + \op(\veps^2) \right) \:.
\end{equation}

Another result was obtained in~\cite{Gimenez-Grau:2022ebb} in dimension $d=4-\veps$ and for the specific coefficients $(a,b,c)=(1-\frac{\veps}{2},1-\frac{\veps}{2},2-2\veps)$
\begin{equation}
     \int \frac{\ud x_3 \ud^{3-\veps}y_3}{\big(y_{13}^2+x_{13}^2\big)^{1-\frac\veps2}\big(y_3^2+x_{23}^2\big)^{1-\frac \veps 2}y_3^{2-2\veps}}= \frac{\pi^2}{y_1^{2\veps}(y_1^2+x_{12}^2)^{1-\frac{5\veps}{2}}}\left(\frac2\veps + 7-\aleph - 8\log 2 + \op(\veps) \right)\:.
\end{equation}

The proof also relies on Schwinger integration, but the naive computation fails here: an analytic regularization is needed, developed in~\cite{Panzer_2014,Panzer_2015} and implemented in \textbf{HyperInt}~\cite{Panzer_PhD}.

\item When the two operators are in the bulk, a result was also obtained in~\cite{Gimenez-Grau:2022ebb} in dimension $d=4-\veps$ and for the specific coefficients $(a,b,c)=(2-\veps,1-\frac{\veps}{2},1-\veps)$
\begin{equation}
\begin{aligned}
     \int \frac{\ud x_3 \ud^{3-\veps}y_3}{\big(y_{13}^2+x_{13}^2\big)^{2-\veps}\big(y_{23}^2+x_{23}^2\big)^{1-\frac \veps 2}y_3^{1-\veps}}
    = \ \frac{\pi^2}{y_1^{1-2\veps}(y_{12}^2+x_{12}^2)^{1-\veps}\big((y_1+y_2)^2+x_{12}^2\big)^\frac{\veps}{2}}\\ \left(\frac2\veps + 3-\aleph + 2\log 2 + \op(\veps) \right)\:,
\end{aligned}  
\end{equation}
with a similar analytic regularization. As a cross-check the limit $\vec{x}_2=0$ of this result gives back~\eqref{eq:Phi2-phi-integral}.

\end{itemize}
\section{Further results for the pinning line defect}\label{app:morepinning}
\subsection{The correlator \texorpdfstring{$\langle(\Phi_I\Phi_J)\hat{\phi}_K\rangle$}{<Phi2 phi>}}
\label{ssec:Phi2phi}

At $O(\veps)$, the correlator $\langle(\Phi_I\Phi_J)\hat{\phi}_K\rangle$ receives contributions from four Feynman diagrams:
\begin{equation}
   \langle(\Phi_I\Phi_J)(y_1,x_1)\hat{\phi}_K(x_2)\rangle \quad=\quad 
  \begin{tikzpicture}[baseline,valign]
  \draw[thick] (0, 0) -- (1.2, 0);
  \draw[dashed](0.3, 0) -- (0.6, 1);
  \draw[dashed](0.9, 0) -- (0.6, 1);
  \node at (0.9, 0) [dcirc] {};
  \end{tikzpicture}\quad +\quad
\begin{tikzpicture}[baseline,valign]
  \draw[thick] (0, 0) -- (1.5, 0);
  \draw[dashed](0.3, 0) -- (0.7, 1);
  \draw[dashed](0.6, 0) -- (0.9, 0.6);
  \draw[dashed](0.9, 0) -- (0.9, 0.6);
  \draw[dashed](1.2, 0) -- (0.9, 0.6);
  \draw[dashed](0.7, 1) -- (0.9, 0.6);
  \node at (0.6, 0) [dcirc] {};
  \node at (0.9, 0) [dcirc] {};
  \node at (1.2, 0) [dcirc] {};
  \node at (0.9, 0.6) [bcirc] {};
  \end{tikzpicture}\quad +\quad
  \begin{tikzpicture}[baseline,valign]
  \draw[thick] (0, 0) -- (1.5, 0);
  \draw[dashed](0.3, 0) -- (0.7, 1);
  \draw[dashed](0.6, 0) -- (0.9, 0.6);
  \draw[dashed](0.9, 0) -- (0.9, 0.6);
  \draw[dashed](1.2, 0) -- (0.9, 0.6);
  \draw[dashed](0.7, 1) -- (0.9, 0.6);
  \node at (0.3, 0) [dcirc] {};
  \node at (0.6, 0) [dcirc] {};
  \node at (1.2, 0) [dcirc] {};
  \node at (0.9, 0.6) [bcirc] {};
  \end{tikzpicture}\quad +\quad
\begin{tikzpicture}[baseline,valign]
  \draw[thick] (-0.5, 0) -- (0.5, 0);
  \draw[dashed] (-0.2, 0) -- (0, 0.4);
  \draw[dashed] (0.2, 0) -- (0, 0.4);
  \node at (0.2,0) [dcirc] {};
  \node at (0,0.4) [bcirc] {}; 
  \draw[dashed] (0,0.4) arc (230:130:0.4);
  \draw[dashed] (0,0.4) arc (310:410:0.4);
  \end{tikzpicture}\quad+\quad \dots  
\end{equation}
The first three diagrams are direct combinations of diagrams already considered in~\cite{Allais:2014fqa,Gimenez-Grau:2022ebb}. 
The last diagram is new and contains a non-trivial integral over the bulk vertex:
\begin{equation}
\begin{aligned}
    \begin{tikzpicture}[baseline,valign]
  \draw[thick] (-0.5, 0) -- (0.5, 0);
  \draw[dashed] (-0.2, 0) -- (0, 0.4);
  \draw[dashed] (0.2, 0) -- (0, 0.4);
  \node at (0.2,0) [dcirc] {};
  \node at (0,0.4) [bcirc] {}; 
  \draw[dashed] (0,0.4) arc (230:130:0.4);
  \draw[dashed] (0,0.4) arc (310:410:0.4);
\end{tikzpicture} \ =\ & \kappa^{4}(\delta_{IJ}\delta_{K1}+\delta_{IK}\delta_{J1}+\delta_{JK}\delta_{I1}) \frac{\lambda_0 h_0}{3}\left(\frac{\sqrt{\pi}\,\Gamma\left(\frac{1-\veps}{2}\right)}{\Gamma\left(1-\frac{\veps}{2}\right)}\right) \\
\ &\int \frac{\mathrm{d}^{4-\veps} X_3}{\big(y_{13}^2+x_{13}^2\big)^{2-\veps}\big(y_{3}^2+x_{23}^2\big)^{1-\frac \veps 2}|y_3|^{1-\veps}}\:,
\end{aligned}
\end{equation}

The integral reduces to the divergence of the Bloch-Wigner function with two coincident external points~\eqref{eq:X1233} integrated over the defect. After using the result~\eqref{eq:Phi2-phi-integral}
\begin{equation}
   \int \frac{\mathrm{d}^{4-\veps} X_3}{\big(y_{13}^2+x_{13}^2\big)^{2-\veps}\big(y_{3}^2+x_{23}^2\big)^{1-\frac \veps 2}|y_3|^{1-\veps}}
    = \ \frac{\pi^2}{|y_1|^{1-2\veps}(y_1^2+x_{12}^2)^{1-\frac{\veps}{2}}} \left(\frac2\veps + 3-\aleph + 2\log 2 + \op(\veps) \right)\:,
\end{equation}
we find the expression of the diagram
\begin{equation}
\begin{tikzpicture}[baseline,valign]
  \draw[thick] (-0.5, 0) -- (0.5, 0);
  \draw[dashed] (-0.2, 0) -- (0, 0.4);
  \draw[dashed] (0.2, 0) -- (0, 0.4);
  \node at (0.2,0) [dcirc] {};
  \node at (0,0.4) [bcirc] {}; 
  \draw[dashed] (0,0.4) arc (230:130:0.4);
  \draw[dashed] (0,0.4) arc (310:410:0.4);
  \end{tikzpicture} \quad=\quad \frac{\kappa^{\frac32}(\delta_{IJ}\delta_{K1}+\delta_{IK}\delta_{J1}+\delta_{JK}\delta_{I1})}{|y_1|^{1-2\veps} (y_1^2+x_{12}^2)^{1-\frac\veps2}} \left(\frac{\blambda_0 h_0}{3}\right) \left(\frac 1{\veps}+\frac{3\aleph+1} 4 +2\log2 \right).   
\end{equation}
Adding all the diagrams and using \eqref{eq:general-2pt-Bb}, we can extract the bulk-to-defect coefficients. For instance, $\mu^{T}_{t}$ is given by
\begin{align}
\mu^{T}_{t}  &= -\frac{\sqrt{N+8}}{2\sqrt{2}}\left[1+ \veps\ \frac{4(N+6)(N+8)\log2-N^2-39N-182}{4(N+8)^2} + \op(\veps^2) \right],
\end{align}
while $\mu^{T}_{\hat{\phi}_1}$ and $\mu^{\Phi^2}_{\hat{\phi}_1}$ are given in table \ref{tab:CFT-data}, and the rest is zero due to symmetry reasons.

\subsection{The correlator \texorpdfstring{$\langle(\Phi_I\Phi_J)(\hat\phi_K\hat\phi_L)\rangle$}{<Phi2 phi2>}}
\label{ssec:Phi2phi2}

The last bulk-to-defect correlator we need to compute is $\langle(\Phi_I\Phi_J)(\hat\phi_K\hat\phi_L)\rangle$. At $O(\veps)$ it contains contributions from four Feynman diagrams:
\begin{equation}
\langle(\Phi_I\Phi_J)(y_1,x_1)(\hat\phi_K\hat\phi_L)(x_2)\rangle \quad=\quad 
   \begin{tikzpicture}[baseline,valign]
  \draw[thick] (0, 0) -- (1, 0);
  \draw[dashed] (0.5,0) arc (210:150:0.9);
  \draw[dashed] (0.5,0) arc (330:390:0.9);
  \end{tikzpicture}\quad+\quad  
  \begin{tikzpicture}[baseline,valign]
  \draw[thick] (0, 0) -- (1, 0);
  \draw[dashed] (0.5,0) arc (230:130:0.35);
  \draw[dashed] (0.5,0) arc (310:410:0.35);
  \draw[dashed] (0.5,1) arc (-230:-130:0.35);
  \draw[dashed] (0.5,1) arc (-310:-410:0.35);
  \node at (0.5, 0.5) [bcirc] {};
  \end{tikzpicture}\quad+\quad 
  \begin{tikzpicture}[baseline,valign]
  \draw[thick] (0, 0) -- (1.4, 0);
  \draw[dashed](0.6, 0.5) -- (0.6, 1);
  \draw[dashed] (0.3,0) arc (190:110:0.45);
  \draw[dashed] (0.3,0) arc (310:350:0.8);
  \draw[dashed](0.9, 0) -- (0.6, 0.5);
  \draw[dashed](1.2, 0) -- (0.6, 1);
  \node at (0.9, 0) [dcirc] {};
  \node at (1.2, 0) [dcirc] {};
  \node at (0.6, 0.5) [bcirc] {};
  \end{tikzpicture}\quad+\quad 
  \begin{tikzpicture}[baseline,valign]
  \draw[thick] (0, 0) -- (1.2, 0);
  \draw[dashed](0.6, 0) -- (0.6, 1);
  \draw[dashed](0.3, 0) -- (0.6, 0.5);
  \draw[dashed](0.9, 0) -- (0.6, 0.5);
  \draw[dashed] (0.9,0) arc (340:415:0.82);
  \node at (0.3, 0) [dcirc] {};
  \node at (0.6, 0) [dcirc] {};
  \node at (0.6, 0.5) [bcirc] {};
  \end{tikzpicture}\quad+\quad \dots     \:,
\end{equation}
that are direct combinations of diagrams previously studied. We can write down the bulk-to-defect coefficients immediately without computing additional integrals. For instance, the result for $\mu_{\Phi^2 \hat{s}_{\pm}}$ is
\begin{align}
\mu^{\Phi^2}_{\hat{s}_\pm} &= \frac{\mp(N+16)+\sqrt{S_N}}{2\sqrt{2N}}\sqrt{1\pm\frac{N+18}{\sqrt{S_N}}}\left(1-\veps \frac{3N+32\pm\sqrt{S_N}}{4(N+8)}(1-2\log2) \right)\: , 
\end{align}
while the ones involving $T$, namely $\mu^{T}_{\hat{V}}, \mu^{T}_{\hat{T}}$ and $\mu^{T}_{\hat{s}_\pm}$, are given in table \ref{tab:CFT-data}.

\subsection{The correlator \texorpdfstring{$\langle(\Phi_I\Phi_J)\hat\phi_K\hat\phi_L\rangle$}{<Phi2 phi phi>}}
\label{ssec:Phi2phiphi}

This last correlator we compute here, $\langle(\Phi_I\Phi_J)\hat\phi_K\hat\phi_L\rangle$, is of particular interest as a test of the sum rules derived in section~\ref{ssec:SRcases}. At $\op(\veps)$ there are eight Feynman diagrams which contribute:
\ba
\langle (\Phi_I\Phi_J)(y_1,x_1)\hat\phi_K(x_2)\hat\phi_L(x_3)\rangle \quad &=\quad 
  \begin{tikzpicture}[baseline,valign]
  \draw[thick] (0, 0) -- (1.9, 0);
  \draw[dashed](0.3, 0) -- (0.5, 1);
  \draw[dashed](0.7, 0) -- (0.5, 1);
  \draw[dashed] (1.7,0) arc (20:160:0.4);
  \node at (0.3, 0) [dcirc] {};
  \node at (0.7, 0) [dcirc] {};
  \end{tikzpicture} \quad+\quad \begin{tikzpicture}[baseline,valign]
  \draw[thick] (0, 0) -- (2, 0);
  \draw[dashed](0.3, 0) -- (0.5, 1);
  \draw[dashed](0.7, 0) -- (0.5, 1);
  \draw[dashed] (1.8,0) arc (20:160:0.45);
  \draw[dashed](1.2, 0) -- (1.4, 0.3);
  \draw[dashed](1.6, 0) -- (1.4, 0.3);
  \node at (1.2, 0) [dcirc] {};
  \node at (1.6, 0) [dcirc] {};
  \node at (1.4, 0.3) [bcirc] {};
  \node at (0.3, 0) [dcirc] {};
  \node at (0.7, 0) [dcirc] {};
  \end{tikzpicture} \quad+\quad \begin{tikzpicture}[baseline,valign]
  \draw[thick] (0, 0) -- (2.4, 0);
  \draw[dashed](0.3, 0) -- (0.7, 1);
  \draw[dashed](0.6, 0) -- (0.9, 0.6);
  \draw[dashed](0.9, 0) -- (0.9, 0.6);
  \draw[dashed](1.2, 0) -- (0.9, 0.6);
  \draw[dashed](0.7, 1) -- (0.9, 0.6);
  \node at (0.3, 0) [dcirc] {};
  \node at (0.6, 0) [dcirc] {};
  \node at (0.9, 0) [dcirc] {};
  \node at (1.2, 0) [dcirc] {};
  \node at (0.9, 0.6) [bcirc] {};
  \draw[dashed] (2.2,0) arc (20:160:0.4);
  \end{tikzpicture}  \\[1ex]
  &\hspace{-5cm}+\quad \begin{tikzpicture}[baseline,valign]
  \draw[thick] (-0.5, 0) -- (1.5, 0);
  \draw[dashed] (-0.2, 0) -- (0, 0.4);
  \draw[dashed] (0.2, 0) -- (0, 0.4);
  \node at (0.2,0) [dcirc] {}; 
  \node at (-0.2,0) [dcirc] {}; 
  \node at (0,0.4) [bcirc] {}; 
  \draw[dashed] (0,0.4) arc (230:130:0.4);
  \draw[dashed] (0,0.4) arc (310:410:0.4);
  \draw[dashed] (1.3,0) arc (20:160:0.4);
  \end{tikzpicture} \quad+\quad \begin{tikzpicture}[baseline,valign]
  \draw[thick] (0, 0) -- (1, 0);
  \draw[dashed](0.2, 0) -- (0.5, 1);
  \draw[dashed](0.8, 0) -- (0.5, 1);
  \end{tikzpicture} \quad+\quad \begin{tikzpicture}[baseline,valign]
  \draw[thick] (0, 0) -- (1.5, 0);
  \draw[dashed](0.3, 0) -- (0.7, 1);
  \draw[dashed](0.6, 0) -- (0.9, 0.6);
  \draw[dashed](0.9, 0) -- (0.9, 0.6);
  \draw[dashed](1.2, 0) -- (0.9, 0.6);
  \draw[dashed](0.7, 1) -- (0.9, 0.6);
  \node at (0.3, 0) [dcirc] {};
  \node at (0.9, 0) [dcirc] {};
  \node at (0.9, 0.6) [bcirc] {};
  \end{tikzpicture} \quad+\quad \begin{tikzpicture}[baseline,valign]
  \draw[thick] (0, 0) -- (1.5, 0);
  \draw[dashed](0.3, 0) -- (0.7, 1);
  \draw[dashed](0.6, 0) -- (0.9, 0.6);
  \draw[dashed](0.9, 0) -- (0.9, 0.6);
  \draw[dashed](1.2, 0) -- (0.9, 0.6);
  \draw[dashed](0.7, 1) -- (0.9, 0.6);
  \node at (0.6, 0) [dcirc] {};
  \node at (1.2, 0) [dcirc] {};
  \node at (0.9, 0.6) [bcirc] {};
  \end{tikzpicture} \quad+\quad \begin{tikzpicture}[baseline,valign]
  \draw[thick] (-0.5, 0) -- (0.5, 0);
  \draw[dashed] (-0.2, 0) -- (0, 0.4);
  \draw[dashed] (0.2, 0) -- (0, 0.4);
  \node at (0,0.4) [bcirc] {}; 
  \draw[dashed] (0,0.4) arc (230:130:0.4);
  \draw[dashed] (0,0.4) arc (310:410:0.4);
  \end{tikzpicture} \quad+\quad \dots
\ea
All the diagrams are built from diagrams previously computed, the last one being precisely the divergence of the Bloch-Wigner function~\eqref{eq:X1233}. We immediately obtain the expressions
\begin{align}
    \mathcal H^{\Phi^2}_{tt}(w)\, &= \,\frac{N+8}{4\sqrt{2N}}\Bigg(1+ \frac{8w}{N+8} - \veps \frac{\pi^2+4\big[\cosh^{-1}(\sqrt{w})\big]^2}{N+8} + \frac{4\veps (N+2)w\log w}{(N+8)^2} \nonumber\\
    & +\, \frac{2\veps(3+4w)\log2}{N+8}-\veps \frac{13N+38+24w(N+6)}{2(N+8)^2} \Bigg)\:, \\
   \mathcal H^{\Phi^2}_{\hat{\phi}_1\hat{\phi}_1}(w) \,&=\, \frac{N+8}{4\sqrt{2N}}\Bigg(1+ \frac{8w}{N+8} - 3\veps \frac{\pi^2+4\big[\cosh^{-1}(\sqrt{w})\big]^2}{N+8} + \frac{12\veps(N+6)w\log w}{(N+8)^2} \nonumber\\
   & +\, \frac{6\veps(1+4w)\log2}{N+8} -\veps \frac{13N+38+8w(7N+50)}{2(N+8)^2} \Bigg)\:, \\
    \mathcal H^{T,(1)}_{tt}(w) \,&=\, \frac{N+8}{4\sqrt{2}}\Bigg(1 - \frac{8w}{(N-1)(N+8)} - \veps \frac{\pi^2+4\big[\cosh^{-1}(\sqrt{w})\big]^2}{N+8} - \frac{8\veps\,w\log w}{(N-1)(N+8)^2}   \nonumber \\
    &+\, \veps\frac{(N+6)(N-1)-8w}{(N-1)(N+8)}\log2 + \veps \frac{(N-1)(N^2-5N-38)+16w(N+9)}{2(N-1)(N+8)^2} \Bigg),\\
    \mathcal H^{T,(2)}_{tt}(w) \,&=\,\sqrt{2}w\left(1 + \veps \frac{\log w}{N+8} +\veps\log2 -\veps \frac{N+9}{N+8}\right)\:,  \\
   \mathcal H^{T}_{t\hat{\phi}_1}(w) &= \frac{w}{\sqrt{2}}\left(1 - \veps \frac{\pi^2+4\big[\cosh^{-1}(\sqrt{w})\big]^2}{8w} + \frac{\veps(N+10)\log w}{2(N+8)} + 2\veps \log2 -\veps \frac{2N+17}{N+8} \right)\:, \\
   \mathcal H^{T}_{\hat{\phi}_1\hat{\phi}_1}(w) \,&=\, \frac{N+8}{4\sqrt{2}w}\Bigg(1+ \frac{8 w}{N+8} - 3\veps \frac{\pi^2+4\big[\cosh^{-1}(\sqrt{w})\big]^2}{N+8} + \veps\frac{8w(N+9)}{(N+8)^2}\log w \nonumber\\
    &+\, \frac{\veps(N+6+24w)\log2}{N+8} -\, \veps \frac{38+5N-N^2+16w(3N+25)}{2(N+8)^2} \Bigg)\:.
\end{align}

Note that there are two functions $\mathcal H^{T,(i)}_{tt}$ for the correlator $\langle T t t \rangle$, coming from two independent tensor structures
\begin{equation}
\label{eq:Tphiphi-structures}
\mathbb{T}^{(1)} = \frac{N-1}{N}\left(\delta_{I1}\delta_{J1}-\frac{P_{IJ}}{N-1}\right)\delta_{kl}\:, \qquad \mathbb{T}^{(2)} = \frac{\delta_{I(k}\delta_{l)J}}2-\frac{P_{IJ}\delta_{kl}}{N-1}\:.
\end{equation}
where we set $P_{IJ}=\delta_{IJ}-\delta_{I}^1 \delta_{J}^1$\,.
The expansion of these functions in conformal blocks gives us another check of the CFT data in Table~\ref{tab:CFT-data}, as well as new relations between the CFT data averaged over degenerate operators:
\begin{equation}
\begin{aligned}
\label{eq:mulambda-Phi2phiphi}
\sqrt{N}&\llangle \mu^{\Phi^2}_{\hat{\phi}^{2n}}\lambda_{tt\hat{\phi}^{2n}}\rrangle = \sqrt{N}\llangle \mu^{\Phi^2}_{\hat{\phi}^{2n}}\lambda_{\hat{\phi}_1\hat{\phi}_1\hat{\phi}^{2n}}\rrangle = -(N-1)\llangle \mu^{T}_{\hat{\phi}^{2n}}\lambda_{tt\hat{\phi}^{2n}}\rrangle^{(1)} \\ 
=\, &\llangle \mu^{T}_{\hat{\phi}^{2n}}\lambda_{tt\hat{\phi}^{2n}}\rrangle^{(2)} = \llangle \mu^{T}_{\hat{\phi}^{2n}}\lambda_{t\hat{\phi}_1\hat{\phi}^{2n}}\rrangle = \llangle \mu^{T}_{\hat{\phi}^{2n}}\lambda_{\hat{\phi}_1\hat{\phi}_1\hat{\phi}^{2n}}\rrangle = -\sqrt{8\pi}\left( \frac{-1}4\right)^{n} \frac{\Gamma(2n)}{\Gamma(2n-\frac12)}\:.
\end{aligned}    
\end{equation}

The form factor expansion even gives us the $\veps$ correction (that differs for each one), but we were unable to find a closed expression for all $n$. The $n=1$ state matches the CFT data of the operators $\hat{V}$, $\hat{T}$ and $\hat{s}_\pm$, and gives us non-trivial identities between the $\hat{s}_+$ and $\hat{s}_-$ contributions
\begin{align}
\llangle \mu_{\Phi^2\hat{\phi}^{2}}\lambda_{tt\hat{\phi}^{2}}\rrangle &= \mu_{\Phi^2 \hat{s}_+}\lambda_{tt \hat{s}_+} + \mu_{\Phi^2 \hat{s}_-}\lambda_{tt \hat{s}_-} = \sqrt{\frac{2}{N}}\left(1+\veps\log2 - \veps\frac{N+5}{N+8} \right)\,, \\
\llangle \mu_{\Phi^2\hat{\phi}^{2}}\lambda_{\hat{\phi}_1\hat{\phi}_1\hat{\phi}^{2}}\rrangle &= \mu_{\Phi^2 \hat{s}_+}\lambda_{\hat{\phi}_1\hat{\phi}_1 \hat{s}_+} + \mu_{\Phi^2 \hat{s}_-}\lambda_{\hat{\phi}_1\hat{\phi}_1 \hat{s}_-} = \sqrt{\frac{2}{N}}\left(1+3\veps\log2 - \veps\frac{2N+13}{N+8} \right)\,, \\
\llangle \mu_{T\hat{\phi}^{2}}\lambda_{tt\hat{\phi}^{2}}\rrangle^{(1)} &= \mu_{T \hat{s}_+}\lambda_{tt \hat{s}_+} + \mu_{T \hat{s}_-}\lambda_{tt\hat{s}_-} = -\frac{\sqrt{2}}{N-1}\left(1+\veps\log2 - \veps\frac{N^2+9N+10}{2(N+8)} \right)\,, \\
\llangle \mu_{T\hat{\phi}^{2}}\lambda_{\hat{\phi}_1\hat{\phi}_1\hat{\phi}^{2}}\rrangle &= \mu_{T \hat{s}_+}\lambda_{\hat{\phi}_1\hat{\phi}_1 \hat{s}_+} + \mu_{T \hat{s}_-}\lambda_{\hat{\phi}_1\hat{\phi}_1 \hat{s}_-} = \sqrt{2}\left( 1+3\veps\log 2 - \veps\frac{3N+26}{2(N+8)} \right)\,.
\end{align}

Moreover, we also find information on the conformal dimension of the tower operators
\begin{align}
\frac{\llangle \mu_{\Phi^2\hat{\phi}^{2n}}\lambda_{tt\hat{\phi}^{2n}}\Delta_{\hat{\phi}^{2n}}\rrangle}{\llangle \mu_{\Phi^2\hat{\phi}^{2n}}\lambda_{tt\hat{\phi}^{2n}}\rrangle} &=2n+\veps\frac{N+2}{N+8}, \\
\frac{\llangle \mu_{\Phi^2\hat{\phi}^{2n}}\lambda_{\hat{\phi}_1\hat{\phi}_1\hat{\phi}^{2n}}\Delta_{\hat{\phi}^{2n}}\rrangle}{\llangle \mu_{\Phi^2\hat{\phi}^{2n}}\lambda_{\hat{\phi}_1\hat{\phi}_1\hat{\phi}^{2n}}\rrangle} &=2n+3\veps\frac{N+6}{N+8}, \\   
\frac{\llangle \mu_{T\hat{\phi}^{2n}}\lambda_{tt\hat{\phi}^{2n}}\Delta_{\hat{\phi}^{2n}}\rrangle^{(1)}}{\llangle \mu_{T\hat{\phi}^{2n}}\lambda_{tt\hat{\phi}^{2n}}\rrangle^{(1)}} &=\frac{\llangle \mu_{T\hat{\phi}^{2n}}\lambda_{tt\hat{\phi}^{2n}}\Delta_{\hat{\phi}^{2n}}\rrangle^{(2)}}{\llangle \mu_{T\hat{\phi}^{2n}}\lambda_{tt\hat{\phi}^{2n}}\rrangle^{(2)}} =2n+\frac{2\veps}{N+8}, \\
\frac{\llangle \mu_{T\hat{\phi}^{2n}}\lambda_{t\hat{\phi}_1\hat{\phi}^{2n}}\Delta_{\hat{\phi}^{2n}}\rrangle}{\llangle \mu_{T\hat{\phi}^{2n}}\lambda_{t\hat{\phi}_1\hat{\phi}^{2n}}\rrangle} &=2n+\veps\frac{N+10}{N+8}, \\
\frac{\llangle \mu_{T\hat{\phi}^{2n}}\lambda_{\hat{\phi}_1\hat{\phi}_1\hat{\phi}^{2n}}\Delta_{\hat{\phi}^{2n}}\rrangle}{\llangle \mu_{T\hat{\phi}^{2n}}\lambda_{\hat{\phi}_1\hat{\phi}_1\hat{\phi}^{2n}}\rrangle} &=2n+2\veps\frac{N+9}{N+8}.
\end{align}

Similarly as above, the $n=1$ state provides a check of the conformal dimension of the operators $\hat{V}$, $\hat{T}$ and $\hat{s}_\pm$, and gives us non-trivial identities between the $\hat{s}_+$ and $\hat{s}_-$ contributions
\begin{align}
\frac{\llangle \mu_{\Phi^2\hat{\phi}^{2}}\lambda_{tt\hat{\phi}^{2}}\Delta_{\hat{\phi}^{2}}\rrangle}{\llangle \mu_{\Phi^2\hat{\phi}^{2}}\lambda_{tt\hat{\phi}^{2}}\rrangle} &=\frac{\mu_{\Phi^2 \hat{s}_+}\lambda_{tt \hat{s}_+}\Delta_{\hat{s}_+} + \mu_{\Phi^2 \hat{s}_-}\lambda_{tt \hat{s}_-}\Delta_{\hat{s}_-}}{\mu_{\Phi^2 \hat{s}_+}\lambda_{tt \hat{s}_+} + \mu_{\Phi^2 \hat{s}_-}\lambda_{tt \hat{s}_-}} = 2+\veps\frac{N+2}{N+8}, \\
\frac{\llangle \mu_{\Phi^2\hat{\phi}^{2}}\lambda_{\hat{\phi}_1\hat{\phi}_1\hat{\phi}^{2}}\Delta_{\hat{\phi}^{2}}\rrangle}{\llangle \mu_{\Phi^2\hat{\phi}^{2}}\lambda_{\hat{\phi}_1\hat{\phi}_1\hat{\phi}^{2}}\rrangle} &= \frac{\mu_{\Phi^2 \hat{s}_+}\lambda_{\hat{\phi}_1\hat{\phi}_1 \hat{s}_+}\Delta_{\hat{s}_+} + \mu_{\Phi^2 \hat{s}_-}\lambda_{\hat{\phi}_1\hat{\phi}_1 \hat{s}_-}\Delta_{\hat{s}_-}}{\mu_{\Phi^2 \hat{s}_+}\lambda_{\hat{\phi}_1\hat{\phi}_1 \hat{s}_+} + \mu_{\Phi^2 \hat{s}_-}\lambda_{\hat{\phi}_1\hat{\phi}_1 \hat{s}_-}} = 2+3\veps\frac{N+6}{N+8}, \\   
\frac{\llangle \mu_{T\hat{\phi}^{2}}\lambda_{tt\hat{\phi}^{2}}\Delta_{\hat{\phi}^{2}}\rrangle^{(1)}}{\llangle \mu_{T\hat{\phi}^{2}}\lambda_{tt\hat{\phi}^{2}}\rrangle^{(1)}} & = \frac{\mu_{T \hat{s}_+}\lambda_{tt\hat{s}_+}\Delta_{\hat{s}_+} + \mu_{T \hat{s}_-}\lambda_{tt \hat{s}_-}\Delta_{\hat{s}_-}}{\mu_{T \hat{s}_+}\lambda_{tt\hat{s}_+} + \mu_{T \hat{s}_-}\lambda_{tt\hat{s}_-}} = 2+\frac{2\veps}{N+8}, \\
\frac{\llangle \mu_{T\hat{\phi}^{2}}\lambda_{\hat{\phi}_1\hat{\phi}_1\hat{\phi}^{2}}\Delta_{\hat{\phi}^{2}}\rrangle}{\llangle \mu_{T\hat{\phi}^{2}}\lambda_{\hat{\phi}_1\hat{\phi}_1\hat{\phi}^{2}}\rrangle} &= \frac{\mu_{T \hat{s}_+}\lambda_{\hat{\phi}_1\hat{\phi}_1 \hat{s}_+}\Delta_{\hat{s}_+} + \mu_{T \hat{s}_-}\lambda_{\hat{\phi}_1\hat{\phi}_1 \hat{s}_-}\Delta_{\hat{s}_-}}{\mu_{T \hat{s}_+}\lambda_{\hat{\phi}_1\hat{\phi}_1 \hat{s}_+} + \mu_{T \hat{s}_-}\lambda_{\hat{\phi}_1\hat{\phi}_1 \hat{s}_-}} = 2+2\veps\frac{N+9}{N+8}.
\end{align}

\subsection{Bulk current operator}

\subsubsection{Operator renormalization}
\label{sssec:current-renorm}

As before, we first renormalize the new operators before computing their CFT data. 

\begin{itemize}
    \item For the bulk current, the tree diagram is
\begin{equation}
\label{eq:2pt-J}
    \langle J_{\mu IJ}(x_1)J_{\nu KL}(x_2) \rangle = \frac{2\kappa^2(2-\veps)}{(x_{12}^2)^{3-\veps}} (\delta_{IK}\delta_{JL}-\delta_{IL}\delta_{JK})\left(\delta_{\mu\nu}- 2 \frac{x_{12\mu}x_{12\nu}}{x_{12}^2}\right).
\end{equation}

The correlator comes with the tensor structure $\left(\delta_{\mu\nu}- 2 \frac{x_{12\mu}x_{12\nu}}{x_{12}^2}\right)$, due to the spin of the current. There is no diagram at one-loop, so the renormalization is directly
\begin{equation}
    Z_{J} = 1, \quad \Delta_{J} = 3-\veps, \quad \mathcal{N}_{J}^2 = 2\kappa^2(2-\veps).
\end{equation}

    \item For the new operators $\hat{A}$ and $\hat{B}$, the tree-level diagram is simply the limit $y_1,y_2\rightarrow0$ of~\eqref{eq:2pt-J} with $\mu=\nu=x$. Note that for $p=1$, the tensor structure collapses to~$(-1)$. There is also a second diagram at one-loop
\begin{multline}
\hspace{-1cm}\begin{tikzpicture}[baseline,valign]
  \draw[thick] (-0.15, 0) -- (1.1, 0);
  \draw[dashed] (0.9,0) arc (20:160:0.45);
  \draw[dashed] (0.9,0) arc (0:180:0.42);
  \draw[dashed](0.3, 0) -- (0.5, 0.3);
  \draw[dashed](0.7, 0) -- (0.5, 0.3);
  \node at (0.3, 0) [dcirc] {};
  \node at (0.7, 0) [dcirc] {};
  \node at (0.5, 0.3) [bcirc] {};
\end{tikzpicture} = -\frac{8\kappa^2(1-\veps+\veps^2)}{(x_{12}^2)^{3-2\veps}}
\left(\frac{-\blambda h^2}{6}\right)\left(\frac1\veps + \aleph \right)\big(\delta_{K[I}\delta_{J]L}+\delta_{1[I}\delta_{J]L}\delta_{K1}+\delta_{K[I}\delta_{J]1}\delta_{L1}\big) \:.
\end{multline}

Summing the two, this yields the renormalization
\begin{align}
    Z_{\hat{A}} &= 1- \frac{\blambda h^2}{6\veps}, \quad \Delta_{\hat{A}} =3, \qquad \mathcal{N}^2_{\hat{A}} = 4\kappa^2(1-\aleph\veps), \\
    Z_{\hat{B}} &= 1- \frac{\blambda h^2}{3\veps},\quad \Delta_{\hat{B}} =3+\veps, \quad \mathcal{N}^2_{\hat{B}} = 4\kappa^2\left(1+\frac{\veps}2(1-4\aleph)\right).
\end{align}
\end{itemize}

\subsubsection{The correlator \texorpdfstring{$\langle (\hat{\phi}_I\overset{\leftrightarrow}{\partial}_x\hat{\phi}_J)\hat{\phi}_K \hat{\phi}_L  \rangle$}{<A phi phi> and <B phi phi>}}
\label{sssec:A/Bphiphi}

We have two contributions at one-loop
\begin{equation}
\langle (\hat{\phi}_I\overset{\leftrightarrow}{\partial}_x\hat{\phi}_J)(x_1)\hat{\phi}_K(x_2)\hat{\phi}_L(x_3)\rangle \quad=\quad 
  \begin{tikzpicture}[baseline,valign]
  \draw[thick] (-0.15, 0) -- (1.3, 0);
  \draw[dashed] (0.85,0) arc (20:160:0.45);
  \draw[dashed] (1.1,0) arc (0:180:0.55);
  \end{tikzpicture} \quad+\quad
  \begin{tikzpicture}[baseline,valign]
  \draw[thick] (-0.15, 0) -- (1.3, 0);
  \draw[dashed] (0.85,0) arc (20:160:0.45);
  \draw[dashed] (1.1,0) arc (0:180:0.55);
  \draw[dashed](0.25, 0) -- (0.45, 0.3);
  \draw[dashed](0.65, 0) -- (0.45, 0.3);
  \node at (0.25, 0) [dcirc] {};
  \node at (0.65, 0) [dcirc] {};
  \node at (0.45, 0.3) [bcirc] {};
  \end{tikzpicture} \quad+\quad \dots \ ,
\end{equation}
that give the following contributions
\begin{align}
\begin{tikzpicture}[baseline,valign]
  \draw[thick] (-0.15, 0) -- (1.3, 0);
  \draw[dashed] (0.85,0) arc (20:160:0.45);
  \draw[dashed] (1.1,0) arc (0:180:0.55);
\end{tikzpicture} &\ =\  \frac{\kappa^2(2-\veps)\operatorname{sgn}(x_{12}x_{23}x_{31})}{(x_{12}^2 x_{13}^2)^\frac{3-\veps}{2}(x_{23}^2)^{-\frac12}} (\delta_{IK}\delta_{JL}-\delta_{IL}\delta_{JK}) \: , \\
\begin{tikzpicture}[baseline,valign]
  \draw[thick] (-0.15, 0) -- (1.3, 0);
  \draw[dashed] (0.85,0) arc (20:160:0.45);
  \draw[dashed] (1.1,0) arc (0:180:0.55);
  \draw[dashed](0.25, 0) -- (0.45, 0.3);
  \draw[dashed](0.65, 0) -- (0.45, 0.3);
  \node at (0.25, 0) [dcirc] {};
  \node at (0.65, 0) [dcirc] {};
  \node at (0.45, 0.3) [bcirc] {};
  \end{tikzpicture} & \ =\ \frac{\kappa^2(2-\veps)\operatorname{sgn}(x_{12}x_{23}x_{31})}{(x_{12}^2 x_{13}^2)^\frac{3-\veps}{2}(x_{23}^2)^{-\frac12}}\left(\frac{-\blambda h^2}{6}\right) \left(\frac1{\veps} + \aleph\right) \left[\begin{array}{c} \big(\delta_{K[I}\delta_{J]L}+2\delta_{1[I}\delta_{J]L}\delta_{K1}\big)(x_{12}^2)^\veps \\
  \big(\delta_{K[I}\delta_{J]L}+2\delta_{K[I}\delta_{J]1}\delta_{L1}\big)(x_{13}^2)^\veps 
  \end{array} \right] \\
  &  \hspace{-1cm}-\frac{2\kappa^2\operatorname{sgn}(x_{12}x_{23}x_{31})}{(x_{12}^2 x_{13}^2)^\frac{3}{2}(x_{23}^2)^{-\frac12}} \left(\frac{-\blambda h^2}{6}\right)\big(\delta_{K[I}\delta_{J]L}+\delta_{1[I}\delta_{J]L}\delta_{K1}+\delta_{K[I}\delta_{J]1}\delta_{L1}\big) \ +\ \text{(descendants)} \:.\nonumber 
\end{align}

We can extract the OPE coefficients:
\begin{equation}
    \lambda_{\hat A tt}=1\ , \qquad \lambda_{\hat{B}t\hat{\phi}_1}=1+\frac\veps4\:.
\end{equation}

Note that there is a subtlety in the computation: when $\hat\op_1$ and $\hat\op_2$ do not have the same conformal dimensions, $\hat{\op}_1\overset{\leftrightarrow}{\partial}_x\hat{\op}_2$ is not a conformal primary but it has a correction proportional to the descendant $\Delta_{12}\partial_x(\hat\op_1\hat\op_2)$. This is not an issue for $\hat A$, but it is for~$\hat B$ at order $\op(\veps)$. The (descendants) term we have written up comes precisely from this mismatch: it will not appear in the correlator $\langle \hat Bt\hat\phi_1\rangle$ with the proper operator $\hat B$. Note that this correction would affect $\Delta_{\hat B}$ and $\mu^J_{\hat B}$  only at order $\op(\veps^2)$.

\subsubsection{The correlator \texorpdfstring{$\langle J_{\mu IJ}\, t_k \rangle$}{<Jt>}}
\label{sssec:Jt}

The correlator has the form~\eqref{eq:Jt} with the tensor structure~\eqref{eq:structure-JO}.
There are three diagrams at order $\op(\veps)$, that look like the diagrams of $\langle \Phi^2 t\rangle$ that have been computed in section~\ref{ssec:Phi2phi}. In this case $J_{\mu IJ}$ contains a derivative, leading to the different expansions
\begin{align}
\begin{tikzpicture}[baseline,valign]
  \draw[thick] (0, 0) -- (1.2, 0);
  \draw[dashed](0.3, 0) -- (0.6, 1);
  \draw[dashed](0.9, 0) -- (0.6, 1);
  \node at (0.9, 0) [dcirc] {};
\end{tikzpicture} & \ =\ - \frac{\kappa^{\frac32}h_0(2-\veps)}{|y_1|^{2-\veps}(y_1^2+x_{12}^2)^{1-\frac\veps2}}(\delta_{Ik}\delta_{J1}-\delta_{Jk}\delta_{I1}) \left(\frac12 + \frac\veps8 (\aleph-1+4\log2) \right) \nonumber \\
& \hspace{3cm} \left(\mathbb{T}_\mu^{J\op} + \veps\delta_{\mu a}\frac{y_{1a}}{|y_1|}\ \frac{x_{12}^2}{y_1^2+x_{12}^2}\right), \\
\begin{tikzpicture}[baseline,valign]
  \draw[thick] (0, 0) -- (1.5, 0);
  \draw[dashed](0.3, 0) -- (0.7, 1);
  \draw[dashed](0.6, 0) -- (0.9, 0.6);
  \draw[dashed](0.9, 0) -- (0.9, 0.6);
  \draw[dashed](1.2, 0) -- (0.9, 0.6);
  \draw[dashed](0.7, 1) -- (0.9, 0.6);
  \node at (0.6, 0) [dcirc] {};
  \node at (0.9, 0) [dcirc] {};
  \node at (1.2, 0) [dcirc] {};
  \node at (0.9, 0.6) [bcirc] {};
\end{tikzpicture}& \ =\ \frac{\kappa^{\frac32}h_0(2-\veps)}{|y_1|^{2-3\veps}(y_1^2+x_{12}^2)^{1-\frac\veps2}}(\delta_{Ik}\delta_{J1}-\delta_{Jk}\delta_{I1})\left(\frac{\blambda h^2}{6}\right) \left(\frac1{4\veps} + \frac{7+5\aleph+12\log2}{16} \right) \nonumber \\
&\hspace{3cm}\left(\mathbb{T}_\mu^{J\op} + \veps\delta_{\mu a}\frac{y_{1a}}{|y_1|}\ \frac{2|y_1|^2+3x_{12}^2}{y_1^2+x_{12}^2}\right), \\ 
\begin{tikzpicture}[baseline,valign]
  \draw[thick] (0, 0) -- (1.5, 0);
  \draw[dashed](0.3, 0) -- (0.7, 1);
  \draw[dashed](0.6, 0) -- (0.9, 0.6);
  \draw[dashed](0.9, 0) -- (0.9, 0.6);
  \draw[dashed](1.2, 0) -- (0.9, 0.6);
  \draw[dashed](0.7, 1) -- (0.9, 0.6);
  \node at (0.3, 0) [dcirc] {};
  \node at (0.6, 0) [dcirc] {};
  \node at (1.2, 0) [dcirc] {};
  \node at (0.9, 0.6) [bcirc] {};
\end{tikzpicture}& \ =\ \frac{\kappa^{\frac32}h_0(2-5\veps)}{|y_1|^{2+\veps}(y_1^2+x_{12}^2)^{1-\frac{5\veps}2}}(\delta_{Ik}\delta_{J1}-\delta_{Jk}\delta_{I1})\left(\frac{\blambda h^2}{6}\right) \left(\frac1{4\veps} + \frac{7+5\aleph-4\log2}{16} \right) \nonumber \\
&\hspace{3cm} \left(\mathbb{T}_\mu^{J\op} + \veps\delta_{\mu a}\frac{y_{1a}}{|y_1|}\ \frac{-2|y_1|^2+3x_{12}^2}{y_1^2+x_{12}^2}\right)\:.
\end{align}

We see that the transverse part structure~$\mathbb{T}_a^{J\op}$ is corrected at order $\op(\veps)$ in each diagram individually. The cancellation of these corrections in the final correlator provides a good check of the computations.

\subsubsection{The correlator \texorpdfstring{$\langle J_{\mu IJ}\, (\hat{\phi}_K\overset{\leftrightarrow}{\partial}_x\hat{\phi}_L) \rangle$}{<JA> and <JB>}}
\label{sssec:JA/B}

The second type of DOE coefficient we need come from the correlator $\langle J_{\mu IJ}\, (\hat{\phi}_K\overset{\leftrightarrow}{\partial}_x\hat{\phi}_L) \rangle$, which has the form~\eqref{eq:JOm} with the tensor structure~\eqref{eq:structure-JOm}.
At $O(\veps)$, there are two contributing diagrams:
\begin{equation}
\langle J_{\mu IJ}(y_1,x_1)(\hat{\phi}_K\overset{\leftrightarrow}{\partial}_x\hat{\phi}_L)(x_2)\rangle \quad=\quad 
  \begin{tikzpicture}[baseline,valign]
  \draw[thick] (0, 0) -- (1, 0);
  \draw[dashed] (0.5,0) arc (210:150:0.9);
  \draw[dashed] (0.5,0) arc (330:390:0.9);
  \end{tikzpicture}\quad+\quad
  \begin{tikzpicture}[baseline,valign]
  \draw[thick] (0, 0) -- (1.2, 0);
  \draw[dashed](0.6, 0) -- (0.6, 1);
  \draw[dashed](0.3, 0) -- (0.6, 0.5);
  \draw[dashed](0.9, 0) -- (0.6, 0.5);
  \draw[dashed] (0.9,0) arc (340:415:0.82);
  \node at (0.3, 0) [dcirc] {};
  \node at (0.6, 0) [dcirc] {};
  \node at (0.6, 0.5) [bcirc] {};
  \end{tikzpicture}\quad+\quad \dots
\end{equation}
The first diagram is simply the $y_2 \rightarrow0$ limit of the two-point function of two bulk currents~\eqref{eq:2pt-J}. The second diagram has the following expansion:
\begin{equation}
\begin{aligned}
\begin{tikzpicture}[baseline,valign]
  \draw[thick] (0, 0) -- (1.2, 0);
  \draw[dashed](0.6, 0) -- (0.6, 1);
  \draw[dashed](0.3, 0) -- (0.6, 0.5);
  \draw[dashed](0.9, 0) -- (0.6, 0.5);
  \draw[dashed] (0.9,0) arc (340:415:0.82);
  \node at (0.3, 0) [dcirc] {};
  \node at (0.6, 0) [dcirc] {};
  \node at (0.6, 0.5) [bcirc] {};
\end{tikzpicture} \ =\ &\frac{4\kappa^2(2-3\veps)}{|y_1|^{2\veps}(y_1^2+x_{12}^2)^{3-3\veps}}
  \left(\frac{-\blambda h^2}{6}\right) \left(\frac1{2\veps} + 1+\frac\aleph2-\log2 \right) \mathbb{T}_{\mu0}^{J\op_m} \\
  &\big(\delta_{K[I}\delta_{J]L}+\delta_{1[I}\delta_{J]L}\delta_{K1}+\delta_{K[I}\delta_{J]1}\delta_{L1}\big)       
\end{aligned}
\end{equation}

\subsubsection{The correlator \texorpdfstring{$\langle J_{\mu IJ}\, \hat{\phi}_K\hat{\phi}_L \rangle$}{<Jt>}}
\label{sssec:JOO}

There are three diagrams at order $\op(\veps)$, that have the expressions \begin{align}
\begin{tikzpicture}[baseline,valign]
  \draw[thick] (0, 0) -- (1, 0);
  \draw[dashed](0.2, 0) -- (0.5, 1);
  \draw[dashed](0.8, 0) -- (0.5, 1);
\end{tikzpicture} &\ =\ \frac{\kappa^2(2-\veps)\mathbb{T}_\mu^{(-)}}{|y_1|(y_1^2+x_{12}^2)(y_1^2+x_{13}^2)^{1-\frac{\veps}{2}}} (\delta_{IK}\delta_{JL}-\delta_{IL}\delta_{JK})  \:, \\
\begin{tikzpicture}[baseline,valign]
  \draw[thick] (0, 0) -- (1.5, 0);
  \draw[dashed](0.3, 0) -- (0.7, 1);
  \draw[dashed](0.6, 0) -- (0.9, 0.6);
  \draw[dashed](0.9, 0) -- (0.9, 0.6);
  \draw[dashed](1.2, 0) -- (0.9, 0.6);
  \draw[dashed](0.7, 1) -- (0.9, 0.6);
  \node at (0.3, 0) [dcirc] {};
  \node at (0.9, 0) [dcirc] {};
  \node at (0.9, 0.6) [bcirc] {};
\end{tikzpicture}&\ =\ \frac{2\kappa^2 \mathbb{T}_\mu^{(+)}}{|y_1|(y_1^2+x_{12}^2)(y_1^2+x_{13}^2)}\left(\frac{-\blambda h^2}{6}\right) \frac{\cosh^{-1}\sqrt{w}}{\sqrt{w(w-1)}} \big(\delta_{K[I}\delta_{J]1}\delta_{L1}-\delta_{1[I}\delta_{J]L}\delta_{K1}\big) \\
\begin{tikzpicture}[baseline,valign]
  \draw[thick] (0, 0) -- (1.5, 0);
  \draw[dashed](0.3, 0) -- (0.7, 1);
  \draw[dashed](0.6, 0) -- (0.9, 0.6);
  \draw[dashed](0.9, 0) -- (0.9, 0.6);
  \draw[dashed](1.2, 0) -- (0.9, 0.6);
  \draw[dashed](0.7, 1) -- (0.9, 0.6);
  \node at (0.6, 0) [dcirc] {};
  \node at (1.2, 0) [dcirc] {};
  \node at (0.9, 0.6) [bcirc] {};
\end{tikzpicture}&\ =\ \frac{\kappa^2(2-\veps) \mathbb{T}_\mu^{(-)}}{|y_1|^{1+2\veps}(y_1^2+x_{12}^2)(y_1^2+x_{13}^2)^{1-\frac{\veps}{2}}} 
\left(\frac{-\blambda h^2}{6}\right) \left(\frac1{2\veps} + 1+ \frac\aleph2 -\log2\right) \nonumber \\
&\Big[\big(\delta_{K[I}\delta_{J]L}+2\delta_{1[I}\delta_{J]L}\delta_{K1}\big)(y_1^2+x_{12}^2)^\veps + \big(\delta_{K[I}\delta_{J]L}+2\delta_{K[I}\delta_{J]1}\delta_{L1}\big)(y_1^2+x_{13}^2)^\veps \Big] \\
&\quad -\frac{2\kappa^2}{|y_1|(y_1^2+x_{12}^2)(y_1^2+x_{13}^2)}\left(\frac{-\blambda h^2}{6}\right)
\left[\begin{array}{rl}
    \mathbb{T}_\mu^{(-)}&\hspace{-0.3cm}\big(\delta_{K[I}\delta_{J]L}+\delta_{1[I}\delta_{J]L}\delta_{K1}+\delta_{K[I}\delta_{J]1}\delta_{L1}\big) \\
    +\ \mathbb{T}_\mu^{(+)}&\hspace{-0.3cm}\big(\delta_{1[I}\delta_{J]L}\delta_{K1}-\delta_{K[I}\delta_{J]1}\delta_{L1}\big)
  \end{array}\right] \: . \nonumber 
\end{align}

\newpage

%----- Bibliography ----------------------

\providecommand{\href}[2]{#2}\begingroup\raggedright\endgroup

\end{document}